\documentclass[manuscript,screen, natbib=True]{acmart}

\usepackage{multirow}
\usepackage[table]{xcolor}

\usepackage{tikz}
\usepackage{pgfplots}
\pgfplotsset{compat=1.18}
\usetikzlibrary{shapes.misc}

\usepackage{algorithm2e}

\usepackage{pifont}
\definecolor{dark-green}{rgb}{0.0, 0.5, 0.0}
\newcommand{\cm}{{\color{dark-green}\checkmark}}
\newcommand{\xm}{{\color{red}\ding{55}}}

\usepackage{cleveref}
\Crefname{appendix}{Appendix}{Appendices}

\theoremstyle{definition}
\newtheorem{definition}{Definition}

\newcommand{\joint}{\textsc{Joint}}
\newcommand{\rel}{\textsc{Rel}}
\newcommand{\fair}{\textsc{Fair}}

\makeatletter
\AddToHook{cmd/appendix/before}{\def\cref@section@alias{appendix}\def\cref@subsection@alias{appendix}}
\makeatother

\DeclareMathOperator*{\argminA}{arg\,min}
\begin{document}

\title{Robustness Analysis of Pareto-based Joint Evaluation of Fairness and Relevance in Recommender Systems}

\author{Theresia Veronika Rampisela}
\orcid{0000-0003-1233-7690}
\email{thra@di.ku.dk}
\affiliation{%
  \institution{University of Copenhagen}
  \city{Copenhagen}
  \country{Denmark}}

\author{Maria Maistro}
\orcid{0000-0002-7001-4817}
\email{mm@di.ku.dk}
\affiliation{%
 \institution{University of Copenhagen}
 \city{Copenhagen}
 \country{Denmark}
}

\author{Tuukka Ruotsalo}
\email{tr@di.ku.dk}
\orcid{0000-0002-2203-4928}
\affiliation{%
 \institution{University of Copenhagen}
 \city{Copenhagen}
 \country{Denmark}
}
\affiliation{%
 \institution{LUT University}
 \city{Lahti}
 \country{Finland}
}

\author{Christina Lioma}
\email{c.lioma@di.ku.dk}
\orcid{0000-0003-2600-2701}
\affiliation{%
 \institution{University of Copenhagen}
 \city{Copenhagen}
 \country{Denmark}
}

\begin{abstract}
Recommender Systems (RSs) can be evaluated for many aspects, two of them being recommendation relevance and fairness. These two aspects can be evaluated either (i) in isolation, using a measure for each aspect; or (ii) jointly with a single `relevance-aware' fairness measure, i.e., a measure that quantifies fairness with respect to relevance. 
Yet, neither approach provides a simultaneous evaluation of both relevance and fairness: the first approach may result in two different rankings of models, each corresponding to an evaluation aspect; the second approach often provides an imbalanced evaluation of the two aspects. To address this gap, we propose Distance to Pareto Frontier (DPFR), a new approach to jointly evaluate RS fairness and relevance. DPFR is based on a Pareto frontier of relevance and fairness scores from existing measures. The frontier is generated from user-item interaction data and represents various Pareto-optimal trade-offs between the two aspects. Thus, the distance from a point in the frontier provides a joint score of fairness and relevance. As DPFR can be used with existing measures, it is modular and intuitive. Our experiments with 4 recommenders and 3 fairness-aware re-rankers on 6 real-world datasets reveal inconsistent associations between existing measures and DPFR, showing that they are not reliable proxies for evaluations based on Pareto optimality. We further examine the robustness of DPFR as a joint evaluation approach for fairness and relevance by conducting experiments with artificial recommendations and with various fairness-relevance trade-offs, as well as by analysing DPFR's agreement to simple measures of fairness and relevance. These analyses show that DPFR is more robust than existing joint evaluation approaches. Finally, we provide guidelines for evaluating fairness and relevance jointly with DPFR.
Our code: \href{https://github.com/theresiavr/robustness-DPFR-recsys-evaluation}{github.com/theresiavr/robustness-DPFR-recsys-evaluation}.
\end{abstract}



\maketitle

\section{Introduction}
\label{s:intro}

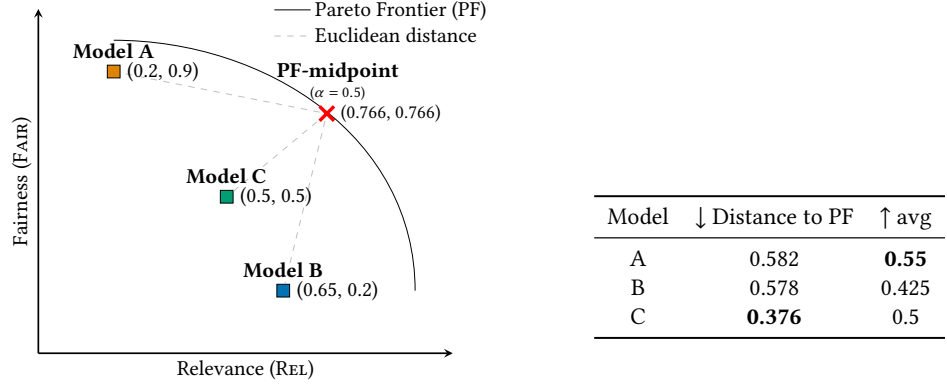
\begin{figure}[!h]
    \centering
    \begin{minipage}{.6\columnwidth}
        \centering
        \scalebox{.8}{
        \begin{tikzpicture}
            \tikzstyle{blue rectangle}=[fill={rgb,255: red,1; green,115; blue,178}, draw=black, shape=rectangle]
            \tikzstyle{green circle pareto}=[fill={rgb,255: red,2; green,158; blue,115}, draw=black, shape=rectangle]
            \tikzstyle{orange rectangle}=[fill={rgb,255: red,222; green,132; blue,5}, draw=black, shape=rectangle]
            \tikzstyle{pink rectangle}=[fill={rgb,255: red,204; green,120; blue,188}, draw=black, shape=rectangle]
            \tikzset{cross/.style={cross out, draw=black, minimum size=2*(#1-\pgflinewidth), inner sep=0pt, outer sep=0pt}, 
            cross/.default={5pt}}
            \begin{axis}
            [
                xmin = 0,
                ymin = 0,
                xmax = 1+0.1,
                ymax = 1+0.1,
                xlabel={\large Relevance (\rel{})}, 
                ylabel={\large Fairness (\fair{})}, 
                ticks=none,
                clip=false,
                axis lines=left,
                axis line style=thick,
                x label style={at={(axis description cs:0.5,0)},anchor=north},
                y label style={at={(axis description cs:0,.5)},anchor=south},
                legend style={at={(0.83,0.87)}, anchor=south,legend columns=1},
                legend style={font=\small, draw=none},
                legend cell align={left}
            ]

            \addlegendimage{}
            \addlegendimage{dashed,gray!50}

            \node (0) at (0.2, 1) {};
            \node (1) at (1, 0.2) {};
            \node [style=green circle pareto] (9) at (0.5, 0.5) {};
            \node [style=orange rectangle] (bestF) at (0.2, 0.9) {};
            \node [style=blue rectangle] (bestR) at (0.65, 0.2) {};
            \draw [in=90, out=0] (0.center) to (1.center);
            \addlegendentry{\large Pareto Frontier (PF)};
            \draw (0.7656854249492, 0.7656854249492) 
            node[cross,red, ultra thick,
            label={[align=center]0:{\normalsize(0.766, 0.766)}}, 
            label={[align=center, xshift=5pt]88:{\large\textbf{PF-midpoint}}\\[-2pt]{\footnotesize($\alpha=0.5$)}}] (10) {};
            \draw[dashed, color=gray!50] (9) -- (10);
            \addlegendentry{\large Euclidean distance};
            \draw[dashed, color=gray!50] (10) -- (bestF);
            \draw[dashed, color=gray!50] (10) -- (bestR);
            \node[label={\large \textbf{Model C}}, label={0:{\large(0.5, 0.5)}}]  at (9) {};
            \node[label={\large \textbf{Model A}}, label={0:{\large(0.2, 0.9)}}]  at (bestF) {};
            \node[label={\large \textbf{Model B}}, label={0:{\large(0.65, 0.2)}}]  at (bestR) {};
            \end{axis}%
        \end{tikzpicture}
        }
    \end{minipage}%
    \begin{minipage}[t][][b]{0.4\columnwidth}

        \begin{tabular}{ccc}
            \toprule
            Model & $\downarrow$ Distance to PF & $\uparrow$ avg\\
            \midrule
            A & 0.582 & \textbf{0.55} \\
            B & 0.578 & 0.425 \\
            C & \textbf{0.376} & 0.5 \\
            \bottomrule
        \end{tabular}
    \end{minipage}
    
    \caption{
    $(x, y)$ denotes the pair of relevance and fairness scores. 
    Example: 
    Model A is best for fairness, 
    Model B is best for relevance, and Model C is the closest 
    to the Pareto Frontier (PF) midpoint, when relevance and fairness are equally weighted ($\alpha=0.5$). 
    Averaging relevance and fairness (avg) leads to falsely concluding that Model A is best for both aspects. Note that distance to PF also quantifies both aspects better than other existing measures of fairness and relevance (see \Cref{ss:corr}).
    }
    \label{fig:pareto_teaser}

\end{figure}

When evaluating recommender systems (RSs), it is important to know how relevant and fair their recommendations are. These two evaluation concepts, RS relevance and fairness, are sometimes studied together \cite{Rampisela2024CanRelevance}, however, they are inherently different from one another. Relevance evaluation aims to capture the proximity of a recommended item to the user need (as represented in the user profile). Fairness evaluation, on the other hand, aims to capture how impartial a recommendation is, in the sense that no items or users should be subjected to consistently favourable or unfavourable recommendations, compared to the other items or users in the RS.  We focus solely on \textbf{individual item fairness} measures, which quantify the exposure of items in the recommendations. Exposure typically refers to the frequency of item appearance in the recommendation list across all users~\cite{Patro2020FairRec:Platforms, Mansoury2020FairMatch:Systems, Rampisela2024EvaluationStudy}. Individual item fairness is important in ensuring that each item in the system has a chance to be recommended to any user.

The inherent difference between relevance and fairness makes their simultaneous evaluation challenging. One approach is to use a single evaluation measure that accounts for both fairness and relevance; however, such \textit{joint measures} tend to be ad hoc, unstable, and oftentimes imbalanced in their quantification of relevance versus fairness \cite{Rampisela2024CanRelevance,Rampisela2025relevanceguidelines}. An alternative approach is to measure relevance and fairness separately, and then aggregate the scores into a single numerical score. There are at least three issues with this approach. Firstly, it is not trivial to combine scores that have different distributions and different scales, which is often the case with existing relevance and fairness evaluation measures. Secondly, it is not trivial to combine scores produced by measures that may not even be computed on the same input. This is often the case with relevance and fairness measures, because relevance scores are computed for individual users and then averaged, while fairness measures for individual items are typically based on individual item recommendation frequency. Finally, even if one merges relevance and fairness scores into a single numerical score, it is not clear how this single score should be interpreted, for instance how well this score reflects the distance from an ideal balance of relevance and fairness, or from an acceptable trade-off of relevance and fairness. 

We contribute a novel approach of evaluating relevance and fairness, which uses Pareto-optimality \cite{Censor1977ParetoProblems}, a well-studied concept in multi-objective optimisation problems across domains, including RSs \cite{Ribeiro2015MultiobjectiveSystems}. A recommendation is said to be Pareto-optimal if there is no other possible recommendation with the same relevance (\rel{}) score that achieves better fairness. 
Practically, this means that when a recommendation is Pareto-optimal, it is impossible to have other recommendations that empirically perform better in terms of fairness unless relevance is compromised. Note that the opposite is also true: a recommendation is Pareto-optimal if there is no other recommendation with the same fairness (\fair{}) score and a better \rel{} score.
Our approach, called \emph{Distance to PF of Fairness and Relevance} (DPFR), combines existing \fair{} measures and \rel{} measures in the following way: first, we build a Pareto Frontier (PF) that maximises relevance; then we find the best fairness achievable under this relevance constraint; and finally, we jointly quantify fairness and relevance as the distance from an optimal solution. The evaluation process with DPFR is illustrated in \Cref{fig:pareto_teaser}. 

Our approach has several advantages. While it can be used with existing evaluation measures of relevance and fairness, it avoids the problems arising from directly combining their numerical scores. Instead, it computes the distance of these scores to a desired fairness-relevance balance, which is empirically optimal. Furthermore, the weight of fairness w.r.t.~relevance in this desired fairness-relevance balance is controlled by a parameter ($\alpha$). The above aspects make our approach modular, tractable, and intuitive in its interpretation as a quantification of relevance and fairness. 

In this work, we ask the following research questions (RQs):

\begin{itemize}
    \item[\textbf{RQ1}] How does our generated PF compare to the true PF (i.e., the PF obtained from exhaustive search of all solutions)?
    \item[\textbf{RQ2}] How do existing single-aspect and joint evaluation measures of relevance and fairness relate to DPFR?
    \item[\textbf{RQ3}] How does the choice of distance measure affect DPFR?
    \item[\textbf{RQ4}] How does the choice of $\alpha$, the fairness-relevance trade-off weight, affect DPFR?
    \item[\textbf{RQ5}] How does DPFR relate to simple measures of effectiveness and fairness that are not used for computing DPFR?
    \item[\textbf{RQ6}] How can DPFR be efficiently approximated without significant loss in effectiveness?
\end{itemize}

We present thorough experiments with various RS models, re-ranking approaches, and datasets, using different evaluation measures of relevance and fairness. Our experiments reveal a noticeable gap between current joint evaluation approaches for relevance and fairness and our Pareto-optimal joint evaluation. This gap increases for larger datasets and when using rank-based relevance measures (i.e., MAP, NDCG), as opposed to set-based relevance measures (i.e., Precision, Recall). 

This work extends our previous publication on this topic \cite{Rampisela2025Pareto} by contributing further analyses and discussions on the robustness of existing approaches for joint fairness-relevance evaluation, including DPFR. Specifically:

\begin{itemize}
    \item We evaluate the closeness of our algorithm's generated PF to the true PF, which is the set of Pareto-optimal solutions obtained through exhaustive enumerations (\Cref{ss:true_pf});

    \item We conduct experiments with an artificial recommender to complement experiments with recommenders trained on real-world data (\Cref{sss:best_model_disagreement}); 
    \item We discuss and empirically investigate the use of various distance measures for computing DPFR (\Cref{ss:sensitivity_distance});
    
    \item We study the effect of varying the fairness-relevance trade-off weight (\Cref{ss:varying_alpha}); and 
    
    \item We analyse their correlation to simple measures of relevance and fairness (\Cref{ss:simple}).
\end{itemize}

Our results show that DPFR can quantify fairness and relevance jointly in a more robust way, compared to existing joint evaluation approaches. Beyond robustness-related analyses, we also discuss theoretical considerations as to why existing multi-aspect evaluation approaches do not suit the specific evaluation problem that we tackle
(\Cref{ss:multi_aspect_evaluation}, summarised in \Cref{tab:eval_approach}). Finally, we contribute a set of guidelines on when and how to use DPFR, summarising several consideration factors for selecting single- and multi-aspect evaluation approaches (\Cref{s:guidelines}).

\section{Related Work}
\label{s:prev_work}

\subsection{Fairness Evaluation in Recommender Systems}
\label{ss:fairness_evaluation}

In RSs, fairness is usually defined as an equal treatment of users and/or items, i.e., avoiding discrimination of users and/or items~\cite{Biega2018EquityRankings}.
Fairness definitions are commonly categorised based on the subjects, e.g., items or users, and the partition granularity of the subjects, e.g., individuals or groups~\cite{Amigo2023ASystems,Wang2022}.
Item fairness is typically defined as the exposure disparity across items or types of items.
The exposure of an item is estimated based on how frequently the item is recommended and at which rank positions.
The more frequently an item is recommended and the closer the item is to the top of the ranking, the higher the amount of exposure associated with that item.
Item fairness can also be defined by accounting for exposure and relevance simultaneously.
The intuition is that an RS should not only provide equal exposure to items, but also recommend items that are relevant to the users.
While item fairness is defined on exposure disparity, user fairness is usually defined on utility disparity, where utility is estimated with the effectiveness of the RS.
In an ideally fair case, all users, or groups of users, should receive recommendations with the same level of effectiveness.

Fairness granularity, i.e., individual vs.~groups, refers to how items or users are considered in the fairness definition.
As the name suggests, individual fairness considers each item or user individually, and requires that equal treatment is provided to similar items or users~\cite{DworkEtAl2012}.
Similarity can be computed based on features for items~\cite{GeEtAl2021,Wang2022ProvidingSystems} or the users' interaction history for users~\cite{WuEtAl2022,Rampisela2026PUF}. 
For example, given two items with similar features,
an unfair RS will provide high exposure to only one of the two items.
Analogously, given two users with similar past interactions, an unfair RS will provide effective/relevant recommendations to only one of the two users.
Group fairness simplifies the problem by grouping items and users, so that each group of items or users will receive equal treatment~\cite{EkstrandEtAl20222}.
Groups are usually formed with the attributes of items and users, e.g., popularity~\cite{FuEtAl2020}, or sensitive attributes, such as gender and ethnicity~\cite{BarocasEtAl2023}.
In group fairness, an example of an unfair RS, is a system that provides more exposure to the group of highly popular items, or is more effective for a group of users, e.g., white male users.

This paper focuses on individual item fairness.
Next, we review measures in this category, focusing solely on item exposure.
We consider all measures that could be corrected for the non-realisability limitation according to~\citet{Rampisela2024EvaluationStudy}.
The non-realisability limitation occurs when a measure cannot empirically reach its theoretical minimum/maximum score. Correcting a measure for non-realisability ensures that there exists a set of top-$k$ recommendation lists such that the measure can empirically reach its theoretical minimum and maximum scores, e.g., 0 and 1, respectively.

\paragraph{Jain Index (Jain)~\cite{jain1984quantitative, Zhu2020FARM:APPs}} Jain index quantifies fairness in terms of recommendation opportunity. It measures how consistently an item is exposed within the top-$k$, with respect to the number of times other items are recommended in the top-$k$. 
The higher the Jain index, the fairer the RS, i.e., the more consistently the exposure is allocated across items.

\paragraph{Qualification Fairness (QF)~\cite{Zhu2020FARM:APPs}} Qualification Fairness is equivalent to coverage, e.g., the number of unique items recommended in the top-$k$ divided by the total number of items. 
When $\text{QF} = 1$, 
each item is recommended at least once in the top-$k$,\footnote{If the total number of items exceeds the total number of recommendation slots available across all users, `each item' is limited to the maximum number of items allowed by those slots.} therefore, the higher the Qualification Fairness, the fairer the RS. 

\paragraph{Entropy (Ent)~\cite{Shannon1948ACommunication, Patro2020FairRec:Platforms}} Entropy quantifies how uniform the exposure probability is across items.
The exposure probability is defined as the number of times an item is recommended in the top-$k$, divided by the number of recommendation slots across all users.
The more uniform the exposure probability, the higher the entropy, and the fairer the RS.

\paragraph{Fraction of Satisfied Items (FSat)~\cite{Patro2020FairRec:Platforms}} Fraction of Satisfied Items is based on the maximin share, i.e., a threshold on the number of times an item needs to be recommended.
The maximin share is computed as the greatest integer less than or equal to the ratio between the total number of recommendation slots within the top-$k$ across all users and the number of items. 
Basically, it is the minimum number of times an item needs to be recommended to ensure that all items have equal exposure within the top-$k$.
Then, Fraction of Satisfied Items is the proportion of items with exposure above or equal to the maximin share.
The higher the FSat, the fairer the RS, and when all items are equally exposed, $\text{FSat} = 1$.

\paragraph{Gini Index (Gini)~\cite{Gini1912VariabilitaMutabilita, Mansoury2020FairMatch:Systems}} Gini Index measures the inequality of the exposure allocation across items.
If a single item receives all the exposure, i.e., it is always recommended in the top-$k$, then $\text{Gini} = 1$; if $\text{Gini} = 0$, then all items are exposed uniformly within the top-$k$.
Thus, Gini Index measures unfairness; the higher the score, the more unfair the RS.\\

All the above-mentioned measures quantify fairness (or unfairness) in terms of exposure only.
Next, we present measures that jointly consider exposure and item relevance when quantifying fairness, which we broadly refer to as relevance-aware fairness measures (\joint{} measures). 
These measures are originally defined for a series of rankings, or recommendation rounds, with a stochastic ranking policy. 
In this paper, we consider a single recommendation round.

\paragraph{Item Better-Off (IBO) and Item Worse-Off (IWO)~\cite{Saito2022FairRanking}} 
Item Better-Off is based on the principle of ``dominance over a uniform ranking''.
First, an impact score is computed for each item based on its exposure in the top-$k$ and its relevance to the users.
Then, IBO is the percentage of items for which the current exposure allocation increases the impact score by at least $10\%$ compared to the exposure allocation provided by a random uniform RS.
The higher the IBO, the higher the gain in terms of impact score over the uniform exposure, thus the fairer the RS.
Item Worse-Off (IWO)~\cite{Saito2022FairRanking} has a specular definition: it is the percentage of items for which the current exposure allocation decreases the impact score by at least $10\%$ compared to the exposure allocation provided by a random uniform RS.
Therefore, IWO measures unfairness; the higher the IWO, the more unfair the RS.

\paragraph{Mean Max Envy (MME)~\cite{Saito2022FairRanking}} Mean Max Envy is based on the concept of envy-freeness.
Given an item, its exposure allocation is swapped with that of all the other items, one item at a time; then, the difference in the impact score before and after the swap is computed. 
The impact score is the same as in IBO/IWO.
MME is the average, over the items, of the maximum difference in the impact scores obtained after performing all swaps.
The intuition is that, given two items $i$ and $j$, if the impact score of $i$ increases when allocated the exposure of $j$, then the item $i$ would benefit more if allocated the exposure of $j$. 
MME is a measure of unfairness: the lower the MME, the lower the impact difference when swapping exposure allocations, and the fairer the RS.

\paragraph{Inequity of Amortized Attention (IAA)~\cite{Biega2018EquityRankings, Borges2019EnhancingAutoencoders}} Inequity of Amortized Attention is defined as the average absolute difference between item exposure and relevance. 
Exposure is computed with a linear examination function, i.e., a decay is assigned linearly with the item rank position.
Exposure and relevance are min-max normalised so they both range in $[0, 1]$, and can be subtracted.
If an item has high relevance but low exposure, the difference between exposure and relevance will be high.
Therefore, IAA is an unfairness measure, where a high score means that relevant items are less exposed than non-relevant ones, thus, the system is unfair.

\paragraph{Individual-user-to-individual-item fairness (II-F)~\cite{Diaz2020EvaluatingExposure, Wu2022JointRecommendation}} Individual-user-to-individual-item fairness quantifies unfairness by comparing the exposure distribution of an RS against an ideal distribution.
The ideal distribution assumes that relevant items get equal expected exposure. 
II-F measures unfairness; the higher the score, the further the RS exposure distribution from the ideal distribution, hence, the more unfair the RS.

\paragraph{All-users-to-individual-item fairness (AI-F)~\cite{Diaz2020EvaluatingExposure}} All-users-to-individual-item fairness is similar to II-F, it compares the exposure distribution of an RS against the same ideal distribution used by II-F.
The two measures differ in the way they aggregate the scores/distributions: II-F compares the RS and ideal distributions for each user individually and then aggregates the scores, while AI-F first estimates the RS and ideal distributions by aggregating across users, and then compares the two aggregated distributions. 
AI-F is also an unfairness measure, and the higher the AI-F score, the more unfair the RS.

\subsection{Multi-aspect Evaluation of Rankings}
\label{ss:multi_aspect_evaluation}

\begin{table*}
\centering
\caption{Types and (non-exhaustive) examples of existing approaches that can be used for multi-aspect evaluation of rankings. We indicate approaches used in our work and justify why some are not suitable. \joint{} measures are relevance-aware fairness measures.}
\label{tab:eval_approach}
\resizebox{\textwidth}{!}{
\begin{tabular}{p{2.5cm}p{5cm}cp{5.25cm}}
\toprule
\midrule
 Type & Existing approach (Examples) &  Used in our paper? & Justification  \\
\midrule
\midrule
\multicolumn{4}{l}{\textit{1. Assign single-aspect human labels or automatic scores per item, then aggregate into a multi-aspect score.$^*$}}\\
\hline
\quad human label & CAM \cite{Lioma2017EvaluationLists}, WHAM \cite{Lioma2017EvaluationLists}, TOMA \cite{Maistro2021PrincipledRankings}, MDCU \cite{JarvelinS24,FaveriF0J25}  & \xm & Public recommender system datasets have no fairness label per item  \\
\quad automatic score \rule{0pt}{4ex} & EPC \cite{vargascastello}, MM \cite{Palotti2018MM:Engines}, TOMA \cite{Maistro2021PrincipledRankings}&  \xm & Unlike other aspects, e.g., item novelty or credibility, item fairness is typically not scored per item, but across all users and all items at once \\
\midrule
\multicolumn{4}{l}{\textit{2.  Compute single-aspect measures (averaged across all users/rankings) for each model/run/recommender, then aggregate}}\\
\hline
\quad average & (weighted) arithmetic mean, geometric mean, harmonic mean \cite{vargascastello,Cherumanal2025Diversification} & \cm & - \\
\quad rank fusion$^\dagger$ \rule{0pt}{4ex} & COMBMAX, COMBMIN  \cite{Fox1993CombinationSearches} & \xm & May collapse to one evaluation dimension as single-aspect measures may have different empirical score range\\
 \rule{0pt}{4ex} & COMBSUM \cite{Fox1993CombinationSearches}, COMBMNZ \cite{Fox1993CombinationSearches,Lee1997AnalysesCombination} & \xm & May give equivalent results as  average \\
 \rule{0pt}{4ex} & COMBMED \cite{Fox1993CombinationSearches} & \xm & The median of two scores (i.e., relevance and fairness) equals their arithmetic mean  \\
 \rule{0pt}{4ex} & Borda Fuse \cite{Aslam2001BordaFuse}, Reciprocal Rank Fusion (RRF) \cite{Cormack2009RRF}  & \xm & Single-aspect scores are individually mapped into another score based on the ranking of models (e.g., from fairest to unfairest); this score does not quantify performance gaps across models when evaluating multiple aspects jointly \\
\midrule
\multicolumn{4}{l}{\textit{3. Compute a multi-aspect score for each ranking or run/recommender, without single-aspect scores per item.}}\\
\hline
\quad \joint{} measures & IBO \cite{Saito2022FairRanking}, MME \cite{Saito2022FairRanking}, IAA~\cite{Borges2019EnhancingAutoencoders,Biega2018EquityRankings}, {II-F}~\cite{Wu2022JointRecommendation}, AI-F \cite{Wu2022JointRecommendation}, FAIR \cite{Gao2022FAIR:Evaluation}, FEIR \cite{LiEtAl2024}, PUF \cite{Rampisela2026PUF} & \cm & - \\
\parbox[t]{2.5cm}{\quad non-fairness  \rule{0pt}{4ex}\\ \phantom{\quad}measures} & EILD \cite{vargascastello}, $\alpha\beta$-NDCG \cite{ParaparAndRadlinski2021} & \xm & Not intended to measure fairness \\
 \bottomrule
 \bottomrule
 \multicolumn{4}{l}{$^*$Technically can be used with either human labels or automatic scores; we categorise based on what the original work used.} \\
 \multicolumn{4}{l}{$^\dagger$Usually combines document/item rankings; here we propose fusing rankings of models based on their single-aspect scores.} \\
\end{tabular}}
\end{table*}

The problem of evaluating relevance and fairness simultaneously falls into the category of multi-aspect evaluation. \Cref{tab:eval_approach} summarises existing approaches for multi-aspect evaluation of rankings and why some do not suit our work.

Several approaches have been proposed to handle multi-aspect or multi-dimensional evaluation~\cite{Lioma2017EvaluationLists,Palotti2018MM:Engines,vargascastello,JarvelinS24}.
\citet{Lioma2017EvaluationLists} propose two measures to evaluate the relevance and credibility of ranked lists, motivated by issues such as misinformation.
One measure is computed based on the error between the item's actual rank and its ideal rank, i.e., its rank in the best possible ordering of items.
The other measure is an extension of  Normalised Discounted Cumulative
Gain (NDCG)~\cite{Jarvelin2002NDCG} that accounts for both item relevance and credibility.
\citet{Palotti2018MM:Engines} presents a framework to evaluate rankings when relevance is defined by several aspects.
It is motivated by consumer health search, where not only it is important to retrieve relevant items, but those items should also be easily understood by users.
The proposed framework takes the harmonic mean of Rank-biased Precision (RBP)~\cite{Moffat2008RBP} computed for topical relevance and understandability independently. 
\citet{Maistro2021PrincipledRankings} proposes a principled framework for multi-aspect evaluation, which can work for any type and number of aspects.
The framework defines an order relation among items labelled with different aspects and then computes standard Information Retrieval (IR) effectiveness measures, e.g., NDCG. 
\citet{JarvelinS24} extends NDCG to consider multidimensional relevance, item usability, and the content overlap among items in the ranking.
The practical feasibility and validity of this measure is evaluated experimentally in~\cite{FaveriF0J25}. In RSs, \citet{vargascastello} propose Expected Popularity Complement (EPC), a metric to evaluate item novelty based on three concepts: choice, discovery, and relevance. Relevance is approximated through user ratings, while choice and discovery are automatically computed based on a browsing model and what has been seen, respectively.

None of the multi-aspect evaluation approaches mentioned above can be applied to the case of relevance and fairness.
This is because they assume that each item can be labelled or scored for each aspect independently of other users or items.
However, this is not possible for fairness, which is defined in relation to other users/items and can not be assessed in isolation. For example, the ``fairness'' of an item or a group of items is often defined based on exposure, i.e., how frequently an item is exposed at the top of the ranking across all users.

Another line of work applies an aggregate measure (e.g., average) on the scores of existing single-aspect measures: 
\citet{vargascastello} compute the harmonic mean of recommendation relevance and novelty to have a single measure that represents both; likewise, \citet{Cherumanal2025Diversification} compute the harmonic mean of NDCG, $\alpha$-NDCG~\cite{Clarke2008AlphaNDCG}, and Attention-Weighted Rank Fairness (AWRF)~\cite{ekstrand2023overview} to aggregate relevance, diversity, and fairness. A different way of aggregating single-aspect measures is to combine or fuse the rankings of models/recommenders based on various evaluation aspects. For example, one may rank models from the best to worst relevance according to their NDCG score and separately rank them from the fairest to the unfairest based on their Gini score. A rank fusion method, e.g., COMBMNZ \cite{Lee1997AnalysesCombination,Fox1993CombinationSearches} can then be applied to combine these rankings, resulting in a ranking of models that represents the best to worst models when considering both relevance and fairness. While rank fusion methods are often applied on rankings of items/documents, they can technically be applied on any rankings, including model rankings.

Our proposed evaluation approach is similar to the above approaches, as it uses common relevance and individual item fairness measures in RSs. Yet, our approach has an added advantage over these existing approaches: it allows a comparison to the empirically optimal relevance-fairness scores, which are not obtainable by simply averaging single-aspect scores or fusing rankings. Furthermore, when looking into the different rank fusion approaches, several issues render them unsuitable for our goal. 
COMBMAX and COMBMIN \cite{Fox1993CombinationSearches} would rank models based on the maximum/minimum of the relevance and fairness scores. As these two measure types typically have different ranges of observed scores \cite{Rampisela2024EvaluationStudy}, using COMBMAX/COMBMIN may collapse to a single aspect. On the other hand, 
COMBSUM \cite{Fox1993CombinationSearches} and COMBMNZ \cite{Fox1993CombinationSearches,Lee1997AnalysesCombination} may give equivalent results to the arithmetic mean: ranking models by the sum of their relevance and fairness scores (COMBSUM) yields the same ranking as averaging them. As the list of models to be ranked is identical across the two aspects, COMBMNZ would also give the same ranking as COMBSUM and averaging. COMBMED \cite{Fox1993CombinationSearches} will also produce the same ranking as averaging, as the median of two scores (relevance and fairness) equals their arithmetic mean. Rank-based fusion methods like Borda Fuse \cite{Aslam2001BordaFuse} and Reciprocal Rank Fusion (RRF) \cite{Cormack2009RRF} work by mapping a single-aspect score to another score based on the model's rank position. The resulting score cannot be used to quantify the gap in model performance when considering multiple aspects concurrently, which is what our approach aims to achieve.

Aside from the relevance-aware fairness measures for individual items (presented in \Cref{ss:fairness_evaluation}), several other evaluation measures also directly combine effectiveness and fairness. 
The Fairness-Aware IR (FAIR) measure~\cite{Gao2022FAIR:Evaluation} combines group item fairness with standard IR effectiveness measures.
It is defined as the ratio between the scores of the IR measure and of the fairness measure cumulated over the rank positions.
Fairness is computed with the Kullback–Leibler (KL) divergence between the item group distribution in the ranking and the desired target distribution.
Our approach is different because it is defined for individual item fairness; it does not require a desired target distribution, which can be hard to choose; it considers the empirically achievable optimal solution, while the target distribution may not be an achievable solution; and it can incorporate any individual item fairness measure.
Fairness through Envy and Inferiority Reduction (FEIR)~\cite{LiEtAl2024} and Pairwise User unFairness (PUF)~\cite{Rampisela2026PUF} are measures that combine individual user fairness and effectiveness. 
Instead, our approach focuses on individual item fairness, and we propose an evaluation approach that is compatible with existing fairness and effectiveness/relevance measures.

Beyond the strict domain of relevance-fairness joint evaluation, there exist composite measures that integrate two or more aspects in a single metric. Similar to the relevance-aware fairness measures, these measures do not require a score for each item and for each aspect. Instead, each item is assessed in relation to other items. For example, relevance-aware diversity measures \cite{vargascastello,ParaparAndRadlinski2021} are defined in terms of similarity to other items in the same ranking. 
\citet{vargascastello} generalise the average intra-list distance metric \cite{Zhang2008AvoidingMonotony,Ziegler2005ImprovingDiversification} by integrating a rank-aware and a relevance component into the measure. The proposed measure thus accounts for position bias (where top-ranked items get more user attention/interaction) and aims for effective diversification of recommendation. 
\citet{ParaparAndRadlinski2021} extend NDCG to consider relevance and diversity.
The proposed measure is tailored to RSs and is based on the intuition that users prefer relevant and diverse items over relevant but similar items. 
As these measures do not evaluate fairness, they cannot be compared to our approach. Finally, some multi-aspect evaluation approaches have been proposed for Retrieval Augmented Generation (RAG)~\cite{GienappEtAl2024,MerkerEtAl2025}, but they take as input the textual output produced by a generative model instead of a ranking of items.

\subsection{Assessing robustness of evaluation measures for rankings}
\label{ss:assess_robustness}

The robustness of evaluation measures for rankings in IR and RS has been assessed through multiple lenses, among others, in terms of how incomplete data or biased data affect their results (e.g., \cite{Valcarce2018Robustness,Valcarce2020AssessingRecommendation,Sakai2008OnAssessments,Buckley2004IncompleteInformation,ParaparAndRadlinski2021}).  

Traditionally, metric robustness to data incompleteness is evaluated by comparing IR system results based on an incomplete set of relevance data to those based on a less incomplete one, e.g., \cite{Buckley2004IncompleteInformation, Sakai2008OnAssessments}. First, they synthetically remove relevance judgments, producing several sets of judgments with varying levels of incompleteness. They then evaluate various runs using IR measures computed on these judgments and rank these runs based on their scores. Robustness is assessed in terms of how stable the ranking of runs is against data incompleteness. 
More recent work has also investigated such robustness of ranking measures in RS evaluation \cite{Valcarce2018Robustness,Valcarce2020AssessingRecommendation,ParaparAndRadlinski2021}. \citet{ParaparAndRadlinski2021} evaluate the robustness of their relevance-aware diversity measure against random missing items. \citet{Valcarce2018Robustness,Valcarce2020AssessingRecommendation} analyse how the removal of random items and users from the test split affects the ranking stability of RS effectiveness measures.  They also study robustness against non-random missing data, focusing on biased (uneven) distribution of ratings across items and users. From the test set, they remove popular items and users with the most interactions. The results show that, among other measures, Precision tends to be the most robust against both random and non-random missing data.

Since our proposed evaluation approach, DPFR, is instantiated with existing relevance and fairness measures, we posit that its robustness would largely depend on these single-aspect measures. While the robustness of the relevance measures in RS has been assessed in prior work \cite{Valcarce2018Robustness,Valcarce2020AssessingRecommendation}, to our knowledge, no such work has been done for individual item fairness measures. 

\subsection{Multi-objective Optimisation in Recommender Systems}
\label{ss:multi_optimization}

The challenge of jointly evaluating RS effectiveness (through recommendation relevance) and fairness is exacerbated by the fact that effectiveness and fairness are not completely independent. 
A trade-off between effectiveness and fairness has been empirically shown, with effectiveness towards users decreasing when fairness towards items increases~\cite{Mehrotra2018TowardsSystems}.
Therefore, multi-aspect optimisation has been proposed as a solution to find the balance between effectiveness and fairness. 
The well-known Pareto optimality has been used in the optimisation functions of RSs to learn the trade-off between fairness and relevance, and recommend items to users.
For example,~\citet{Ge2022TowardLearning} present a recommendation algorithm that uses multi-objective reinforcement learning and the Pareto-optimal solution to control the trade-off between effectiveness and fairness~\cite{Zheng2022AOptimization}.
\citet{Ribeiro2015MultiobjectiveSystems} use the concept of Pareto efficiency, a.k.a.~Pareto-optimal solution, to merge ranked lists generated by different optimisation aspects, e.g., effectiveness, novelty, diversity, and to create a weighted combination of algorithms optimised for different aspects. 

Computing the true PF can be unfeasible due to its  complexity~\cite{Laszczyk2019SurveyMeasures,Audet2020PerformanceOptimization}.
Therefore, some prior work could only show that their proposed algorithm was better in terms of Pareto dominance (see \Cref{ss:motivation}), without computing the actual Pareto-optimal frontier~\cite{PangEtAl20219,Xu2023P-MMF:System,JainEtAl2020,ZouEtAl2021}. For example,~\citet{Xu2023P-MMF:System} show with the PF that their recommender model for fairness and effectiveness is better than state-of-the-art baselines, but they do not compute to what extent the proposed model is far from the Pareto-optimal solution.  
Other work used scores computed with a single or several models to generate the PF~\cite{Lin2019ARecommendation,Nia2022RethinkingNetworks,Paparella2023Post-hocRecommendation}, instead of using the optimal scores (from ground truth labels).
For example,~\citet{Lin2019ARecommendation} propose an approach to generate Pareto-optimal recommendations balancing Click Through Rate (CTR) and Gross Merchandise Volume (GMV). 
The proposed approach uses the scores of the training losses and assumes that those loss functions are differentiable.
\citet{Paparella2023Post-hocRecommendation} generate the PF with the scores of different objective functions as inputs, and define an efficient way to select the best Pareto-optimal solution from the PF. 
Using such optimal solution during training leads to better performance in search and recommendation. 

Our work is inherently different from the above, both in terms of the purpose and also in the way we compute the PF.
First, our aim is to use the PF as the basis of a theoretically grounded solution for the \textit{evaluation} of rankings, not for multi-objective optimisation of RSs.
Second, because or aim is to evaluate and compare rankings, we compute the PF using the ground truth, i.e., the labels in the test set, as opposed to using the models' scores, e.g., training loss, which represent an estimation of the ground truth and are prone to errors.
Therefore, our PF is model agnostic and can be used to compare different RSs.
Finally, our work is different from~\citet{Paparella2023Post-hocRecommendation}, where the Pareto-optimal solution is the point in the PF with minimum distance from the utopia point, i.e., the ideal optimisation target.
Such a utopia point might not be achievable depending on the datasets or measure characteristics~\cite{Rampisela2024EvaluationStudy,Moffat2013SevenMetrics}, instead, since we generate the PF from the test data, the PF will always be a set of Pareto-optimal solutions that are actually achievable.

\section{Distance to Pareto Frontier (DPFR)}
\label{s:our_method}

We present our Pareto-based joint evaluation approach, Distance to PF of Fairness and Relevance (DPFR). 
\Cref{ss:motivation} contains our problem definitions. The remaining subsections present the methodological steps of DPFR: 
given a \fair{} and a \rel{} measure, (i) how PF is 
generated based on the ground truth data in the test set (\Cref{ss:generation}); how to choose a reference point in the PF based on $\alpha$ (e.g., the midpoint for $\alpha=0.5$) (\Cref{ss:pareto-for-eval}); and how to compute the distance of the \rel{} and \fair{} scores to the reference point with a distance measure $d$ (\Cref{ss:pareto-for-eval}). 
Finally, \Cref{ss:compute-eff} contains our proposed computationally efficient adaptation of DPFR. 

\subsection{Definitions}
\label{ss:motivation}

We use the following Pareto-optimality definition \cite{vanVeldhuizen1999MultiobjectiveInnovations}: the multi-objective problem is finding the optimum \fair{} score $s_f$, and \rel{} score, $s_r$ from a list of possible recommendations across all users. We define the tuple $s = (s_r, s_f) \in S$, where $S$ denotes the Cartesian product of all possible \rel{} and \fair{} scores. The relation $\geq_A$ ($>_A$) means `better or equal to' (`better to') according to an aspect $A \in \{\rel{}, \fair{}\}$. 

\begin{definition}[Pareto Dominance]
A tuple $s = (s_r, s_f)$ dominates $s'=(s_r', s_f')$ iff $s$ is partially better than $s'$, i.e., $s_r \geq_\rel{} s'_r$ and $s_f \geq_\fair{} s'_f$, in addition to $s_r >_\rel{} s'_r$ or $s_f >_\fair{} s'_f$. 
\end{definition}

\begin{definition}[Pareto Optimality]
A solution (in our case, recommendation list) that has \rel{} and \fair{} scores of $x = (x_r, x_f) \in S$ is Pareto-optimal iff there is no other solution with $x'=(x_r', x_f') \in S$ that dominates $x$.
\end{definition}

\begin{definition}[Pareto Frontier]
The set of all Pareto-optimal tuples.
\end{definition}

\subsection{Pareto Frontier Generation}
\label{ss:generation}
Given some user-item preference data (e.g., a test set), individual item fairness is usually computed across items while relevance scores are computed per user and then averaged.
Our aim is to find the empirically maximum fairness towards individual items based on the given test set, such that the recommendation satisfies Pareto-optimality w.r.t.~fairness scores across all items and an average relevance score across users, e.g., MAP@$10=0.9$. 
We do this by measuring how far a model performance is from these Pareto-optimal solutions.
The process of comparing all possible recommendations for users and items to find the complete set of Pareto-optimal solutions is computationally expensive and therefore infeasible, and there is no analytical solution either. We address this by proposing an algorithm that iteratively builds upon a maximally relevant initial recommendation list. Our algorithm iteratively finds Pareto-optimal recommendations by prioritising relevance over fairness, because, in practice, recommendations are usually optimised for relevance (with or without fairness). This prioritisation is known as lexicographic optimisation, in which one objective is optimised first, and the other is optimised while keeping the first objective at its optimal value \cite{Ryu2018Multi-objectiveWeight}. 
Our algorithm is called {\sc Oracle2Fair}, and the full technical description can be found in \Cref{app:algo}. 

Our algorithm generates the PF of fairness and relevance in two steps: 
 \textbf{(1) initialisation} of the recommendations with an \textit{Oracle} (\Cref{app:algo}, \Cref{alg:oracle}). The \textit{Oracle} generates a recommendation with the highest empirical score for relevance, based on user interactions that are part of the test set. This step is followed by \textbf{(2) replacements} to make the recommendations as \textit{Fair} as possible; at the end of this algorithm, the \fair{} scores should reach the empirically fairest score while maintaining as much relevance as possible. 
Note that throughout the PF generation, we make sure that items in a user's train/val split are not recommended to the same user. Henceforth, \emph{relevant items} refers to the items in a user's test split.

\paragraph{\textbf{(1) Initialisation}} 
    The Oracle recommends at most $k$ (e.g., 10) relevant items, from the $n$ items in the dataset, to each of the $m$ users in the test split, one user at a time. 
    The recommendation begins with users having exactly $k$ items in the test split; only these items can be recommended to those users to gain the maximum relevance. Recommendations to other users are made maximally relevant while also aiming to be as fair as possible, through the following steps: if a user has $>k$ relevant items, we pick $k$ items with the least exposure among them. Item exposure is computed based on what has been recommended to other users who already have exactly $k$ items; a more even distribution of exposure across all items leads to higher fairness. Note that this process is not trivial (see \Cref{app:algo}, 
    \Cref{alg:oracle}, ll.~\ref{ln:startgtk}--\ref{ln:endgtk}). If a user has $<k$ relevant items, we recommend those items at the top (to maximise top-weighted \rel{} measures) and fill the rest of their recommendation slots with the least exposed items in the dataset (\Cref{alg:oracle}, ll.~\ref{ln:startltk}--\ref{ln:endltk}). This least-exposure prioritisation strategy aims to find solutions that are Pareto-optimal. 
    
    \paragraph{\textbf{(2) Replacements}} To achieve maximum fairness at the end of the process, the algorithm iteratively replaces the recommended items, such that each replacement results in a fairer recommendation than the previous. 
    We compute the \rel{} and \fair{} measures after each replacement as follows. The most popular item, which is recommended most often, is replaced with one of these item types, 
    in decreasing order of priority: an unexposed item, then the least popular item in the recommendation; this increases fairness from the previous recommendations. We do this one item and one user at a time, starting with the users who have the most popular item at the bottom of their recommendation list, to strive for minimal decrease in relevance, as the replacement item is mostly not relevant to that user. Nonetheless, the \textsc{Oracle2Fair} prioritises replacing the recommendations of users for whom the replacement item is relevant (if any). In most cases, fairness increases, and relevance decreases or stays the same from the previous recommendation: if relevance decreases, the new recommendation should be Pareto-optimal; if relevance remains the same, to maintain Pareto optimality, we remove duplicates of \rel{} scores by keeping \rel{}-\fair{} score pairs with the best \fair{} score for a single value of \rel{}. 
    We continue the replacement until the maximum number of times any item is recommended is $\left\lceil km/n \right\rceil$, i.e., the upper bound of how many times an item can be recommended, if all items in the dataset must appear in the recommendation as uniformly as possible. We explicitly used $\left\lceil km/n \right\rceil$ as a stopping condition for \textsc{Oracle2Fair}. 
    To ensure maximum \rel{} scores (especially in top-weighted measures), each time a replacement takes place, we rerank the recommendations based on descending relevance.
    
    The resulting (\textsc{Rel, Fair}) scores reflecting Pareto-optimal recommendations from this process make up the PF. While it cannot be reasonably verified that the resulting PF always matches the true PF, we evaluate their closeness in \Cref{ss:true_pf}.

\subsection{Distance Computation} 
\label{ss:pareto-for-eval}

The distance computation consists of two steps: (1) determining a reference point in the PF; and (2) computing the distance from the reference point to the model scores.

\paragraph{(1) Determining a PF reference point} For each pair of \rel{} and \fair{} measures, we find a reference point using a tunable parameter $\alpha \in [0,1]$; 
$\alpha=0$ means giving maximum priority to relevance (i.e., the point with the highest possible \fair{} score for a maximally relevant recommendation), and $\alpha=1$ means giving maximum priority to fairness (i.e., the point with the highest possible \rel{} score for a maximally fair recommendation). Note that giving maximum priority to one aspect does not mean that the other aspect is completely discarded in the joint evaluation. For example, at $\alpha=0$, the \fair{} score will still influence the joint evaluation, but its impact will not be as large as it would be at a higher $\alpha$. We explore what $\alpha$ practically means in \Cref{ss:varying_alpha}.

Below, we present how to compute the reference point. We first use the following equation to find the length of a subset $T$ of the PF: 
$
    lenPF(T) = \sum_{t=1}^{|T|-1} d_E(x^t, x^{t+1})
$.
 Given that $P$ is the set of all Pareto-optimal solutions, $x^t = (x_r^{t}, x_f^{t})$ is the pair of Pareto-optimal solutions $(x_r,x_f)$ with the $t$-th highest $x_r$ in $P$, and $d_E$ is the Euclidean distance. The overall PF length is $lenPF(P)$ or simply $lenPF$. 
 
 The reference point is $s_{\alpha} = x^{t'}$, where $t'$ is computed as follows:
 $
    t' = \argminA_{j \in \left[1,\dots,|P|-1\right]}\left|lenPF(T^j)-\ \alpha \cdot lenPF \right|
$. $T^{j}$ is a subset of $P$ containing the $j$ highest $x_r$ scores. So, the reference point is a point in the PF whose cumulative traversal distance is closest to the $\alpha$-weighted PF length. In other words, the reference point $s_\alpha$ is based on the PF length from the pair with the best \rel{} score to the one with the best \fair{} score, multiplied by $\alpha$. As the PFs may have different densities of points along the frontiers, the reference point is not computed based on a percentile (e.g., median) to avoid bias towards the denser part. 

\paragraph{(2) Computing distance from the reference point to the model scores} Next, the distance between the reference point $s_\alpha$ and each model's ($x_r, x_f$) scores is computed with a distance measure $d$ that accommodates 2D-vectors (e.g., the Euclidean distance $d_E$). The model with the closest distance is the best in terms of both relevance and fairness, given the weight $\alpha$. 
We call this \emph{Distance to Pareto frontier of Fairness and Relevance} (DPFR). 

\subsection{Efficient Computation of Pareto Frontier}
\label{ss:compute-eff}
  
Generating the PF, as described in \Cref{ss:generation}, is computationally costly. As an efficient alternative, we propose computing a subset of the PF. This can be done by selecting a fixed number of Pareto-optimal solutions to compute, $p$ (e.g., 10). However, in order to reliably approximate the PF, these solutions should be selected so that they are spread according to the PF distribution, as opposed to, for instance, only computing the first $p$ points of the PF. The spread of the points is important, because the reference point in DPFR is computed based on the overall estimated PF. 
In the estimated PF, the first point corresponds to the measure scores of the initial recommendation given by the Oracle, and the rest are spread evenly throughout the PF generation. 
To select at which point of the \textsc{Oracle2Fair} algorithm the measures should be computed, we first estimate the total number of replacements needed by examining the distribution of recommended items frequency. We do this by computing the individual frequency count of all items in the recommendation, and subtracting the ideal upper bound of item count $\lceil km/n \rceil$ (\Cref{ss:generation}) from each count. 
The number of expected replacements is computed as the sum of the difference between the item frequency count and the ideal upper bound of item count in \Cref{ss:generation}. Items with recommendation frequency counts less than the upper bound are excluded. With the estimated total number of replacements $numRep$, we compute the measures every $numRep\ \text{div} (p-1)$ replacements done by \textsc{Oracle2Fair}, such that the measures are computed a total of $p-1$ times during the replacement process + 1 time before the replacement starts. As a result, these $p$ points are spread evenly in terms of distance in the PF, which is important because DPFR is a distance-based measurement. 

\section{Experimental Setup}
\label{s:experiments}

We empirically compare our joint evaluation approach, DPFR, to existing single- and multi-aspect evaluation measures of relevance and fairness in RS. This section presents our experimental setup.\footnote{The repository of our previous publication on the topic \cite{Rampisela2025Pareto} includes more details about hyperparameter tuning, links to download the datasets, and scripts for the results: \url{https://github.com/theresiavr/DPFR-recsys-evaluation}.} 
The experiments were conducted in a cluster of various CPUs and GPUs (e.g., Intel(R) Xeon(R) Silver 4214R CPU @ 2.40GHz, AMD EPYC 7413 and 7443, Titan X/Xp/V, Titan RTX, Quadro RTX 6000, A40, A100, and H100).

\subsubsection*{Datasets} We use the following six publicly available and commonly used RS datasets. From the domain of e-commerce: A\-ma\-zon Luxury Beauty (Amazon-lb) \cite{Ni2019JustifyingAspects}; from the domain of movies: ML-10M and ML-20M \cite{Harper2015TheContext}; from the domain of music: Lastfm \cite{Cantador20112nd2011}; from the domain of videos: QK-video \cite{Yuan2022Tenrec:Systems}; and from the domain of jokes: Jester \cite{Goldberg2001Eigentaste:Algorithm}. 
We use all datasets as provided by \cite{Zhao2021RecBole:Algorithms}, except for QK-video, which we obtain from \citet{Yuan2022Tenrec:Systems}. 
The statistics of these datasets are displayed in \Cref{tab:stats}, and extended statistics are included in \Cref{app:stat}.

\begin{table}[tb]
\caption{Statistics of the preprocessed datasets.}
\label{tab:stats}

\begin{tabular}{lrrrr}
\toprule
\textbf{dataset} & \multicolumn{1}{l}{\textbf{\#users ($m$)}} & \multicolumn{1}{l}{\textbf{\#items ($n$)}} & \multicolumn{1}{l}{\textbf{\#interactions}} & \multicolumn{1}{l}{\textbf{sparsity}} \\ 
\midrule                                                                   
Lastfm \cite{Cantador20112nd2011}& 1,859 & 2,823 & 71,355 & 98.64\% \\
Amazon-lb \cite{Ni2019JustifyingAspects} & 1,054 & 791 & 12,397 & 98.51\% \\
QK-video \cite{Yuan2022Tenrec:Systems} & 4,656 & 6,423 & 51,777 & 99.83\% \\
Jester \cite{Goldberg2001Eigentaste:Algorithm} & 63,724 & 100 & 2,150,060 & 66.26\% \\ 
ML-10M \cite{Harper2015TheContext} & 49,378 & 9,821 & 5,362,685 & 98.89\% \\
ML-20M \cite{Harper2015TheContext} & 89,917 & 16,404 & 10,588,141 & 99.28\% \\
\bottomrule
\end{tabular}
\end{table}

\subsubsection*{Preprocessing} As per standard practices, we remove duplicate interactions and keep the most recent ones. We also keep only users and items with $\geq5$ interactions. 
We convert ratings equal/above the following threshold to $1$ and discard the rest: for Amazon-lb and ML-*, the threshold is 3, because their ratings range between $[1,5]$ and $[0.5, 5]$ respectively; the threshold for Jester is $0$, because the ratings range in $[-10, 10]$. Lastfm and QK-video have no ratings. QK-video has several interaction types; we use the `sharing' interactions. 
For Jester, we remove users with $>80$ interactions, to provide a large enough number of item candidates for recommendation during testing. We do this because some users in Jester have interacted with almost all 100 items. If a user has 80 items in the train/val set, there would only be 20 candidate items to recommend during test, which makes it easier to achieve higher relevance.

\subsubsection*{Data splits} To obtain the train/val/test sets for Amazon-lb and ML-*, we use global temporal splits \cite{Meng2020ExploringModels} with a ratio of 
6:2:2 on the preprocessed datasets. Global random splits with the same ratio are used for the other datasets, which have no timestamps. 
From all splits, we remove users with $<5$ interactions in the train set.

\subsubsection*{Evaluation measures}
We use the following evaluation measures of relevance (\rel{}): Hit Rate (HR), Mean Reciprocal Rank (MRR), Precision (P), Recall (R), Mean Average Precision (MAP), and Normalised Discounted Cumulative Gain (NDCG \cite{Jarvelin2002NDCG}). 
For individual item fairness (\fair{}), we use the following evaluation measures: Jain's Index (Jain) \cite{jain1984quantitative, Zhu2020FARM:APPs}, Qualification Fairness (QF) \cite{Zhu2020FARM:APPs}, Gini Index (Gini) \cite{Gini1912VariabilitaMutabilita, Mansoury2020FairMatch:Systems}, Fraction of Satisfied Items (FSat) \cite{Patro2020FairRec:Platforms}, and Entropy (Ent) \cite{Shannon1948ACommunication, Patro2020FairRec:Platforms}. 
These are set-based measures, but we do not expect the conclusions to differ for rank-based measures (see~\Cref{app:discussion}).

We also use relevance-aware fairness measures, i.e., joint measures of both relevance and fairness (\joint{}) for individual items, as per \cite{Rampisela2024CanRelevance}. These are: 
Item Better-Off (IBO) \cite{Saito2022FairRanking},\footnote{
The measure Item Worse-Off is not used as its formulation is highly similar to IBO.}
Mean Max Envy (MME) \cite{Saito2022FairRanking},
Inequity of Amortized Attention (IAA) \cite{Biega2018EquityRankings, Borges2019EnhancingAutoencoders}, Individual-user-to-individual-item fairness (II-F) \cite{Diaz2020EvaluatingExposure, Wu2022JointRecommendation}, and 
All-users-to-individual-item fairness (AI-F) \cite{Diaz2020EvaluatingExposure}. 
We denote by $\uparrow$/$\downarrow$ measures where the higher/lower the score, the better.
Unless otherwise stated, DPFR is computed with Euclidean distance and $\alpha=0.5$ (PF midpoint) for all datasets; this midpoint is based on the empirically achievable scores, per dataset and measure pair.
We study the effect of varying $\alpha$ in \Cref{ss:varying_alpha} and \Cref{ss:simple}. For all runs, we use $k=10$. 

\subsubsection*{Recommenders} We use the following four common collaborative filtering-based recommenders: Item-based K-Nearest Neighbour (ItemKNN) \cite{Deshpande2004Item-basedAlgorithms},
Bayesian Personalised Ranking (BPR) \cite{Rendle2009BPR:Feedback},
Variational Autoencoder with multinomial likelihood (MultiVAE) \cite{Liang2018VariationalFiltering}, and 
Neighbourhood-enriched Contrastive Learning (NCL) \cite{Lin2022ImprovingLearning}, 
with RecBole \cite{Zhao2021RecBole:Algorithms} and tune their hyperparameters. We train for 300 epochs with early stopping, and keep the configuration with the best NDCG@10 during validation. Each user's train/val items are excluded from their recommendations during testing.

\subsubsection*{Fairness-aware re-rankers} 
In order to have fairer recommendations, we reorder the top-$k'$ items that are pre-optimised for relevance. Ideally, $k'>k$ to allow exposing items that are not in the top-$k$. In RS data, there are generally few relevant items per user.
For the datasets in this paper, the median number of relevant items per user is 2--53 (see \Cref{app:stat}).
Therefore, $k'$ should not be too big (e.g., 100). So, we re-rank the top $k'=25$ items per dataset and model using three methods: GS, CM, and BC (explained below). 
We re-rank separately per user for CM and BC, or altogether for GS, for all $k'm$ recommended items, where $m$ is the number of users. 
Note that other fair ranking methods exist, but they cannot be used here because they apply to group or two-sided fairness only (e.g., \cite{Zehlike2017FAIR:Algorithm, Zehlike2020ReducingApproach, Patro2020FairRec:Platforms}), or to stochastic rankings only (e.g., \cite{Wu2022JointRecommendation, Oosterhuis2021ComputationallyFairness}), or do not scale to larger datasets (e.g., \cite{Biega2018EquityRankings, Saito2022FairRanking}).

\begin{itemize}
    \item Greedy Substitution (GS) \cite{Wang2022ProvidingSystems} is a re-ranker for individual item fairness. We modify the GS algorithm, to replace the most popular items with the least popular ones, both considering how many times an item is at the top-$k'$ recommendations for all users (\Cref{app:gs}). As such, items can be swapped across users. To determine which items are most popular (i.e., to be replaced) and least popular (replacement items), the parameter $\beta=0.05$ is used. We pair these two item types, and for each pair, we calculate the loss of (predicted) relevance if the items are swapped. We then replace up to 25\% of the initial recommendations, starting from item pairs with the least loss. 
    
    \item COMBMNZ (CM) \cite{Fox1993CombinationSearches, Lee1997AnalysesCombination} is a common rank fusion method. Two rankings are fused for each user: one based on the (min-max) normalised predicted relevance score and another based on the coverage of each top-$k'$ item (to approximate fairness). We calculate item coverage only based on their appearance in the top-$k$ across all users and min-max normalise the score across all users. 
    As favouring items with higher coverage would boost unfairness, we generate the ranking using $1$ minus the normalised coverage. CM uses a multiplier based on the item appearance count in the two rankings above; this count is also only based on the top-$k$. 
    The resulting ranking is a fused ranking of fairness and relevance.
    
    \item Borda Count (BC) \cite{Aslam2001BordaFuse} is a common rank fusion method. For each user, we combine the original recommendation list and the rankings based on increasing item coverage, as in CM. Unlike CM, BC uses points. Higher points are given to items placed at the top. The result is a fused ranking of fairness and relevance. 

\end{itemize}

\section{Experimental Results}
\label{ss:result-pareto}

This section describes the experimental findings and is organised as follows. 
\Cref{ss:groundwork} presents the evaluation scores of 16 runs (using 4 recommenders $\times$ 3 re-rankers, including no reranking). Note that the relevance and fairness scores of these runs are the input to our DPFR approach. 
Not all combinations of evaluation measures are suited for PF. We discuss this in \Cref{sss:no-fit}.
\Cref{sss:result-pf} presents the generated PFs of our approach and \Cref{ss:true_pf} compares our PFs to the true PFs. 
Then, we compare existing measures to DPFR (\Cref{ss:corr}). We study the effect of using another distance measure for computing DPFR (\Cref{ss:sensitivity_distance}); the effect of varying $\alpha$, the DPFR fairness-relevance trade-off weight (\Cref{ss:varying_alpha}); and the correlation of DPFR to simple measures of relevance and fairness (\Cref{ss:simple}). 
Lastly, the results of efficient DPFR are compared to other joint evaluation approaches (\Cref{ss:compare-eff}).

\subsection{Groundwork Runs} 
\label{ss:groundwork}

The scores of \rel{}, \fair{}, and \joint{} measures for our 16 runs are shown in \Cref{tab:base-rerank-all-1,tab:base-rerank-all-2}. Two main findings emerge from these tables. First, for all six datasets, \textbf{none of the best models according to \rel{} are also the best according to \fair{} measures}. This is similar to our toy example (\Cref{fig:pareto_teaser}), where one model ranks highest for fairness and another for relevance. 
Second, \textbf{the five \joint{} measures have no unanimous agreement on the best model}. 
IBO has a different best model from the others in 4/6 times, but sometimes agrees with one or more \fair{} measures. 
MME and AI-F agree on the best model 5/6 times, and sometimes agree on the best model with \fair{} measures.
The best model according to IAA and II-F is always the same, and 4/6 times the same as the best model based on the \rel{} measures. 
The overall picture is inconclusive, with some \joint{} measures aligning more with \fair{} measures, and others aligning more with \rel{} measures. 

\begin{table*}[tbp]
\caption{Relevance (\textsc{Rel}), fairness (\textsc{Fair}), as well as joint fairness and relevance (\joint{}) scores at $k=10$ of the recommender models for Lastfm, Amazon-lb, and QK-video, without and with re-ranking the top $k'=25$ items using Borda Count (BC), COMBMNZ (CM), and Greedy Substitution (GS). The most relevant or most fair score per measure is in bold. $\uparrow$ means the higher the better, $\downarrow$ the lower the better.}
\label{tab:base-rerank-all-1}
\resizebox{0.975\textwidth}{!}{
\begin{tabular}{lll*{4}{r}|*{4}{r}|*{4}{r}|*{4}{r}}
\toprule
 &  & model & \multicolumn{4}{c|}{ItemKNN} & \multicolumn{4}{c|}{BPR} & \multicolumn{4}{c|}{MultiVAE} & \multicolumn{4}{c}{NCL} \\ 
\midrule
 &  & reranking & - & BC & CM & GS & - & BC & CM & GS & - & BC & CM & GS & - & BC & CM & GS \\
\midrule
\multirow[c]{16}{*}{\rotatebox[origin=c]{90}{Lastfm}} & \multirow[c]{6}{*}{\rotatebox[origin=c]{90}{\textsc{Rel}}} & $\uparrow$ $\text{HR}$ & 0.765 & 0.742 & 0.581 & 0.750 & 0.773 & 0.729 & 0.587 & 0.751 & 0.778 & 0.693 & 0.523 & 0.734 & \bfseries 0.793 & 0.726 & 0.571 & 0.765 \\
 &  & $\uparrow$ $\text{MRR}$ & 0.484 & 0.333 & 0.270 & 0.481 & 0.492 & 0.323 & 0.280 & 0.488 & 0.476 & 0.285 & 0.232 & 0.470 & \bfseries 0.503 & 0.311 & 0.260 & 0.499 \\
 &  & $\uparrow$ $\text{P}$ & 0.172 & 0.147 & 0.089 & 0.167 & 0.178 & 0.140 & 0.092 & 0.169 & 0.176 & 0.129 & 0.076 & 0.161 & \bfseries 0.184 & 0.141 & 0.087 & 0.173 \\
 &  & $\uparrow$ $\text{MAP}$ & 0.137 & 0.085 & 0.053 & 0.135 & 0.141 & 0.080 & 0.058 & 0.138 & 0.138 & 0.070 & 0.045 & 0.132 & \bfseries 0.148 & 0.079 & 0.050 & 0.144 \\
 &  & $\uparrow$ $\text{R}$ & 0.218 & 0.186 & 0.114 & 0.211 & 0.224 & 0.180 & 0.119 & 0.211 & 0.224 & 0.163 & 0.098 & 0.205 & \bfseries 0.234 & 0.180 & 0.110 & 0.220 \\
 &  & $\uparrow$ $\text{NDCG}$ & 0.245 & 0.181 & 0.119 & 0.241 & 0.252 & 0.173 & 0.126 & 0.244 & 0.247 & 0.155 & 0.102 & 0.235 & \bfseries 0.261 & 0.170 & 0.115 & 0.252 \\
\cline{2-19}
 & \multirow[c]{5}{*}{\rotatebox[origin=c]{90}{\textsc{Fair}}} & $\uparrow$ $\text{Jain}$ & 0.042 & 0.101 & 0.094 & 0.046 & 0.058 & 0.151 & 0.140 & 0.067 & 0.097 & \bfseries 0.236 & 0.222 & 0.115 & 0.082 & 0.216 & 0.215 & 0.095 \\
 &  & $\uparrow$ $\text{QF}$ & 0.474 & 0.642 & \bfseries 0.679 & 0.533 & 0.362 & 0.491 & 0.528 & 0.402 & 0.517 & 0.658 & 0.678 & 0.554 & 0.453 & 0.622 & 0.657 & 0.502 \\
 &  & $\uparrow$ $\text{Ent}$ & 0.589 & 0.727 & 0.735 & 0.622 & 0.610 & 0.736 & 0.740 & 0.646 & 0.707 & 0.820 & \bfseries 0.826 & 0.740 & 0.671 & 0.801 & 0.810 & 0.705 \\
 &  & $\uparrow$ $\text{FSat}$ & 0.129 & 0.197 & 0.216 & 0.152 & 0.147 & 0.211 & 0.228 & 0.177 & 0.202 & 0.293 & \bfseries 0.321 & 0.249 & 0.178 & 0.269 & 0.286 & 0.221 \\
 &  & $\downarrow$ $\text{Gini}$ & 0.904 & 0.810 & 0.790 & 0.879 & 0.910 & 0.827 & 0.818 & 0.887 & 0.839 & 0.715 & \bfseries 0.696 & 0.803 & 0.872 & 0.748 & 0.728 & 0.840 \\
\cline{2-19}
 & \multirow[c]{5}{*}{\rotatebox[origin=c]{90}{\joint{}}} & $\uparrow$ $\text{IBO}$ & 0.209 & 0.270 & 0.256 & 0.227 & 0.208 & 0.263 & 0.253 & 0.228 & 0.261 & \bfseries 0.314 & 0.278 & 0.281 & 0.242 & 0.308 & 0.292 & 0.265 \\
 &  & $\downarrow$ $\text{MME}$ & 0.001 & 0.001 & 0.001 & 0.001 & 0.001 & 0.001 & 0.001 & 0.001 & 0.001 & 0.000 & 0.001 & 0.001 & 0.001 & \bfseries 0.000 & 0.001 & 0.001 \\
 &  & $\downarrow$ $\text{IAA}$ & 0.004 & 0.004 & 0.004 & 0.004 & 0.004 & 0.004 & 0.004 & 0.004 & 0.004 & 0.004 & 0.004 & 0.004 & \bfseries 0.004 & 0.004 & 0.004 & 0.004 \\
 &  & $\downarrow$ $\text{II-F}$ & 0.001 & 0.001 & 0.002 & 0.001 & 0.001 & 0.001 & 0.002 & 0.001 & 0.001 & 0.001 & 0.002 & 0.001 & \bfseries 0.001 & 0.001 & 0.002 & 0.001 \\
 &  & $\downarrow$ $\text{AI-F}$ & 0.000 & 0.000 & 0.000 & 0.000 & 0.000 & 0.000 & 0.000 & 0.000 & 0.000 & 0.000 & 0.000 & 0.000 & 0.000 & \bfseries 0.000 & 0.000 & 0.000 \\
\cline{1-19}
\multirow[c]{16}{*}{\rotatebox[origin=c]{90}{Amazon-lb}} & \multirow[c]{6}{*}{\rotatebox[origin=c]{90}{\textsc{Rel}}} & $\uparrow$ $\text{HR}$ & \bfseries 0.046 & 0.021 & 0.016 & 0.043 & 0.011 & 0.014 & 0.021 & 0.011 & 0.039 & 0.007 & 0.014 & \bfseries 0.046 & 0.034 & 0.021 & 0.011 & 0.034 \\
 &  & $\uparrow$ $\text{MRR}$ & 0.020 & 0.011 & 0.011 & 0.020 & 0.003 & 0.005 & 0.007 & 0.003 & 0.023 & 0.003 & 0.004 & \bfseries 0.024 & 0.022 & 0.006 & 0.003 & 0.022 \\
 &  & $\uparrow$ $\text{P}$ & 0.005 & 0.002 & 0.002 & 0.005 & 0.001 & 0.001 & 0.002 & 0.001 & 0.004 & 0.001 & 0.002 & \bfseries 0.005 & 0.004 & 0.002 & 0.001 & 0.004 \\
 &  & $\uparrow$ $\text{MAP}$ & 0.006 & 0.004 & 0.004 & 0.006 & 0.002 & 0.003 & 0.004 & 0.002 & 0.006 & 0.002 & 0.003 & 0.006 & \bfseries 0.006 & 0.002 & 0.001 & 0.006 \\
 &  & $\uparrow$ $\text{R}$ & \bfseries 0.013 & 0.007 & 0.005 & 0.013 & 0.005 & 0.008 & 0.010 & 0.005 & 0.010 & 0.005 & 0.008 & 0.012 & 0.012 & 0.007 & 0.003 & 0.011 \\
 &  & $\uparrow$ $\text{NDCG}$ & 0.011 & 0.006 & 0.005 & 0.011 & 0.003 & 0.005 & 0.006 & 0.003 & 0.010 & 0.003 & 0.004 & \bfseries 0.011 & 0.011 & 0.004 & 0.002 & 0.011 \\
\cline{2-19}
 & \multirow[c]{5}{*}{\rotatebox[origin=c]{90}{\textsc{Fair}}} & $\uparrow$ $\text{Jain}$ & 0.271 & \bfseries 0.547 & 0.431 & 0.324 & 0.223 & 0.432 & 0.359 & 0.259 & 0.035 & 0.123 & 0.097 & 0.043 & 0.026 & 0.098 & 0.080 & 0.031 \\
 &  & $\uparrow$ $\text{QF}$ & 0.650 & \bfseries 0.679 & 0.612 & 0.663 & 0.549 & 0.630 & 0.594 & 0.571 & 0.222 & 0.294 & 0.286 & 0.254 & 0.229 & 0.315 & 0.310 & 0.265 \\
 &  & $\uparrow$ $\text{Ent}$ & 0.802 & \bfseries 0.882 & 0.839 & 0.829 & 0.747 & 0.839 & 0.809 & 0.776 & 0.418 & 0.587 & 0.558 & 0.469 & 0.371 & 0.560 & 0.534 & 0.426 \\
 &  & $\uparrow$ $\text{FSat}$ & 0.370 & \bfseries 0.538 & 0.438 & 0.435 & 0.314 & 0.410 & 0.376 & 0.364 & 0.114 & 0.159 & 0.152 & 0.138 & 0.091 & 0.146 & 0.138 & 0.115 \\
 &  & $\downarrow$ $\text{Gini}$ & 0.665 & \bfseries 0.492 & 0.598 & 0.613 & 0.747 & 0.601 & 0.660 & 0.703 & 0.949 & 0.882 & 0.899 & 0.930 & 0.959 & 0.898 & 0.910 & 0.943 \\
\cline{2-19}
 & \multirow[c]{5}{*}{\rotatebox[origin=c]{90}{\joint{}}} & $\uparrow$ $\text{IBO}$ & 0.062 & 0.038 & 0.029 & \bfseries 0.067 & 0.019 & 0.029 & 0.038 & 0.019 & 0.029 & 0.019 & 0.029 & 0.033 & 0.038 & 0.033 & 0.024 & 0.033 \\
 &  & $\downarrow$ $\text{MME}$ & 0.001 & 0.001 & 0.001 & 0.001 & 0.001 & \bfseries 0.001 & 0.001 & 0.001 & 0.003 & 0.001 & 0.001 & 0.003 & 0.004 & 0.001 & 0.001 & 0.004 \\
 &  & $\downarrow$ $\text{IAA}$ & 0.011 & 0.011 & 0.011 & \bfseries 0.011 & 0.011 & 0.011 & 0.011 & 0.011 & 0.011 & 0.011 & 0.011 & 0.011 & 0.011 & 0.011 & 0.011 & 0.011 \\
 &  & $\downarrow$ $\text{II-F}$ & 0.006 & 0.006 & 0.006 & \bfseries 0.006 & 0.006 & 0.006 & 0.006 & 0.006 & 0.006 & 0.006 & 0.006 & 0.006 & 0.006 & 0.006 & 0.006 & 0.006 \\
 &  & $\downarrow$ $\text{AI-F}$ & 0.000 & \bfseries 0.000 & 0.000 & 0.000 & 0.000 & 0.000 & 0.000 & 0.000 & 0.001 & 0.000 & 0.000 & 0.001 & 0.002 & 0.000 & 0.000 & 0.002 \\
\cline{1-19}
\multirow[c]{16}{*}{\rotatebox[origin=c]{90}{QK-video}} & \multirow[c]{6}{*}{\rotatebox[origin=c]{90}{\textsc{Rel}}} & $\uparrow$ $\text{HR}$ & 0.040 & 0.046 & 0.047 & 0.038 & 0.099 & 0.063 & 0.045 & 0.089 & 0.109 & 0.089 & 0.061 & 0.103 & \bfseries 0.130 & 0.102 & 0.077 & 0.124 \\
 &  & $\uparrow$ $\text{MRR}$ & 0.013 & 0.014 & 0.013 & 0.013 & 0.039 & 0.018 & 0.015 & 0.038 & 0.039 & 0.028 & 0.021 & 0.038 & \bfseries 0.048 & 0.030 & 0.024 & 0.047 \\
 &  & $\uparrow$ $\text{P}$ & 0.004 & 0.005 & 0.005 & 0.004 & 0.011 & 0.007 & 0.005 & 0.010 & 0.012 & 0.009 & 0.006 & 0.011 & \bfseries 0.014 & 0.011 & 0.008 & 0.013 \\
 &  & $\uparrow$ $\text{MAP}$ & 0.005 & 0.005 & 0.005 & 0.005 & 0.017 & 0.008 & 0.006 & 0.016 & 0.018 & 0.012 & 0.009 & 0.017 & \bfseries 0.022 & 0.013 & 0.010 & 0.021 \\
 &  & $\uparrow$ $\text{R}$ & 0.014 & 0.018 & 0.019 & 0.014 & 0.043 & 0.028 & 0.019 & 0.039 & 0.051 & 0.039 & 0.027 & 0.047 & \bfseries 0.061 & 0.045 & 0.033 & 0.058 \\
 &  & $\uparrow$ $\text{NDCG}$ & 0.009 & 0.011 & 0.010 & 0.009 & 0.029 & 0.015 & 0.011 & 0.027 & 0.031 & 0.022 & 0.016 & 0.030 & \bfseries 0.038 & 0.025 & 0.019 & 0.037 \\
\cline{2-19}
 & \multirow[c]{5}{*}{\rotatebox[origin=c]{90}{\textsc{Fair}}} & $\uparrow$ $\text{Jain}$ & 0.483 & \bfseries 0.815 & 0.589 & 0.567 & 0.081 & 0.333 & 0.379 & 0.101 & 0.012 & 0.038 & 0.032 & 0.014 & 0.020 & 0.076 & 0.071 & 0.023 \\
 &  & $\uparrow$ $\text{QF}$ & 0.901 & \bfseries 0.956 & 0.790 & 0.924 & 0.625 & 0.809 & 0.823 & 0.678 & 0.100 & 0.155 & 0.163 & 0.127 & 0.201 & 0.331 & 0.365 & 0.253 \\
 &  & $\uparrow$ $\text{Ent}$ & 0.933 & \bfseries 0.979 & 0.937 & 0.950 & 0.755 & 0.888 & 0.903 & 0.792 & 0.420 & 0.557 & 0.547 & 0.458 & 0.507 & 0.667 & 0.674 & 0.549 \\
 &  & $\uparrow$ $\text{FSat}$ & 0.443 & \bfseries 0.659 & 0.547 & 0.522 & 0.212 & 0.346 & 0.382 & 0.259 & 0.052 & 0.089 & 0.090 & 0.070 & 0.077 & 0.140 & 0.150 & 0.104 \\
 &  & $\downarrow$ $\text{Gini}$ & 0.472 & \bfseries 0.235 & 0.442 & 0.397 & 0.807 & 0.613 & 0.570 & 0.761 & 0.982 & 0.957 & 0.959 & 0.976 & 0.966 & 0.909 & 0.902 & 0.952 \\
\cline{2-19}
 & \multirow[c]{5}{*}{\rotatebox[origin=c]{90}{\joint{}}} & $\uparrow$ $\text{IBO}$ & 0.033 & 0.038 & 0.038 & 0.035 & 0.054 & 0.050 & 0.036 & 0.052 & 0.031 & 0.042 & 0.036 & 0.033 & 0.043 & \bfseries 0.060 & 0.054 & 0.047 \\
 &  & $\downarrow$ $\text{MME}$ & 0.000 & \bfseries 0.000 & 0.000 & 0.000 & 0.000 & 0.000 & 0.000 & 0.000 & 0.000 & 0.000 & 0.000 & 0.000 & 0.000 & 0.000 & 0.000 & 0.000 \\
 &  & $\downarrow$ $\text{IAA}$ & 0.001 & 0.001 & 0.001 & 0.001 & 0.001 & 0.001 & 0.001 & 0.001 & 0.001 & 0.001 & 0.001 & 0.001 & \bfseries 0.001 & 0.001 & 0.001 & 0.001 \\
 &  & $\downarrow$ $\text{II-F}$ & 0.001 & 0.001 & 0.001 & 0.001 & 0.001 & 0.001 & 0.001 & 0.001 & 0.001 & 0.001 & 0.001 & 0.001 & \bfseries 0.001 & 0.001 & 0.001 & 0.001 \\
 &  & $\downarrow$ $\text{AI-F}$ & 0.000 & \bfseries 0.000 & 0.000 & 0.000 & 0.000 & 0.000 & 0.000 & 0.000 & 0.000 & 0.000 & 0.000 & 0.000 & 0.000 & 0.000 & 0.000 & 0.000 \\
\bottomrule
\end{tabular}}
\end{table*}

\begin{table*}[tbp]
\caption{Relevance (\textsc{Rel}), fairness (\textsc{Fair}), as well as joint fairness and relevance (\joint{}) scores at $k=10$ of the recommender models for Jester and ML-*, without and with re-ranking the top $k'=25$ items using Borda Count (BC), COMBMNZ (CM), and Greedy Substitution (GS). The most relevant or most fair score per measure is in bold. $\uparrow$ means the higher the better, $\downarrow$ the lower the better.}
\label{tab:base-rerank-all-2}
\resizebox{0.975\textwidth}{!}{
\begin{tabular}{lll*{4}{r}|*{4}{r}|*{4}{r}|*{4}{r}}
\toprule
 &  & model & \multicolumn{4}{c|}{ItemKNN} & \multicolumn{4}{c|}{BPR} & \multicolumn{4}{c|}{MultiVAE} & \multicolumn{4}{c}{NCL} \\ 
\midrule
 &  & reranking & - & BC & CM & GS & - & BC & CM & GS & - & BC & CM & GS & - & BC & CM & GS \\
\midrule
\multirow[c]{16}{*}{\rotatebox[origin=c]{90}{Jester}} & \multirow[c]{6}{*}{\rotatebox[origin=c]{90}{\textsc{Rel}}} & $\uparrow$ $\text{HR}$ & 0.933 & 0.888 & 0.652 & 0.932 & 0.929 & 0.876 & 0.742 & 0.928 & \bfseries 0.944 & 0.899 & 0.818 & 0.944 & 0.939 & 0.893 & 0.804 & 0.939 \\
 &  & $\uparrow$ $\text{MRR}$ & 0.632 & 0.443 & 0.307 & 0.632 & 0.635 & 0.455 & 0.322 & 0.635 & \bfseries 0.661 & 0.465 & 0.370 & 0.661 & 0.651 & 0.479 & 0.349 & 0.651 \\
 &  & $\uparrow$ $\text{P}$ & 0.334 & 0.250 & 0.144 & 0.333 & 0.330 & 0.243 & 0.163 & 0.329 & \bfseries 0.351 & 0.262 & 0.194 & 0.351 & 0.342 & 0.257 & 0.185 & 0.341 \\
 &  & $\uparrow$ $\text{MAP}$ & 0.352 & 0.198 & 0.101 & 0.352 & 0.348 & 0.195 & 0.112 & 0.348 & \bfseries 0.379 & 0.208 & 0.145 & 0.379 & 0.367 & 0.211 & 0.133 & 0.367 \\
 &  & $\uparrow$ $\text{R}$ & 0.529 & 0.393 & 0.197 & 0.529 & 0.524 & 0.377 & 0.255 & 0.523 & \bfseries 0.555 & 0.405 & 0.324 & 0.555 & 0.543 & 0.400 & 0.305 & 0.542 \\
 &  & $\uparrow$ $\text{NDCG}$ & 0.496 & 0.336 & 0.189 & 0.496 & 0.493 & 0.331 & 0.216 & 0.492 & \bfseries 0.525 & 0.350 & 0.265 & 0.524 & 0.512 & 0.352 & 0.249 & 0.511 \\
\cline{2-19}
 & \multirow[c]{5}{*}{\rotatebox[origin=c]{90}{\textsc{Fair}}} & $\uparrow$ $\text{Jain}$ & 0.343 & 0.556 & 0.445 & 0.345 & 0.377 & \bfseries 0.583 & 0.547 & 0.380 & 0.295 & 0.544 & 0.509 & 0.297 & 0.351 & 0.504 & 0.534 & 0.354 \\
 &  & $\uparrow$ $\text{QF}^*$ & \bfseries 1.000 & \bfseries 1.000 & \bfseries 1.000 & \bfseries 1.000 & \bfseries 1.000 & \bfseries 1.000 & \bfseries 1.000 & \bfseries 1.000 & 0.967 & \bfseries 1.000 & \bfseries 1.000 & \bfseries 1.000 & \bfseries 1.000 & \bfseries 1.000 & \bfseries 1.000 & \bfseries 1.000 \\
 &  & $\uparrow$ $\text{Ent}$ & 0.702 & 0.854 & 0.784 & 0.705 & 0.754 & \bfseries 0.875 & 0.857 & 0.757 & 0.648 & 0.852 & 0.839 & 0.651 & 0.722 & 0.838 & 0.855 & 0.725 \\
 &  & $\uparrow$ $\text{FSat}$ & 0.267 & \bfseries 0.378 & 0.289 & 0.267 & 0.244 & 0.344 & 0.333 & 0.244 & 0.256 & 0.344 & 0.300 & 0.256 & 0.222 & 0.344 & 0.311 & 0.222 \\
 &  & $\downarrow$ $\text{Gini}$ & 0.687 & 0.502 & 0.595 & 0.685 & 0.632 & \bfseries 0.467 & 0.495 & 0.629 & 0.738 & 0.506 & 0.520 & 0.735 & 0.668 & 0.528 & 0.502 & 0.665 \\
\cline{2-19}
 & \multirow[c]{5}{*}{\rotatebox[origin=c]{90}{\joint{}}} & $\uparrow$ $\text{IBO}$ & 0.600 & \bfseries 0.930 & 0.740 & 0.600 & 0.840 & 0.910 & 0.780 & 0.840 & 0.500 & 0.870 & 0.810 & 0.500 & 0.740 & 0.920 & 0.780 & 0.740 \\
 &  & $\downarrow$ $\text{MME}$ & 0.003 & 0.003 & 0.006 & 0.003 & 0.004 & \bfseries 0.002 & 0.005 & 0.004 & 0.008 & 0.003 & 0.004 & 0.008 & 0.004 & 0.003 & 0.006 & 0.004 \\
 &  & $\downarrow$ $\text{IAA}$ & 0.081 & 0.093 & 0.104 & 0.081 & 0.081 & 0.094 & 0.103 & 0.081 & 0.078 & 0.092 & 0.100 & \bfseries 0.078 & 0.079 & 0.092 & 0.101 & 0.079 \\
 &  & $\downarrow$ $\text{II-F}$ & 0.028 & 0.035 & 0.040 & 0.028 & 0.029 & 0.035 & 0.040 & 0.029 & 0.027 & 0.035 & 0.038 & \bfseries 0.027 & 0.028 & 0.034 & 0.038 & 0.028 \\
 &  & $\downarrow$ $\text{AI-F}$ & 0.002 & 0.002 & 0.003 & 0.002 & 0.002 & \bfseries 0.001 & 0.002 & 0.002 & 0.003 & 0.002 & 0.002 & 0.003 & 0.002 & 0.002 & 0.002 & 0.002 \\
\cline{1-19}
\multirow[c]{16}{*}{\rotatebox[origin=c]{90}{ML-10M}} & \multirow[c]{6}{*}{\rotatebox[origin=c]{90}{\textsc{Rel}}} & $\uparrow$ $\text{HR}$ & 0.487 & 0.480 & 0.443 & 0.481 & 0.512 & 0.462 & 0.386 & 0.485 & 0.417 & 0.438 & 0.387 & 0.410 & \bfseries 0.521 & 0.473 & 0.402 & 0.513 \\
 &  & $\uparrow$ $\text{MRR}$ & 0.282 & 0.242 & 0.225 & 0.279 & 0.299 & 0.208 & 0.185 & 0.295 & 0.237 & 0.231 & 0.191 & 0.235 & \bfseries 0.302 & 0.216 & 0.203 & 0.301 \\
 &  & $\uparrow$ $\text{P}$ & 0.137 & 0.128 & 0.105 & 0.133 & 0.146 & 0.114 & 0.088 & 0.132 & 0.107 & 0.111 & 0.096 & 0.105 & \bfseries 0.154 & 0.123 & 0.094 & 0.149 \\
 &  & $\uparrow$ $\text{MAP}$ & 0.089 & 0.074 & 0.060 & 0.086 & 0.095 & 0.061 & 0.047 & 0.088 & 0.067 & 0.067 & 0.054 & 0.066 & \bfseries 0.101 & 0.067 & 0.052 & 0.099 \\
 &  & $\uparrow$ $\text{R}$ & 0.022 & 0.022 & 0.018 & 0.022 & 0.025 & 0.019 & 0.012 & 0.023 & 0.020 & 0.021 & 0.016 & 0.021 & \bfseries 0.026 & 0.020 & 0.013 & 0.026 \\
 &  & $\uparrow$ $\text{NDCG}$ & 0.150 & 0.133 & 0.113 & 0.147 & 0.160 & 0.115 & 0.092 & 0.150 & 0.119 & 0.121 & 0.100 & 0.118 & \bfseries 0.167 & 0.123 & 0.100 & 0.164 \\
\cline{2-19}
 & \multirow[c]{5}{*}{\rotatebox[origin=c]{90}{\textsc{Fair}}} & $\uparrow$ $\text{Jain}$ & 0.011 & 0.026 & 0.027 & 0.012 & 0.037 & 0.100 & \bfseries 0.115 & 0.044 & 0.003 & 0.005 & 0.006 & 0.004 & 0.024 & 0.063 & 0.069 & 0.027 \\
 &  & $\uparrow$ $\text{QF}^\dag$ & 0.044 & 0.062 & 0.068 & 0.047 & 0.145 & 0.199 & \bfseries 0.216 & 0.160 & 0.014 & 0.021 & 0.025 & 0.016 & 0.086 & 0.123 & 0.132 & 0.094 \\
 &  & $\uparrow$ $\text{Ent}$ & 0.407 & 0.503 & 0.514 & 0.418 & 0.596 & 0.697 & \bfseries 0.716 & 0.624 & 0.238 & 0.302 & 0.324 & 0.258 & 0.519 & 0.625 & 0.638 & 0.544 \\
 &  & $\uparrow$ $\text{FSat}^\dag$ & 0.044 & 0.062 & 0.068 & 0.047 & 0.145 & 0.199 & \bfseries 0.216 & 0.160 & 0.014 & 0.021 & 0.025 & 0.016 & 0.086 & 0.123 & 0.132 & 0.094 \\
 &  & $\downarrow$ $\text{Gini}$ & 0.987 & 0.973 & 0.971 & 0.985 & 0.945 & 0.895 & \bfseries 0.879 & 0.932 & 0.997 & 0.994 & 0.993 & 0.996 & 0.969 & 0.936 & 0.930 & 0.963 \\
\cline{2-19}
 & \multirow[c]{5}{*}{\rotatebox[origin=c]{90}{\joint{}}} & $\uparrow$ $\text{IBO}$ & 0.031 & 0.043 & 0.046 & 0.034 & 0.069 & 0.089 & \bfseries 0.091 & 0.076 & 0.012 & 0.016 & 0.018 & 0.014 & 0.054 & 0.073 & 0.074 & 0.058 \\
 &  & $\downarrow$ $\text{MME}$ & 0.001 & 0.001 & 0.001 & 0.001 & 0.001 & \bfseries 0.001 & 0.001 & 0.001 & 0.003 & 0.002 & 0.001 & 0.002 & 0.001 & 0.001 & 0.001 & 0.001 \\
 &  & $\downarrow$ $\text{IAA}$ & 0.008 & 0.009 & 0.009 & 0.008 & 0.008 & 0.009 & 0.009 & 0.008 & 0.009 & 0.009 & 0.009 & 0.009 & \bfseries 0.008 & 0.009 & 0.009 & 0.008 \\
 &  & $\downarrow$ $\text{II-F}$ & 0.000 & 0.000 & 0.000 & 0.000 & 0.000 & 0.000 & 0.000 & 0.000 & 0.000 & 0.000 & 0.000 & 0.000 & \bfseries 0.000 & 0.000 & 0.000 & 0.000 \\
 &  & $\downarrow$ $\text{AI-F}$ & 0.000 & 0.000 & 0.000 & 0.000 & 0.000 & \bfseries 0.000 & 0.000 & 0.000 & 0.000 & 0.000 & 0.000 & 0.000 & 0.000 & 0.000 & 0.000 & 0.000 \\
\cline{1-19}
\multirow[c]{16}{*}{\rotatebox[origin=c]{90}{ML-20M}} & \multirow[c]{6}{*}{\rotatebox[origin=c]{90}{\textsc{Rel}}} & $\uparrow$ $\text{HR}$ & 0.488 & 0.473 & 0.420 & 0.483 & \bfseries 0.505 & 0.444 & 0.392 & 0.483 & 0.489 & 0.432 & 0.391 & 0.465 & \bfseries 0.505 & 0.453 & 0.388 & 0.493 \\
 &  & $\uparrow$ $\text{MRR}$ & 0.280 & 0.237 & 0.213 & 0.278 & 0.293 & 0.205 & 0.190 & 0.290 & 0.259 & 0.193 & 0.180 & 0.256 & \bfseries 0.293 & 0.206 & 0.193 & 0.292 \\
 &  & $\uparrow$ $\text{P}$ & 0.142 & 0.131 & 0.106 & 0.139 & 0.145 & 0.116 & 0.094 & 0.136 & 0.141 & 0.112 & 0.091 & 0.128 & \bfseries 0.150 & 0.121 & 0.094 & 0.141 \\
 &  & $\uparrow$ $\text{MAP}$ & 0.092 & 0.077 & 0.061 & 0.090 & 0.096 & 0.063 & 0.052 & 0.092 & 0.089 & 0.060 & 0.049 & 0.082 & \bfseries 0.100 & 0.068 & 0.053 & 0.095 \\
 &  & $\uparrow$ $\text{R}$ & 0.019 & 0.017 & 0.014 & 0.019 & 0.019 & 0.014 & 0.012 & 0.018 & 0.019 & 0.014 & 0.011 & 0.018 & \bfseries 0.020 & 0.016 & 0.011 & 0.020 \\
 &  & $\uparrow$ $\text{NDCG}$ & 0.154 & 0.135 & 0.112 & 0.151 & 0.158 & 0.116 & 0.098 & 0.152 & 0.148 & 0.111 & 0.093 & 0.139 & \bfseries 0.163 & 0.121 & 0.099 & 0.157 \\
\cline{2-19}
 & \multirow[c]{5}{*}{\rotatebox[origin=c]{90}{\textsc{Fair}}} & $\uparrow$ $\text{Jain}$ & 0.008 & 0.017 & 0.018 & 0.009 & 0.028 & 0.068 & \bfseries 0.081 & 0.033 & 0.029 & 0.070 & 0.074 & 0.034 & 0.018 & 0.044 & 0.049 & 0.021 \\
 &  & $\uparrow$ $\text{QF}^\dag$ & 0.035 & 0.047 & 0.051 & 0.037 & 0.114 & 0.154 & \bfseries 0.165 & 0.125 & 0.117 & 0.146 & 0.154 & 0.126 & 0.074 & 0.103 & 0.112 & 0.082 \\
 &  & $\uparrow$ $\text{Ent}$ & 0.399 & 0.483 & 0.491 & 0.411 & 0.581 & 0.670 & \bfseries 0.690 & 0.606 & 0.591 & 0.669 & 0.680 & 0.615 & 0.517 & 0.608 & 0.624 & 0.541 \\
 &  & $\uparrow$ $\text{FSat}^\dag$ & 0.035 & 0.047 & 0.051 & 0.037 & 0.114 & 0.154 & \bfseries 0.165 & 0.125 & 0.117 & 0.146 & 0.154 & 0.126 & 0.074 & 0.103 & 0.112 & 0.082 \\
 &  & $\downarrow$ $\text{Gini}$ & 0.991 & 0.982 & 0.981 & 0.990 & 0.960 & 0.926 & \bfseries 0.914 & 0.951 & 0.957 & 0.927 & 0.920 & 0.948 & 0.976 & 0.953 & 0.947 & 0.971 \\
\cline{2-19}
 & \multirow[c]{5}{*}{\rotatebox[origin=c]{90}{\joint{}}} & $\uparrow$ $\text{IBO}$ & 0.021 & 0.031 & 0.033 & 0.022 & 0.049 & 0.064 & \bfseries 0.067 & 0.054 & 0.052 & 0.064 & 0.065 & 0.056 & 0.039 & 0.051 & 0.054 & 0.042 \\
 &  & $\downarrow$ $\text{MME}$ & 0.001 & 0.001 & 0.001 & 0.001 & 0.001 & 0.000 & \bfseries 0.000 & 0.001 & 0.001 & 0.000 & 0.000 & 0.001 & 0.001 & 0.001 & 0.001 & 0.001 \\
 &  & $\downarrow$ $\text{IAA}$ & 0.007 & 0.007 & 0.007 & 0.007 & 0.007 & 0.007 & 0.007 & 0.007 & 0.007 & 0.007 & 0.007 & 0.007 & \bfseries 0.007 & 0.007 & 0.007 & 0.007 \\
 &  & $\downarrow$ $\text{II-F}$ & 0.000 & 0.000 & 0.000 & 0.000 & 0.000 & 0.000 & 0.000 & 0.000 & 0.000 & 0.000 & 0.000 & 0.000 & \bfseries 0.000 & 0.000 & 0.000 & 0.000 \\
 &  & $\downarrow$ $\text{AI-F}$ & 0.000 & 0.000 & 0.000 & 0.000 & 0.000 & 0.000 & \bfseries 0.000 & 0.000 & 0.000 & 0.000 & 0.000 & 0.000 & 0.000 & 0.000 & 0.000 & 0.000 \\
\bottomrule
\multicolumn{19}{l}{\small *QF $=1$ means that all items in the dataset appear in the recommendation across all users.} \\
\multicolumn{19}{l}{\small $^{\dag}$The scores of QF are the same as FSat for ML-*, as QF is computed based on the percentage of items in the dataset that are recommended, which in these datasets is} \\
\multicolumn{19}{l}{\small equivalent to FSat: the percentage of items in the dataset that are recommended at least $\left \lfloor \frac{km}{n}\right \rfloor=1$ time.}
\end{tabular}}
\end{table*}

\subsection{Measure Compatibility with DPFR}
\label{sss:no-fit}

We study the compatibility of single-aspect measures with DPFR: which pairs of \rel{} and \fair{} measures are suitable to generate the PF? We answer this based on the PF slope. The slope is calculated using the two endpoints of the PF, i.e., the start and end of the  \textsc{Oracle2Fair} algorithm. A slope of zero means the \rel{} scores of the PF vary, but the \fair{} scores do not. As we compute the PF for multiple measures simultaneously, we expect a zero gradient for cases where the initial recommendation according to a \fair{} measure is already the fairest, even if other \fair{} scores are not. An undefined gradient value occurs when the initial recommendation is already the fairest and the most relevant according to a pair of \rel{} and \fair{} measures. Thus, we posit that a PF with a gradient value other than zero or undefined makes the corresponding pair of measures fit for PF generation, as it allows for trade-offs in both aspects. 
The \rel{}-\fair{} measure pairs that are fit for DPFR based on their gradient are: \{P, MAP, R, NDCG\} $\times$ \{Jain, Ent, Gini\} (\Cref{app:actual-scores}, \Cref{tab:app-gradient}). Henceforth, only results from these pairs are shown. Next, we explain what causes an undefined or zero gradient. 

\subsubsection*{Causes of zero/undefined gradient}
Generating the PF requires a ranking of items. Any score that is based on a single relevant item, e.g., HR and MRR, is by design not suitable. 
Out of the \fair{} measures, QF and FSat sometimes behave inconsistently depending on the dataset properties, as follows. 
A dataset with relatively few relevant items can already be made maximally fair at the start of the PF generation, as QF quantifies fairness with ignorance to frequency of exposure; the score does not change as long as the same set of items appears in the top-$k$ recommendations of all users, no matter how many times each. 
When all items in the dataset already occur in the initial recommendations of our Oracle, nothing can be done to improve QF. 
For FSat, in a few cases, the score is already maximum at the start of the PF generation. A maximum FSat score is achieved when all items in the dataset have at least the maximum possible exposure, if the available recommendation slots are shared equally across all items, which is $\left\lfloor km/n\right\rfloor$, the total number of recommendation slots across users divided by the number of items. In principle, QF and FSat can still be used for DPFR when the initial recommendation by Oracle is not yet the maximum. Otherwise, the interpretation would be less meaningful in joint evaluation, as there is no trade-off between different aspects.

\subsection{The Generated PF} 
\label{sss:result-pf}

\begin{figure}[htbp]
    \centering
    \includegraphics[width=0.8\columnwidth]{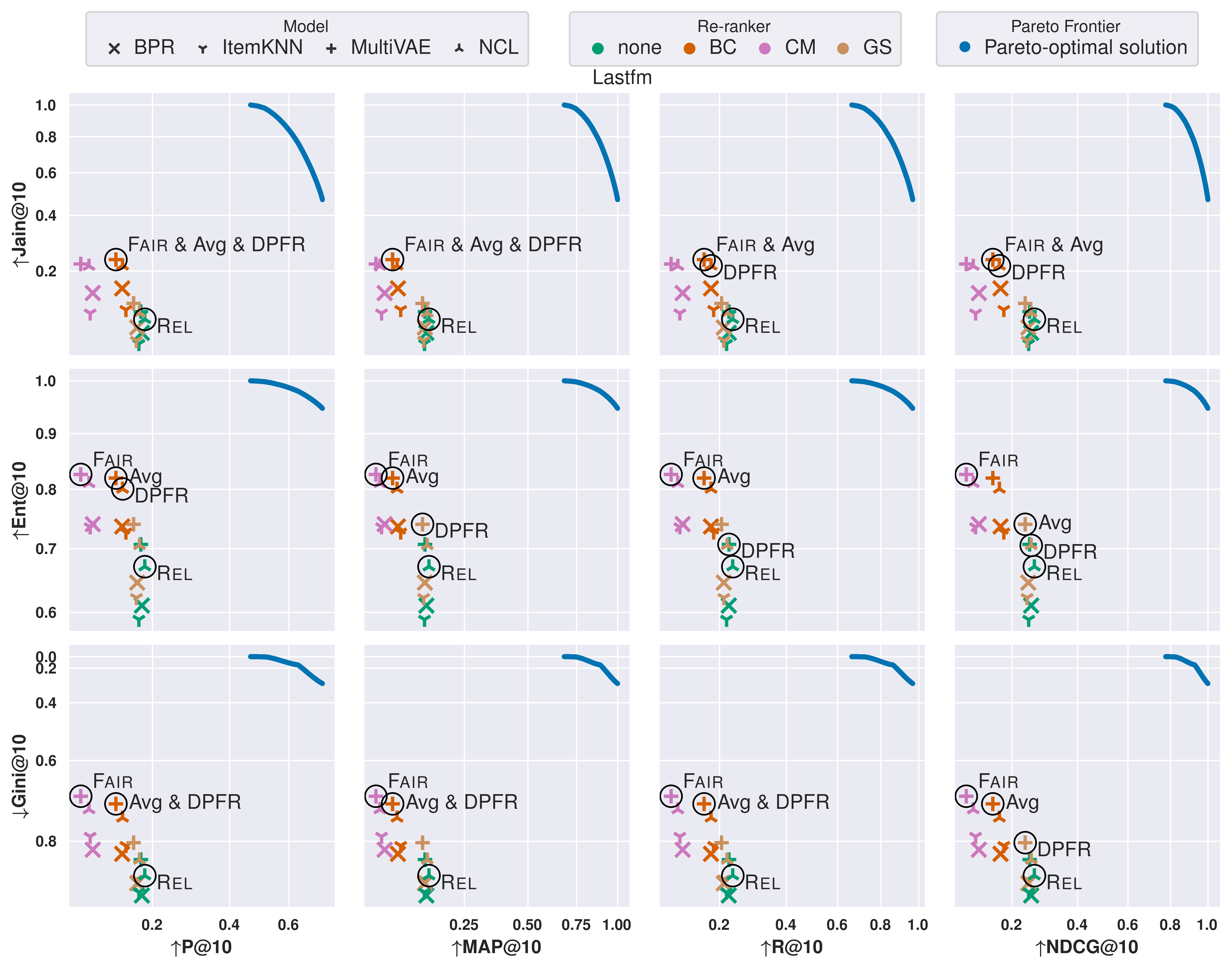}
    \includegraphics[width=0.8\columnwidth]{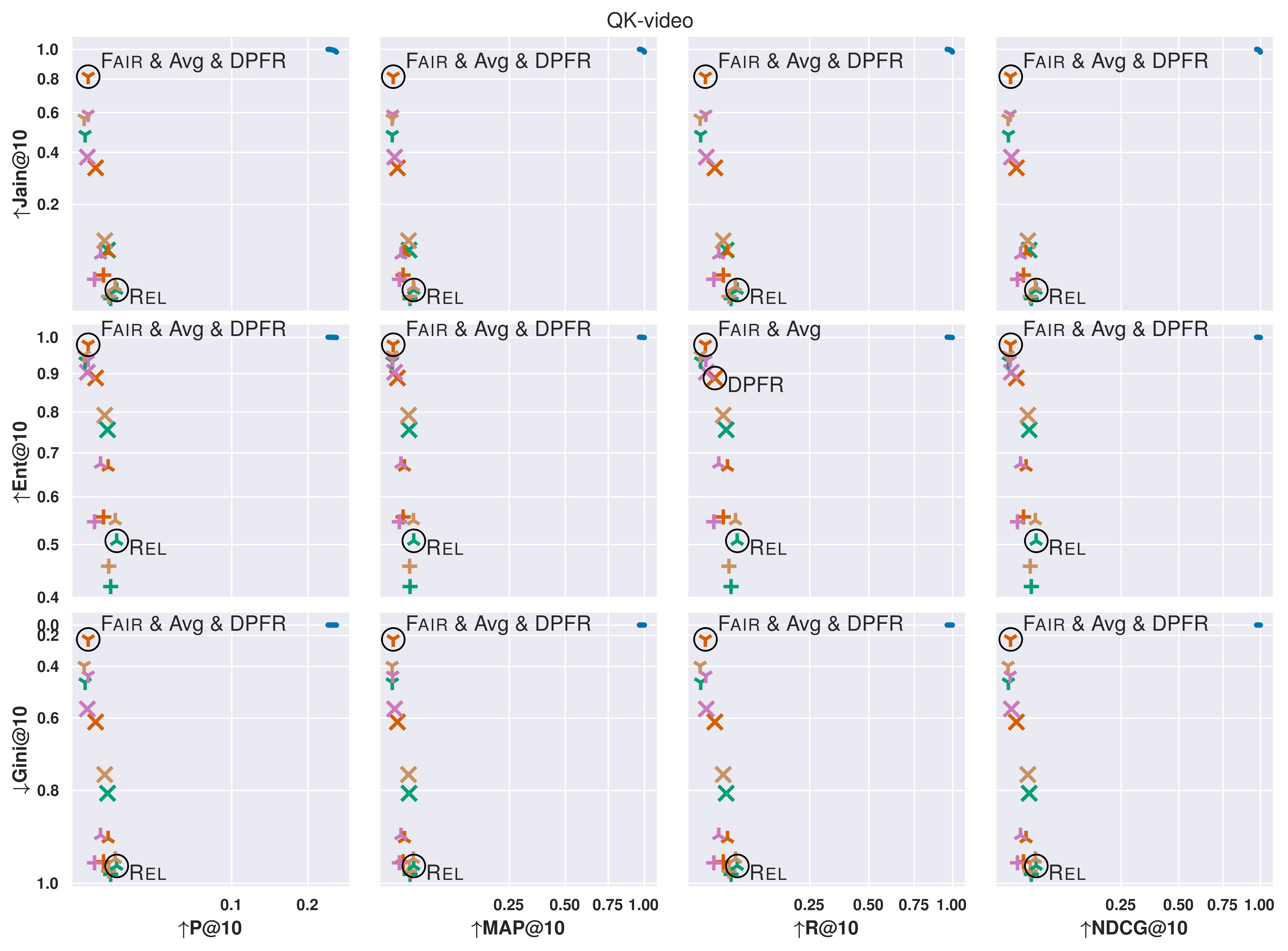}
    \caption{Pareto Frontier of fairness and relevance (in blue) and recommender scores for Lastfm and QK-video on exponential-like scales. \rel{}, \fair{}, avg (mean of \rel{} and \fair{}), and DPFR denote the best model per evaluation approach.}
    \label{fig:pareto-pair-plot}
\end{figure}

\Cref{fig:pareto-pair-plot} shows the PF plots of the pairs of \rel{} and \fair{} measures that are suited for DPFR, only for Lastfm and QK-video, which are representative of the overall trends in all our datasets (see \Cref{app:actual-scores}, \Cref{fig:app-pairplot}). The scores plotted are those computed in \Cref{ss:generation}. The corresponding scores of our recommendation models are in \Cref{app:actual-scores}. 
We see that, as the recommendations are made fairer, the generated PF for all datasets is a series of monotonic scores of \fair{}, specifically monotonic increasing \fair{} scores (except $\downarrow$Gini), and the remaining measures are monotonic decreasing. 
The monotonicity should theoretically and empirically hold for the \fair{} measures, as we replace an item with the most exposure by another item with the least exposure, thereby making the recommendation fairer. 
Note that some users do not have exactly $k$ relevant items in the test set, so the perfect relevance score cannot be reached for Precision@$k$ and Recall@$k$ \cite{Moffat2013SevenMetrics}. NDCG and MAP are implemented with normalisation so that they can still achieve a score of 1 in this situation. Normalisation means that only the first $\min{(|R^*_u|, k)}$ items in a user $u$'s recommendations are considered, where $R^*_u$ is the set of relevant items for user $u$.

\subsubsection*{Short PF} The datasets that were randomly split, as they have no timestamps (QK-video, Jester), have relatively short, compact PF. This happens because the random split results in a uniform distribution of items in each split, which means that items in the test split are quite diverse (64--100\% of all unique items in the dataset). Considering that the \textsc{Oracle2Fair} algorithm starts by recommending items in the test split and stops when the recommendation reaches the fairest, there is not much room for change in \fair{} scores, as the initial recommendation is already rather fair. Additionally, there are not many relevant items per user in these datasets (i.e., the median for both datasets is 6 or less); random non-relevant items were chosen to make up for the remaining recommendations.
Thus, the PF generation decreases relevance only marginally in 2/6 cases. 
Correspondingly, we find that in QK-video and Jester, there exist Pareto-optimal recommendations, that are close to maximally fair and maximally relevant, with the exception of P@10. These can be seen in the measure pairs of \{MAP, R, NDCG\} $\times$ \{Jain, Ent, Gini\}, where the PF is close to the coordinates of $(1,1)$, or $(1,0)$ for 
the measure pairs with Gini. Thus, these results show that \textbf{a fair recommendation does not necessarily have to sacrifice relevance}. 
Note that Lastfm is also split randomly but does not have a short PF because on average it has more relevant items per user compared to QK-video and Jester (see \Cref{app:stat}).

\subsection{Comparison to True PF (RQ1)}
\label{ss:true_pf}

How close is our generated PF to the true PF? We explain the complexity of addressing this question and how we resolve it in our experimental setup.

\subsubsection*{Complexity of computing the true PF} Generally, computing the true PF on real-world datasets is regarded impractical due to its complexity~\cite{Laszczyk2019SurveyMeasures,Audet2020PerformanceOptimization}, and this applies to RS datasets as well. There are $\binom{n}{k}$ possible ways of recommending a set of $k$ items out of $n$ items to a single user, not even considering the order of items. Suppose that we use a dataset with a relatively small number of items, e.g., Jester with $n=100$ and a common choice of $k=10$, there are $\binom{100}{10}\approx$ 17 trillion possible combinations of items for a single user alone. With a machine that can complete 100,000 iterations per minute, it would take around 323 years to go through all combinations, which is not feasible. 

One may also think of enumerating all the possible \textit{scores} instead of the possible items, as two distinct recommendation lists can map to the same \rel{} or \fair{} score. However, this comes with other issues. Particularly, without knowing the exact item combinations, it is hard to identify which \rel{} score can be paired with which \fair{} score for a particular dataset, and vice versa.

\subsubsection*{Experimental setup}
To analyse the closeness of our generated PF to the true PFs, we brute-force the true PF for a small subset of each dataset, and then compare it to our algorithm's generated PF. Next, we detail the process of performing exhaustive search for the true PFs and how the closeness is evaluated.

\paragraph{Exhaustive search for true PFs} Taking into account the complexity of computing the true PF discussed above, we ask: how big of a dataset subset can be used such that the computation is still feasible? With 5 users, 5 items, and $k=3$, there are $\binom{5}{3}^5=100{,}000$ combinations, which can be completed in about a minute. Increasing either the number of users or items to 10 would be infeasible, as it would result in more than 10 billion combinations, which would take about 69 days to complete. Hence, to maintain reasonable computational cost while maintaining robustness, we take 100 samples of 2--5 users for each dataset and ensure that there are no duplicates of samples. The number of users should be at least 2, as evaluating item fairness (with the measures in this paper) for just one user is not meaningful. For the set of items, we take the union of test set items across all users within a sample. However, if users are sampled randomly, they altogether may have more than 5 items. Therefore, we need to constrain the sample to users whose test sets collectively have 4--5 unique, relevant items.

To select the 2--5 users given the above constraint, we first get the $n_{pop}$ most popular items in the full dataset's test sets  (i.e., items that are relevant for many users), as they are more likely to be shared across multiple users than less popular items. Note that the number of $n_{pop}$ differs for each dataset. Depending on the dataset's test item distribution, if $n_{pop}$ is too low, there may not be 100 unique samples (or any samples) of 2--5 users that satisfy the constraints. Through trial and error with manual increments, we set $n_{pop}=5$ for Amazon-lb, QK-video, and Jester; $n_{pop}=15$ for ML-20M; $n_{pop}=25$ for ML-10M; and $n_{pop}=255$ for Lastfm. We then get a set of users whose test set is a subset of the $n_{pop}$ most popular items. From this set of users, we randomly sample 5 and verify that these users altogether satisfy the constraint of having 4--5 unique test items; if not, we perform re-sampling. If no sample fulfils the constraint after 100 iterations, the number of users to be sampled is decreased by 1 (until a minimum of two), and the sampling process is repeated.

For each sample, we enumerate all possible sets of recommendations. To get the maximum \rel{} score for a given set of items, we rank the items based on decreasing relevance, based on the user's test set. For each possible recommendation, we then compute the NDCG@3 (averaged across users) and Ent@3, to represent \rel{} and \fair{} measures, respectively; other \rel{} or \fair{} measures are not computed to reduce computational cost. This results in at most 100,000 non-unique pairs of NDCG-Ent scores. To get the true PF for each sample, we keep only the Pareto-optimal pairs of NDCG-Ent scores, i.e., pairs that fulfill both conditions across all score pairs: (1) the best Ent score for a given NDCG; and (2) the best NDCG for a given Ent score (\Cref{ss:motivation}). This process yields 100 true PFs (one for each sample) per dataset. Separately, we generate our PF for each sample with the {\sc Oracle2Fair} algorithm (\Cref{app:algo}, \Cref{alg:oracle2fair}) and remove the non-Pareto-optimal scores. This results in 100 generated PFs (one per sample) for each dataset. 

\paragraph{Evaluation} To compare how close our generated PFs are to the brute-forced (true) PFs, we compute Generational Distance (GD), the average Euclidean distance from any point in our generated PF to the closest point in the true PF \cite{vanVeldhuizenSAC, VanVeldhuizen:1998:eccpf}. As such, lower GD is desirable, and zero GD means that our generated PF exactly matches the true PF. We use \texttt{pymoo} \cite{pymoo} library's implementation of GD. The results are averaged across the 100 samples and reported in \Cref{tab:GD}.

\begin{table}
    \caption{Mean Generational Distance (GD) and percentage of zero GD between our generated PFs and the true, brute-forced PFs. Results are aggregated across 100 random samples of users from each dataset. Mean GD is rounded to 3 d.p.}
    \label{tab:GD}

    \centering
    \begin{tabular}{lrr}
    \toprule
     & $\downarrow$ Mean GD & $\uparrow$ \% Zero GD\\
    \midrule
    Lastfm & 0.000 & 100 \\
    Amazon-lb & 0.003 & 93 \\
    QK-video & 0.000 & 100 \\
    Jester & 0.002 & 97 \\
    ML-10M & 0.000 & 100 \\
    ML-20M & 0.001 & 98 \\
    \bottomrule
    \end{tabular}
\end{table}

\subsubsection*{Overall findings} We find that \textbf{the generated PFs are generally very close to the true PFs}, with a mean Generational Distance (GD) of less than 0.003 for the 100 small samples of each dataset. In fact, at least 93\% of the generated PFs are exact matches of the true PFs (i.e., zero GD). For cases where our PFs are not exact matches, the maximum GD is 0.122 (for one sample of Jester), but the rest have a GD below 0.08, which means that the difference is small. We posit that the greedy nature of the {\sc Oracle2Fair} algorithm produces suboptimal solutions in some cases. However, we note that our {\sc Oracle2Fair} algorithm is much more efficient than the brute-force approach: in this experiment, each generation with {\sc Oracle2Fair} takes about 2 seconds, which is about 30$\times$ faster than brute-forcing the true PF.

\subsection{Agreement between Measures (RQ2)} 
\label{ss:corr}

We study the agreement between DPFR and other evaluation approaches in ranking our 16 runs (on the full datasets) from best to worst. Low agreement means that the other approaches are not related to the fairness-relevance trade-off that DPFR quantifies. We compare DPFR to (a) existing \rel{} and \fair{} measures, (b) existing \joint{} measures, and also (c) the average (arithmetic mean) of \rel{} and \fair{} scores from the selected measure pairs that are used to generate the PFs. 
To compute the average for a measure where lower values are better (i.e., Gini), we compute $1-$the Gini score instead. Unlike previous work \cite{vargascastello,Cherumanal2025Diversification}, we do not compute the harmonic mean as it is extremely sensitive to small values, which are present in our \rel{} scores.

\subsubsection{Comparison of existing measures to DPFR}
We find that for all datasets and all measure pairs, \textbf{the best model as per DPFR is always different from the best model as per \rel{} measures} (\Cref{fig:pareto-pair-plot} and \Cref{app:actual-scores}, \Cref{fig:app-pairplot}). Moreover, \textbf{half the time, the best model as per DPFR is different from the best model as per a \fair{} measure}. 
Existing \joint{} measures tend to have the same best model as either \fair{} or \rel{} measures (73.3\% of the time; see \Cref{tab:base-rerank-all-1,tab:base-rerank-all-2}), instead of having a more balanced evaluation of both aspects. These findings are expected as existing joint evaluation measures use relevance in their formulation differently than the \rel{} measures. 
Overall, the best model found with DPFR is less skewed towards relevance or fairness.

\subsubsection{Correlation of measures}
\label{sss:corr_measures}
\begin{figure}[htbp]
    \centering   
    \includegraphics[width=\columnwidth]{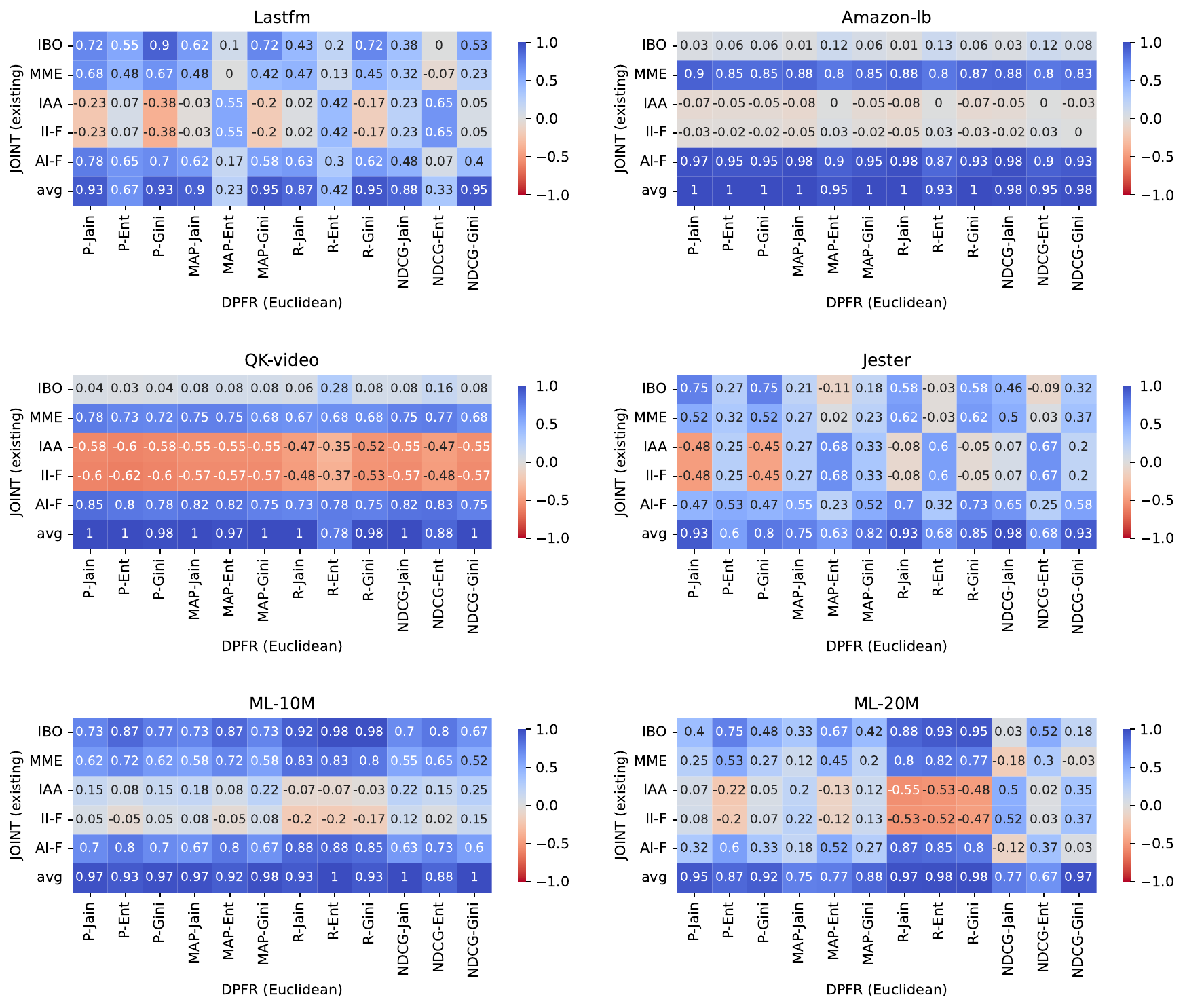}

    \caption{Kendall's $\tau$ correlation heatmap between the rank ordering of RSs based on existing \joint{} evaluation approaches (including the average of \rel{} and \fair{} scores, avg), and DPFR.}
    \label{fig:corrplot}
\end{figure}

For each dataset, we compute the Kendall's $\tau$ \cite{KENDALL1945THEPROBLEMS} correlations between the ranking given by DPFR and by the joint evaluation baselines (see \Cref{fig:corrplot}). 
Kendall's $\tau$ handles ties, unlike Spearman's $\rho$. Rankings are considered equivalent if $\tau \geq 0.9$ \cite{Maistro2021PrincipledRankings, Voorhees2001EvaluationDocuments}. 
We see similar agreement trends in datasets where recommenders have higher \rel{} scores (Lastfm and Jester) or lower (Amazon-lb and QK-video). 

Overall, most times DPFR orders models differently ($\tau<0.9$) than all \joint{} measures except AI-F. We see similar trends between IAA and II-F, and between MME and AI-F. IBO can have similar trends as MME (except for Amazon-lb and QK-video). For all datasets, IAA and II-F have overall either weak or negative $\tau$ with DPFR (e.g., $[-0.2, 0.25]$ for ML-10M and $[-0.62, -0.35]$ for QK-video). A notable exception is DPFR with \{MAP, R, NDCG\} $\times$ {Ent} for Lastfm and Jester, where we see moderate correlations, $\tau \in [0.42,0.68]$.

Ent differs from this trend because DPFR with \{MAP, R, NDCG\} $\times$ \{Jain, Gini\} has PF gradients of greater magnitude. This only affects Lastfm and Jester as they have higher \rel{} scores than the other datasets. DPFR with P has different patterns from other \rel{} measures: the raw DPFR scores of pairs involving P are lower on average, as the scores from Oracle do not start from $1$, but much less, and therefore closer to the models' scores (\Cref{fig:pareto-pair-plot}). 
Meanwhile, IBO has varying $\tau$ across datasets: a huge range of $\tau$, i.e., $[0.00,0.9]$ for Lastfm, weak correlations $[0.01, 0.13]$ for Amazon-lb, and moderate to strong correlations $[0.67, 0.98]$ for ML-10M. These variations might be because IBO is based on the number of items satisfying a certain criterion, rather than an average of scores across users and/or items, i.e., how other \joint{} measures are defined. 

Among the \joint{} measures, AI-F correlates the strongest with DPFR, as both AI-F and DPFR, indirectly or directly, consider 
the recommendation frequency of each item and compare it with that of other items. 
However, the rank orderings given by AI-F are not equivalent to DPFR, as $\tau < 0.9$ for 5/6 datasets (excluding Amazon-lb). 
For the same measure-pair and between datasets, the $\tau$ of AI-F and DPFR also varies a lot, e.g., $\tau=0.07$ for NDCG-Ent for Lastfm, but $\tau=0.9$ 
for Amazon-lb. We thereby do not recommend using any of the \joint{} measures to evaluate the trade-off between fairness and relevance,  as none consistently correlates with DPFR.

Taking the mean of \rel{} and \fair{} scores (avg) at a glance seems to correlate highly with DPFR. However, while it gives equivalent rankings ($\tau\geq 0.9$) in some cases (e.g., for Amazon-lb, most of ML-10M and QK-video, and half of ML-20M), it only does so for (1) datasets with lower \rel{} scores (Amazon-lb, QK-video), i.e., in cases where all models perform poorly, we have low variance in \rel{}, which leads to fairness dominating both avg and DPFR; (2) datasets with low variance in \fair{} scores (ML-*). 
In such cases, quantifying the evaluation jointly is challenging as one aspect dominates over the other. 
In the other datasets, the rank ordering given by the average is inconsistent: sometimes $\tau \geq 0.9$ for one dataset, but not for the others. This inconsistency between datasets holds for all measure pairs, except for P-Jain and NDCG-Gini. Due to these inconsistencies, we discourage using the arithmetic mean for joint fairness-relevance evaluation. 

Overall, our correlation analysis shows that existing \joint{} evaluation measures cannot be used as a reliable proxy for DPFR. 

\subsubsection{Best model disagreement}
\label{sss:best_model_disagreement}

We take a closer look at how DPFR relates to computing averages, as they are similar approaches in terms of combining scores from a measure pair. As comparing the raw scores of DPFR and the average is invalid, we instead count the disagreement between the best model based on DPFR and the mean of \rel{} and \fair{} scores. 
The aim is to study whether one would come to the same conclusion regarding the best model, using the two different joint evaluation approaches. We do this on both groundwork runs (\Cref{ss:groundwork}) and artificial runs with varying recommendation effectiveness. Both results are presented in \Cref{tab:best_model_disagreement}.

\paragraph{Disagreement on groundwork runs} 
Among the 12 measure pairs that are fit for DPFR, we find that \textbf{the best model according to DPFR is not always the same according to the average of \rel{} and \fair{} scores of the same pair}; in one case the disagreement is up to 58.33\% (for Lastfm). The disagreement is generally much higher in the more complex rank-based measures (0--83.33\%) compared to simpler set-based \rel{} measures (0--50.00\%). 
Therefore, there are many cases where the mean of \rel{} and \fair{} scores is not the best case, especially for Lastfm and Jester where \rel{} scores are higher and vary more. 
In these two datasets, more often than not, DPFR leads to different conclusions than a simple average. Yet, sometimes the average agrees with DPFR in the best model: for QK-video, disagreement is low (8.33\%), and there is a perfect agreement on the best model for Amazon-lb; we posit that these are due to equally poor and low variance in the \rel{} scores. This aligns with our correlation analysis. 
As there is a huge range of variability across datasets (0--58.33\%), we do not recommend using a simple average to get the same result as DPFR, as it is unreliable and inconsistent. 
Generally, averaging fails to reach the same conclusion as DPFR almost half the time for the rank-based \rel{} measures, especially when the \rel{} scores vary highly.

\begin{table}[tbp]
\centering
\caption{The percentage of times the best model based on avg (the mean of \rel{} and \fair{} scores) differs from DPFR for groundwork and artificial runs. We separate the results based on the \rel{} measure type: set-based (P@$k$ and R@$k$) and rank-based (NDCG and MAP). Only the 12 suitable measure pairs are considered (see \Cref{sss:no-fit}). 
Darker grey shades indicate a higher percentage of best model disagreement.}
\label{tab:best_model_disagreement}
\begin{tabular}{lrrr|rrr}
\toprule
run type & \multicolumn{3}{c|}{Groundwork} &  \multicolumn{3}{c}{Artificial}\\
\midrule
\rel{} measure type  &                    Set-based (\%) & Rank-based (\%)&    All (\%)& Set-based (\%)& Rank-based (\%)&    All (\%)\\
\midrule
Lastfm    &          \cellcolor[HTML]{E9ECEF} 50.00 &     \cellcolor[HTML]{CED4DA} 66.67 & \cellcolor[HTML]{CED4DA} 58.33  & \cellcolor[HTML]{ADB5BD} 100.00 & \cellcolor[HTML]{ADB5BD} 100.00 & \cellcolor[HTML]{ADB5BD} 100.00\\
Amazon-lb &          0.00 &       0.00 &   0.00 & \cellcolor[HTML]{ADB5BD} 100.00 & \cellcolor[HTML]{ADB5BD} 100.00 & \cellcolor[HTML]{ADB5BD} 100.00 \\
QK-video  &          16.67 &       0.00 &   8.33 &  \cellcolor[HTML]{E9ECEF} 33.33 & 0.00 & 16.67\\
Jester    &          16.67 &      \cellcolor[HTML]{ADB5BD}83.33 &  \cellcolor[HTML]{E9ECEF} 50.00 & \cellcolor[HTML]{E9ECEF} 50.00 & 16.67 & \cellcolor[HTML]{E9ECEF} 33.33\\
ML-10M    &           0.00 &    \cellcolor[HTML]{CED4DA}  66.67 &  \cellcolor[HTML]{E9ECEF} 33.33 & \cellcolor[HTML]{ADB5BD}83.33 & \cellcolor[HTML]{ADB5BD} 100.00 & \cellcolor[HTML]{ADB5BD}91.67\\
ML-20M       &       0.00 &       \cellcolor[HTML]{E9ECEF} 50.00 &  25.00 & \cellcolor[HTML]{ADB5BD} 100.00 & \cellcolor[HTML]{ADB5BD} 100.00 & \cellcolor[HTML]{ADB5BD} 100.00\\
\midrule
All datasets &      13.89 &    \cellcolor[HTML]{E9ECEF}   44.44 &  \cellcolor[HTML]{E9ECEF} 29.17 & \cellcolor[HTML]{ADB5BD}77.78 & \cellcolor[HTML]{CED4DA} 69.44 & \cellcolor[HTML]{CED4DA} 73.61\\
\bottomrule
\end{tabular}
\end{table}

\paragraph{Disagreement on artificial runs with varying recommendation effectiveness}
\label{ss:artificial}

Our previous analyses relied upon the performance of 16 runs, from models that are trained on real-world data. Consequently, their \rel{} and \fair{} scores cannot be controlled, resulting in extremely different score distributions across datasets. For example, in \Cref{fig:pareto-pair-plot} the \rel{} scores for QK-video are all extremely low and have little variation (e.g., in the range of $[0.004, 0.130]$), while they are more spread out for Lastfm ($[0.045, 0.793]$). The severe differences in model score variations cause difficulty in understanding how fairness and relevance are jointly quantified. It is unclear if the low variance of the \rel{} scores causes the joint evaluation approaches to rely mainly on the \fair{} scores that have a higher variance, e.g., $\uparrow$Ent ranges in $[0.420, 0.979]$ for QK-video and in $[0.589, 0.826]$ for Lastfm. 
In this experiment, we aim to analyse the difference between how avg and DPFR quantify fairness and relevance jointly, given a controlled distribution of \rel{} scores. We only focus on artificially varying the \rel{} scores as avg and DPFR have been studied under varying \fair{} scores (\Cref{sss:result-pf}).

We control the score distribution by artificially generating 11 runs of varying \rel{} scores from 0 to the maximum score. In most cases, the maximum will be 1, but there are cases where it may not be 1, e.g., P@10 cannot be 1 if a user only has one relevant item in their test set~\cite{Moffat2013SevenMetrics}. 
The runs are artificially generated as follows: 
First, for each user, at most $k$ (relevant) items are selected randomly from their test set as their recommended items, resulting in the maximum \rel{} score. 
Next, we sample 10\% of the users randomly and replace all of their recommended items with random non-relevant items that are not in the user's train/val split. This is expected to cause the average \rel{} scores to drop approximately by 0.1. We continue the process until all users have non-relevant items. Even though the \fair{} scores are not directly controllable through this process, they are nevertheless expected to vary due to changes in the recommended items. 
Specifically, they are expected to increase because the overall exposure across items increases when more non-relevant items are included.
However, the increase in \fair{} scores is not expected to be monotonic. As the non-relevant item selection is not restricted to only items that would balance the exposure distribution, the \fair{} scores can also become slightly more unfair between one iteration and another.

We attempt to keep the artificial run generation process realistic. While the above procedure selects the recommended items artificially, the item relevance scores are taken from the six real-world datasets. Additionally, by only recommending items that are not in the user's train/val split, our approach aligns with a common recommendation task setup, where items that a user has rated/interacted with (i.e., the ones in their train/val split) are not re-recommended. 

\begin{figure}
    \centering
    \includegraphics[width=0.8\linewidth]{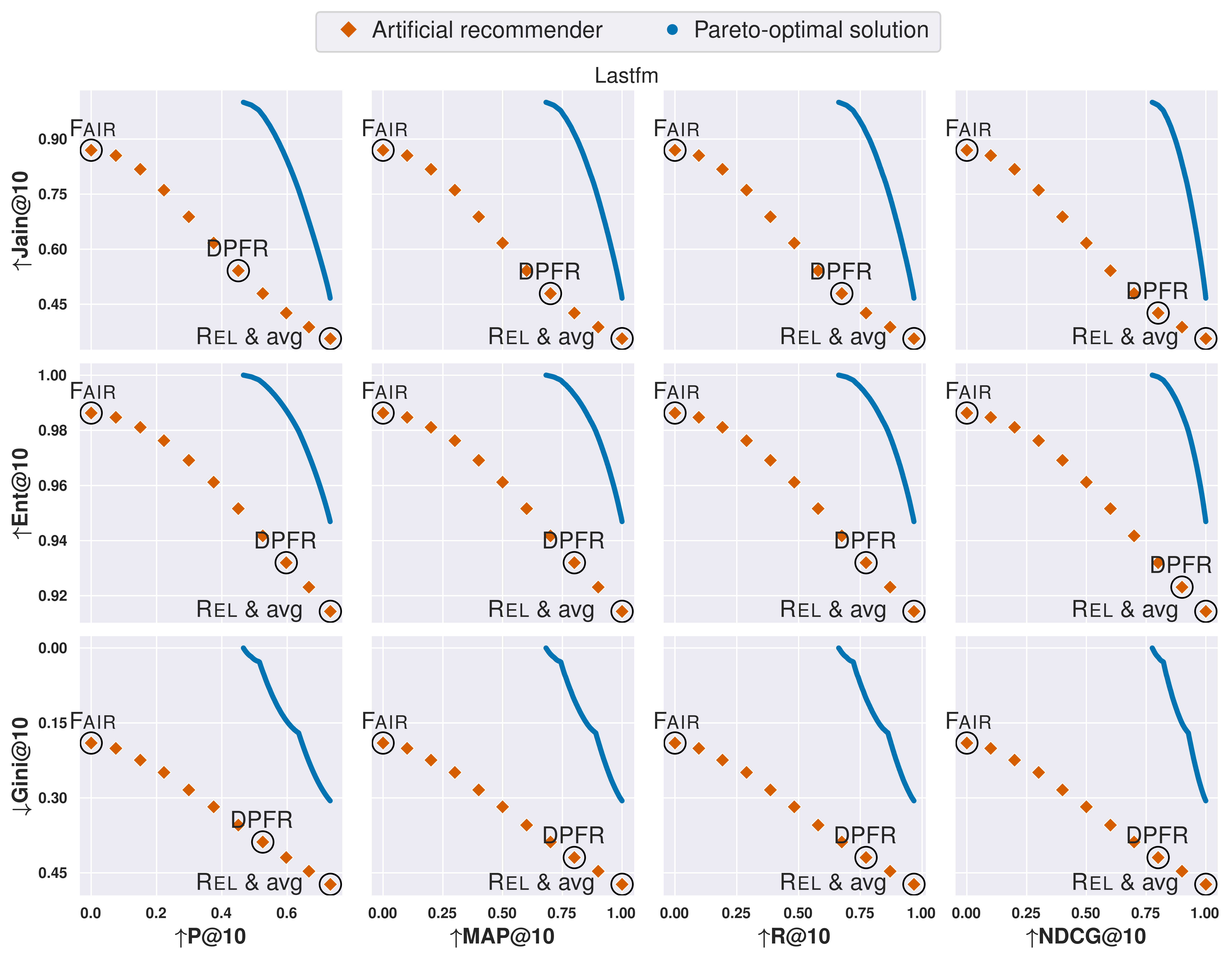}
    \includegraphics[width=0.8\linewidth]{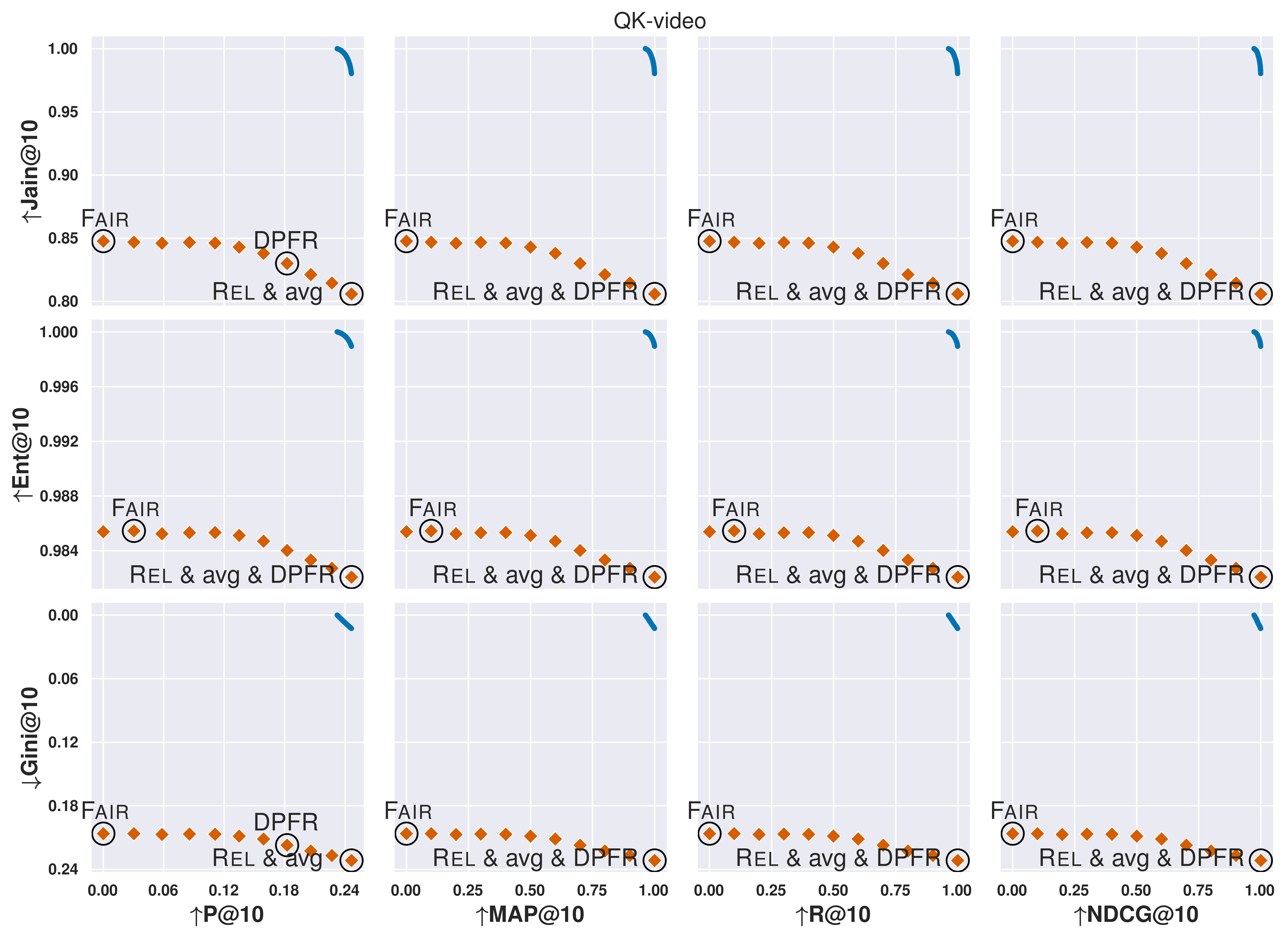}
    \caption{Pareto Frontier of fairness and relevance (in blue) and artificial recommender scores (in orange) for Lastfm and QK-video. The best runs based on \rel{}, \fair{}, avg, or DPFR are circled.}
    \label{fig:artificial}
\end{figure}

\Cref{fig:artificial} shows the artificial run scores along with the generated PF (\Cref{sss:result-pf}) for Lastfm and QK-video. We show the rest in \Cref{app:actual-scores}, \Cref{fig:artificial_app}. For each \rel{}-\fair{} measure pair, we identify the best run according to \rel{}, \fair{}, avg, and DPFR score. Further, we analyse how often the best run based on avg differs from DPFR and present the results in \Cref{tab:best_model_disagreement}. 
We find that \textbf{the best artificial run based on DPFR always differs from the one based on avg for 3/6 datasets and for all 12 \rel{}-\fair{} measure pairs}. This means that in most cases (i.e., 73.61\% across all datasets and measure pairs), avg gives a different conclusion from DPFR. For the rest of the datasets, avg disagrees with DPFR on the best run between 16.67--91.67\% of the time, which further indicates its unreliability and inconsistency. The disagreement happens less frequently for QK-video and Jester than for other datasets, which may be due to their low variation of \fair{} scores. This is similar to our findings in \Cref{sss:best_model_disagreement}, where avg agrees more with DPFR for lower \rel{} score variance.

\paragraph{Overall findings} Combining our findings on both the groundwork and artificial runs, we conclude two things. 
Firstly, \textbf{avg is more skewed towards the aspect with higher score variation}. In \Cref{sss:best_model_disagreement}, avg tends to select the same best model as \fair{} measures, as the \fair{} scores generally have higher variance than the \rel{} scores. Here, avg agrees more with \rel{} measures for similar reasons.
Secondly, while DPFR is also affected by extremely low variance of single-aspect scores, it is more robust than avg.
Altogether, \textbf{this is yet another reason not to compute simple average as a way to jointly evaluate fairness and relevance, considering that its conclusion could differ from DPFR more than 70\% of the time}.

\subsection{Sensitivity to Distance Measures (RQ3)}
\label{ss:sensitivity_distance} 

In the previous experiments, DPFR is computed with Euclidean distance, which we refer to as ``DPFR (Euclidean)''. How does the choice of distance measure affect DPFR? To answer this question, we first discuss alternative distance measures for DPFR and then compare their sensitivity in terms of agreement with DPFR (Euclidean). 

\subsubsection*{Alternative distance measures} Some possible choices include Manhattan, Chebyshev, Minkowski, and cosine distance. We explain below why these distances may not be suitable for DPFR.

The Manhattan distance between two points $(x_1, y_1)$ and $(x_2, y_2)$ is the sum of the absolute differences between each dimension. While in principle Manhattan distance can be used with DPFR, the conclusion obtained may be equivalent to averaging \rel{} and \fair{} scores (avg). Next, we show theoretically how DPFR (Manhattan) could potentially have the same ranking of models as avg. 
Suppose that a model's \rel{} and \fair{} scores are denoted as the tuple $(x_r, x_f)$, and the PF reference point is $(x_r^{t'}, x_f^{t'})$. Averaging the model scores, we have $\frac{x_r + x_f}{2}$. Meanwhile, the DPFR (Manhattan) would be $|x_r^{t'}-x_r|+|x_f^{t'}-x_f|$, which is equivalent to $x_r^{t'} + x_f^{t'} - (x_r + x_f)$ because the reference point represents an ideal trade-off, which should always be greater than or equal to the model scores, i.e., $x_{(.)}^{t'} \geq x_{(.)}$. As the reference points would be the same for all models, $x_r^{t'} + x_f^{t'}$ can be regarded as constant. Hence, for DPFR (Manhattan), higher $x_r+x_f$ would place the points closer to the reference point, which means better in the joint evaluation of relevance and fairness. Likewise, based on avg, higher $x_r+x_f$ would also mean better performance in both aspects. Therefore, DPFR (Manhattan) and avg would yield equivalent rankings of models, which we also verify empirically below.

The Chebyshev distance is the maximum absolute difference between two points across any dimension. In our case, this is equivalent to $\max\left\{{|x_r^{t'}-x_r|, |x_f^{t'}-x_f|}\right\}$. Computing DPFR with Chebyshev distance may not be useful, as it collapses the evaluation to one aspect (either relevance or fairness), defeating the purpose of joint evaluation, where each evaluation aspect can have equal weighting. 
Minkowski distance generalises Manhattan, Euclidean, and Chebyshev distances. However, using the generalised version may harm the interpretability of DPFR because the path taken between two points is hard to visualise. 
Cosine distance is based on the angle between two vectors, and does not make sense to be used for DPFR: if two points lie on the same line (drawn from the origin), their cosine distance would be zero. To give an (extreme) example, let the reference point be $(1,1)$ and the scores of two models are $(0,0)$ and $(0.99, 0.99)$. Under DPFR (Cosine), both models would be regarded as performing equally well, which is counter-intuitive, given that $(0,0)$ here refers to the worst possible \rel{} and \fair{} scores, and $(0.99, 0.99)$ is very close to the best possible scores. In short, even though DPFR can technically be used with \textit{any} distance measure that accommodates 2D-vectors, one should also consider the practical implications, including the possible interpretation of the distance.

\subsubsection*{Sensitivity analysis} 
To compare the sensitivity of DPFR to distance measures, we first compute DPFR with Manhattan distance. We present the raw DPFR (Manhattan) scores in \Cref{app:actual-scores}, \Cref{tab:dpfr-scores-manhattan,tab:dpfr-scores2-manhattan}. Using these scores, for each \rel{}-\fair{} measure pair, we compute: (i) the agreement of DPFR (Euclidean) and DPFR (Manhattan); and (ii) the agreement between DPFR (Manhattan) and avg. The agreement between DPFR (Manhattan) and \joint{} measures is also computed and presented in \Cref{app:actual-scores}, \Cref{fig:corr_manhattan}.

\begin{figure}
    \centering
    \includegraphics[width=\columnwidth]{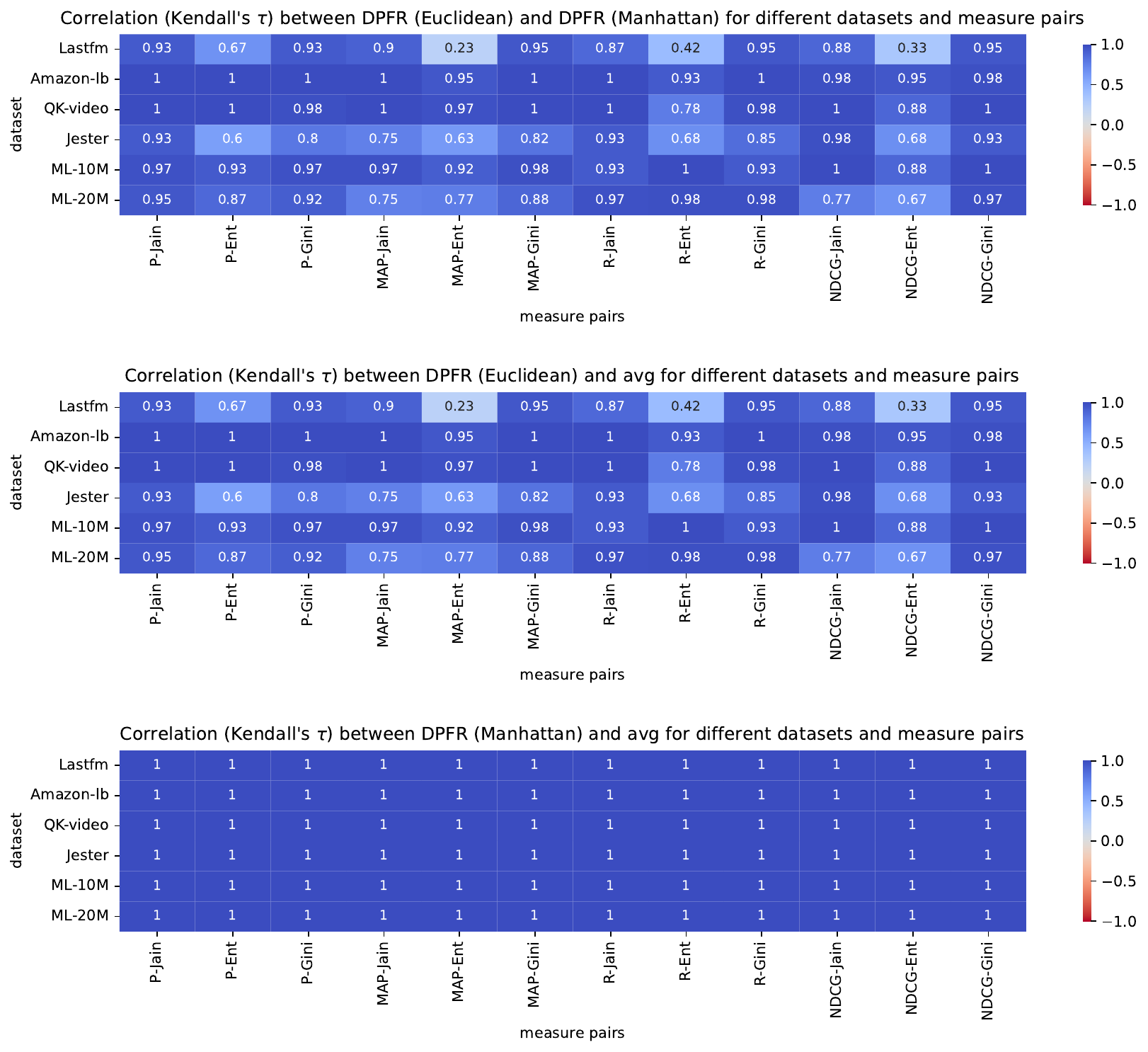}
    \caption{Kendall's $\tau$ correlation heatmap of the rank ordering of RSs (from top to bottom figure): 
    (i) between DPFR computed with Euclidean distance and DPFR computed with Manhattan distance; 
    (ii) DPFR (Euclidean) and avg (average of \rel{} and \fair{} measures); and 
    (iii) DPFR (Manhattan) and avg. The correlations are shown for different datasets ($x$-axis) and measure pairs ($y$-axis).}
    \label{fig:corr_euclid_manhattan}
\end{figure}

\Cref{fig:corr_euclid_manhattan} presents the agreement between: (i) DPFR (Euclidean) and DPFR (Manhattan); (ii) DPFR (Euclidean) and avg, for comparison to (i); and (iii) DPFR (Manhattan) and avg. Across six datasets and 12 measure pairs, DPFR (Euclidean) and DPFR (Manhattan) have varying agreement ($\tau \in [0.23, 1]$), and the agreement is especially low for measure pairs with Ent (because of a lower PF gradient compared to other measure pairs; see \Cref{sss:corr_measures}). However, upon closer inspection, \textbf{the agreement between DPFR (Euclidean) and DPFR (Manhattan) matches exactly the agreement between DPFR (Euclidean) and avg}. This is because, as hypothesized above, \textbf{DPFR (Manhattan) and avg correlate perfectly}. Hence, rather than computing DPFR (Manhattan), which would be computationally expensive, avg could instead be computed as a proxy to obtain similar conclusions. 

\subsection{Varying Weight of Fairness and Relevance (RQ4)}
\label{ss:varying_alpha}

We compare how differently the measures consider the trade-off between fairness and relevance. In the previous analyses, DPFR is computed with $\alpha=0.5$, which selects the PF midpoint as a reference point for computing DPFR. Using a different $\alpha$ would give a different weight between fairness and relevance in their joint evaluation with DPFR, where $\alpha=0$ gives the most weight to relevance and $\alpha=1$ to fairness. Here, we study how varying the selection of the fairness-relevance weight affects DPFR's agreement with existing \rel{}, \fair{}, and \joint{} evaluation measures and approaches, including simple average (avg) and weighted average (weighted-avg). To this end, we compute Kendall's $\tau$ between DPFR and the existing measures to assess their agreement in ranking recommender models for different $\alpha$'s. We vary $\alpha$ from 0 to 1 with increments of 0.01. Through this study, we aim to: 
(i) provide insights that can guide the selection of $\alpha$ in the DPFR computation; and 
(ii) investigate whether there are cases where avg, weighted-avg, or any other \joint{} evaluation approaches could be a good proxy for DPFR across various $\alpha$'s.

\begin{figure}[h]
    \centering

    \begin{tikzpicture}

    \definecolor{forange}{RGB}{222,132,5}
    \definecolor{rblue}{RGB}{1,115,178}
      \begin{axis}[
        width=12cm, height=7cm,
        xmin=0-0.05, xmax=1+0.05,
        ymin=-1-0.1, ymax=1+0.1,
        legend style={at={(1.2, 0.8)}, anchor=north},
            legend style={font=\small, draw=none},
            legend cell align={left},
        xlabel={$\alpha$},
        ylabel={Kendall's $\tau$},
        axis lines=middle,
        axis line style={thick, -},
        y label style={
            at={(axis description cs:-0.025,1.1)}, 
            anchor=north},
        after end axis/.code={
        \path (axis cs:0,0) 
            node [anchor=north west,yshift=-0.075cm] {0}
            node [anchor=south east,xshift=-0.075cm] {0};
            } 
        ]
        \addlegendimage{empty legend}
        \addlegendentry{\hspace{-.7cm}\textbf{Evaluation Measure}}
        \addplot[rblue, thick] coordinates {(0,1) (1,-1)};
        \addlegendentry{\rel}
        \addplot[forange, thick] coordinates {(0,-1) (1,1)};
        \addlegendentry{\fair}
      \end{axis}
    \end{tikzpicture}
    \caption{Expected agreement (Kendall's $\tau$) of DPFR with a \rel{} measure and with a \fair{} measure for varying the fairness-relevance trade-off weight in the DPFR computation ($\alpha$). This assumes that the \rel{} and \fair{} measures have the exact opposite ranking of models ($\tau=-1$).}
    \label{fig:expected_corr}
\end{figure}
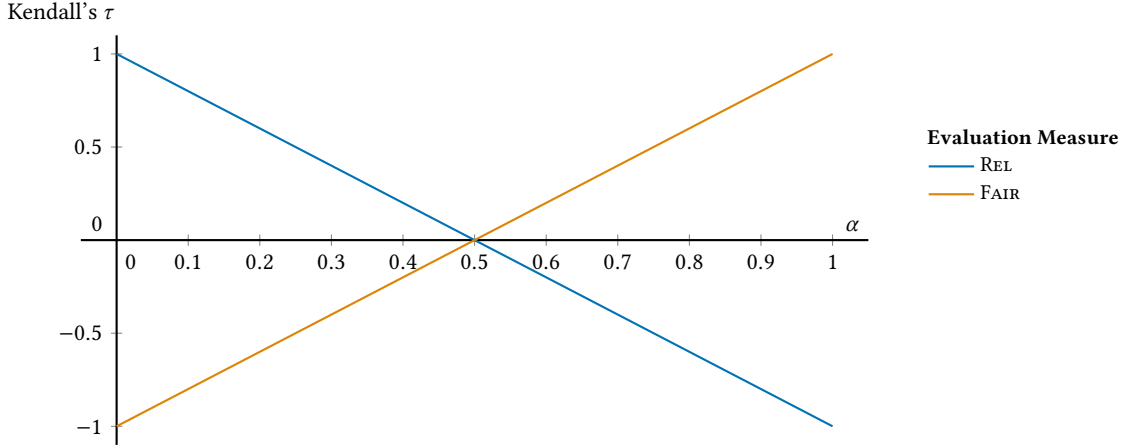

Our hypothesis is as follows: if a \rel{} and a \fair{} measure have the exact opposite model rankings (i.e., $\tau=-1$), we would expect the trend of their agreement with DPFR to resemble \Cref{fig:expected_corr}. Under this assumption, a \rel{} measure would agree perfectly with DPFR for $\alpha=0$, gradually decreasing as $\alpha$ is increased to give more weight to fairness, and eventually having the opposite conclusion with DPFR for $\alpha=1$. In contrast, a \fair{} measure would first have the opposite conclusion with DPFR for $\alpha=0$, but as more weight is given to fairness, its agreement with DPFR would increase, reaching the highest at $\alpha=1$. However, this assumption would most likely not hold, as \rel{} and \fair{} measures tend to either have weak correlations or moderate-to-strong positive correlations instead \cite{Rampisela2024EvaluationStudy}. In addition, \joint{} measures have varying alignments to \rel{} and \fair{} measures: IAA/II-F agrees more with \rel{} measures; MME/AI-F aligns more with \fair{} measures; and IBO has an inconsistent relationship with the single-aspect measures \cite{Rampisela2024CanRelevance,Rampisela2025relevanceguidelines}.  
Therefore, an empirical analysis is needed to understand better the measures' agreement with DPFR, given different fairness-relevance weights.

\begin{figure}
    \centering
    \includegraphics[width=0.85\linewidth]{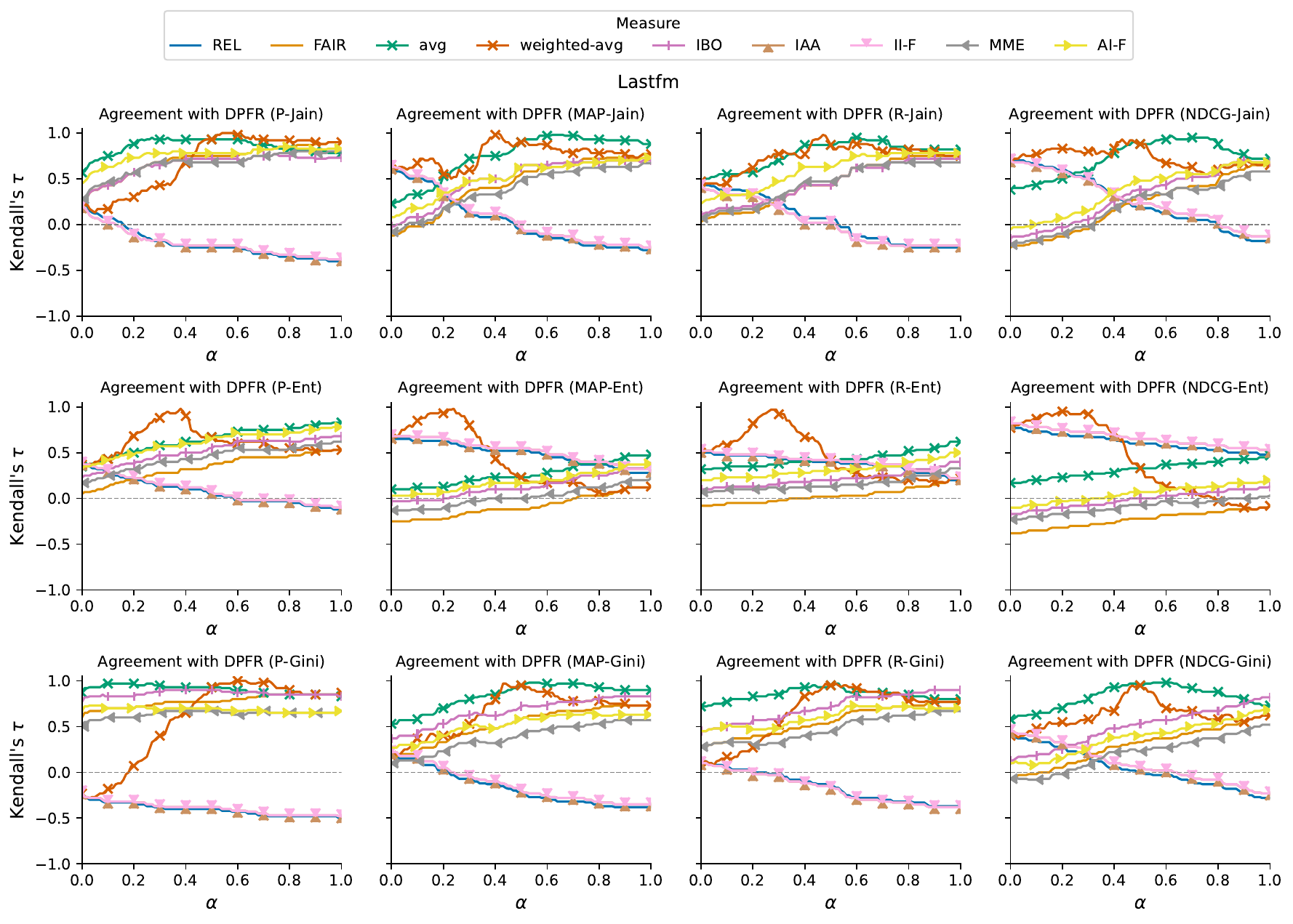}
    \includegraphics[width=0.85\linewidth]{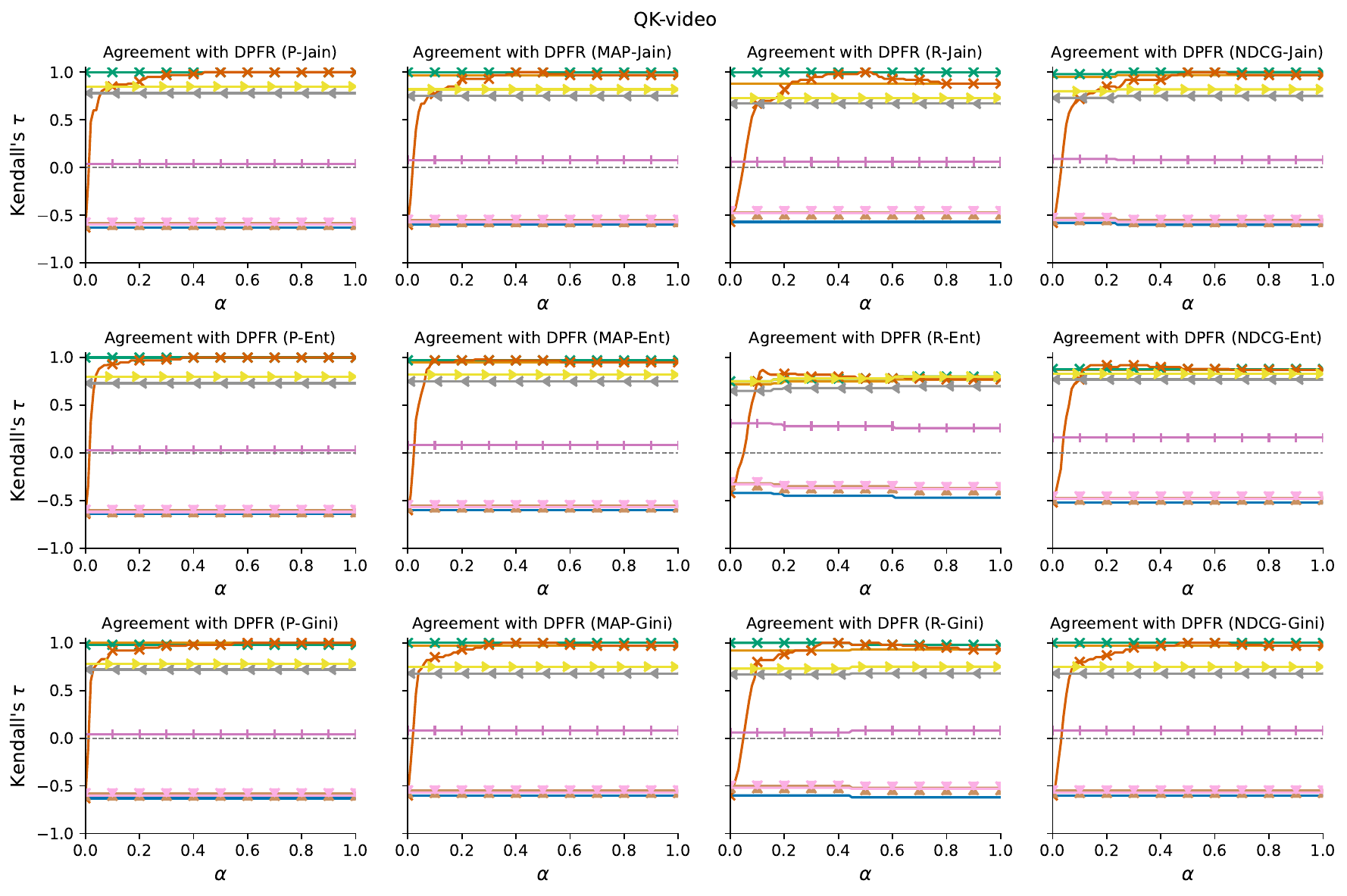}

    \caption{Agreement (Kendall's $\tau$) between DPFR and \rel{}, \fair{}, and \joint{} evaluation approaches (including avg and weighted-avg) when varying $\alpha$ for Lastfm and QK-video. Note that for \rel{}, \fair{}, and avg, we only show the agreement of the measures that are part of the DPFR \rel{}-\fair{} measure pair. For example, in DPFR (P-Jain), \rel{} is P and \fair{} is Jain. Dashed grey lines denote no agreement ($\tau=0$).
    }
    \label{fig:corr_alpha}
\end{figure}

We present the results for Lastfm and QK-video in \Cref{fig:corr_alpha}; we show the rest in \Cref{app:actual-scores}, \Cref{fig:corr_alpha_app1,fig:corr_alpha_app2} as they have similar trends. Specifically, the trends for Lastfm are similar to ML-20M, while QK-video is similar to Amazon-lb, Jester, as well as ML-10M. All measures, in general, agree more than disagree with DPFR ($\tau>0$), even for varying fairness-relevance weights, suggesting that DPFR can capture both evaluation aspects. Exceptions to this (i.e., $\tau<-0.5$) are both \rel{} measures and the \rel{}-aligned \joint{} measures (IAA and II-F) for QK-video, which may be due to its extremely short PF and their low \rel{} score variance. 
Next, we explain the overall agreement of DPFR to single- and multi-aspect measures and provide a more fine-grained analysis on some points of interest.

\subsubsection*{Overall agreement between single-aspect measures and DPFR} 
At a high level, the agreement of \fair{} measures to DPFR is stronger than of \rel{} measures to DPFR. This is because of the generally low \rel{} score variance, which has a weaker effect than the \fair{} scores on the model ranking when considering both aspects. 
For \rel{} measures, as more weight is given to fairness (increasing $\alpha$), their agreement with DPFR either decreases (e.g., for Lastfm) or remains constant (e.g., for QK-video), while for \fair{} measures the agreement either increases at roughly similar rate to the \rel{} measures' rate of change, or stays constant. 

Firstly, we address the generally increasing/decreasing trends across various fairness-relevance weights: even though the trends fulfil our expectations, the correlation strength is lower than expected. As an example, for Lastfm, at $\alpha=0$ the correlation of \fair{} measure to DPFR (P-Ent) is close to 0 instead of -1. This shows that \textbf{even when relevance is given the maximum priority, DPFR still accounts for both relevance and fairness.}. The same applies for \fair{} measures at $\alpha=1$. \textbf{Therefore, DPFR remains a multi-aspect evaluation approach at $\alpha\in\{0,1\}$ and thus may have a different conclusion from single-aspect measures.}

Secondly, we explain why for some datasets (i.e., QK-video, Amazon-lb, and Jester) the single-aspect measures' agreement with DPFR is constant for varying $\alpha$: QK-video and Jester have relatively short PF (\Cref{ss:result-pareto}), which means that the Pareto-optimal solutions do not have much trade-off between relevance and fairness. Amazon-lb does not have a short PF, but the model \rel{} scores have very low variance. If effectiveness does not vary a lot between RSs, then we can choose the most fair recommender, and it will have a comparable effectiveness to the others. 
Hence, \textbf{in cases where the PF is short or the single-aspect scores have low variance, one aspect dominates the joint evaluation and $\alpha$ has little effect on the conclusion}.

\subsubsection*{Overall agreement between \joint{} evaluation approaches and DPFR} We study (i) the agreement of \joint{} measures (IBO, IAA, II-F, MME, and AI-F) and DPFR; and (ii) the agreement of averaging scores (avg and weighted-avg) to DPFR. 

About (i), the agreement of \rel{}-aligned \joint{} measures (IAA and II-F) with DPFR follows closely the \rel{} measures, which have decreasing agreement as $\alpha$ is increased to give more weight to fairness. In contrast, the \fair{}-aligned \joint{} measures (MME and AI-F) have increasing agreement with DPFR with higher $\alpha$. IBO, which is known to have an inconsistent relationship with single-aspect measures, has an agreement trend similar to \fair{} measures for Lastfm, Jester, and ML-*, while its trends for Amazon-lb are similar to \rel{} measures, and the trends for QK-video align with neither \fair{} nor \rel{} measures. All these results fulfil our hypothesis.

About (ii), the overall correlation of averaging scores (avg and weighted-avg) to DPFR is mostly positive, showing that generally the averaging approach produces similar (but not always equivalent) model ranking as DPFR. This follows the trends of the \fair{} measures and \fair{}-aligned \joint{} measures more than the \rel{} measures and \rel{}-aligned \joint{} measures. This is not surprising, as the best model based on avg is occasionally the same as \fair{} measures, but never the same as \rel{} measures in our analysis of the groundwork runs (\Cref{sss:result-pf}). We also find that the agreement of both avg and weighted-avg to DPFR tend to be stronger than the \fair{}-aligned measures' agreement to DPFR, indicating that \textbf{both averaging approaches are better proxies for DPFR than the \fair{}-aligned measures}.

Next, we highlight a few differences between trends of avg and weighted-avg: 
Firstly, based on the correlation strength, avg generally has stronger correlation to DPFR across varying $\alpha$ than weighted-avg. This finding may seem counter-intuitive, as one would expect weighted-avg and DPFR to correlate stronger than avg and DPFR, across varying fairness-relevance trade-offs, which is accounted by weighted-avg. We posit that in the case of weighted-avg, assigning small weights to the \rel{} scores, which tend to be very low, will practically make them negligible when taking their weighted average with \fair{} scores. The weighted-avg ends up accounting mostly for one aspect instead of two.
Secondly, based on the change in avg's and weighted-avg's agreement to DPFR across various $\alpha$'s, avg is generally more stable than weighted-avg, which can range widely (e.g., $\tau \in [-0.64,1]$ for QK-video and for DPFR computed with P-Gini). While weighted-avg frequently exhibits a clear upward trend that is followed by a decrease in agreement  (i.e., for Lastfm, Jester, and ML-*) or by a constant, strong agreement (QK-video and Amazon-lb), avg only shows this for Lastfm and ML-20M. For the rest of the datasets, avg is either flat or slightly increases with higher $\alpha$, further indicating the stability of its agreement to DPFR.
From these differences, considering all ways of selecting $\alpha$, avg would make a better proxy for DPFR than weighted-avg. However, \textbf{avg and weighted-avg do not always give equivalent rankings to DPFR for various $\alpha$'s. Hence, they do not always make reliable proxies for DPFR.}

We also note that the trends for avg and weighted-avg differ between the \fair{} measures used to compute DPFR. Measure pairs with Jain or Gini have similar results (i.e., the agreement trends resemble an inverted v-shape), while the ones with Ent tend to differ (i.e., their agreements mostly increase monotonically with a higher $\alpha$). The similarity/difference between the measure pairs may be because the generated PFs for Jain and Gini have more similar characteristics than those for Ent (\Cref{sss:corr_measures}).

\subsubsection*{Points of interest} We identify several points that may be interesting to analyse $\alpha$ at which: 
(i) the strongest agreement with DPFR occurs; 
(ii) there is a flip from disagreement to agreement, or vice versa (i.e., the $x$-axis intercept); and
(iii) the \rel{}- and \fair{}(-aligned) measures have similar agreement level (i.e., their intersection with one another). For this analysis, we only consider measures that have varying agreement with DPFR for various $\alpha$'s (i.e., the non-flat lines).

\paragraph{Strongest agreement} The existing measures reach the strongest agreement with DPFR at different $\alpha$. For \rel{} and \rel{}-aligned \joint{} measures, this usually happens at $\alpha=0$, i.e., where DPFR prioritises relevances the most. Similarly, \fair{} and \fair{}-aligned \joint{} measures typically reach the highest agreement with DPFR at $\alpha=1$, where DPFR prioritises fairness the most. 

For avg, the $\alpha$ where the highest agreement happens varies per measure pair: pairs involving Jain/Gini tend to peak between $\alpha\approx0.3$ and $\alpha\approx0.8$, while those with Ent usually peaks at $\alpha\approx1$. Notably, the Jain/Gini pairs peak most frequently at $\alpha\approx0.4$ or $\alpha\approx0.6$. We also note that in these cases, avg tends to produce more equivalent model rankings to DPFR for $\alpha$ values close to 0.4 or 0.6, than for other $\alpha$. This means that avg cannot estimate DPFR well for other $\alpha$'s. 
Hence, \textbf{for non-short PFs, while simple averaging could approximate DPFR well for specific $\alpha$ values, the values are highly dependent on both the datasets and measure pairs}. Without prior analysis, it would thus be \textbf{hard to know the specific DPFR combination(s) that can or cannot be estimated with avg}. 

For weighted-avg, the strongest agreement occurs between $\alpha\approx0.2$ and $\alpha\approx1$. Similar to the avg trends, there are differences between DPFR with Jain/Gini and DPFR with Ent. For DPFR with Ent, the peak agreement happens at a lower $\alpha$ than for DPFR with Jain/Gini. Additionally, we find that the peak of agreement for weighted-avg tends to be much more visible than for avg. This means that \textbf{given slight adjustments to $\alpha$, weighted-avg is prone to drastic change in its agreement to DPFR,  especially for $\alpha$ where the strongest agreement with DPFR occurs}.

\paragraph{Agreement-disagremeent flip} With increasing $\alpha$, we observe that some measures occasionally have changes or flip their agreement with DPFR to disagreement, or vice versa. We refer to both cases as `agreement-disagreement flip'. \joint{} evaluation approaches with frequent agreement-disagreement flips are less stable proxies for DPFR, as their conclusions could change from agreeing to disagreeing with DPFR for different $\alpha$'s.

\textbf{Among all measures, weighted-avg flips the most often and thus is the least stable proxy for DPFR, among existing measures} (i.e., at least for one measure pair and for all datasets except for Amazon-lb and ML-10M), and avg never flips. Weighted-avg flips between $\alpha\approx0.05$ and $\alpha\approx0.8$, and most often at $\alpha \leq 0.3$. 
Other \joint{} measures only flip sometimes, which makes them somewhat better than weighted-avg in terms of stability. Both \rel{}- and \fair{}-aligned \joint{} measures flip between $\alpha\approx0.1$ and $\alpha\approx0.9$. However, considering that their agreement strength to DPFR is almost always weaker than avg's, they are not reliable proxies for DPFR either.

\subsubsection*{Similar \rel{}- and \fair{}-aligned measure agreement with DPFR}
When \rel{}-aligned measures agree as strongly as \fair{}-aligned measures with DPFR (i.e., their agreement intersects), the ranking of models given by relevance and fairness in isolation is equally far from the rankings obtained by DPFR.
The intersection is somehow a point of balance between the two aspects. 
Otherwise, one aspect dominates the joint evaluation more than the other.
The \rel{}-aligned measures intersect with \fair{}-aligned measures at different $\alpha$'s for different datasets and measure pairs. 
For Lastfm and ML-* datasets, \textbf{similar agreement between fairness and relevance to DPFR happens across all $\alpha$'s, but most often at $\alpha\in[0.3,0.4]$}. Interestingly, pairs with Precision@$k$ tend to intersect at lower $\alpha$ (i.e., $\alpha \leq 0.4$), while pairs with Recall@$k$ rarely intersect at all.
For QK-video, Amazon-lb, and Jester, there are no intersections for any measure pairs. Since the \rel{} scores have low variance (i.e., for QK-video and Amazon-lb) or the PF is short (i.e., for QK-video and Jester), varying $\alpha$ has little to no effect on the agreement between DPFR and \rel{}/\fair{}-aligned measures. As such, their agreement to DPFR remains (nearly) constant across various $\alpha$'s, resulting in no intersections.

We attribute the differences in results across datasets/measure pairs to differences in their PF length and PF slope. As the length and/or slope differs, the same change of $\alpha$ could correspond to different changes in the \rel{}/\fair{} scores of different PFs. \textbf{Due to the PF differences across datasets and measure pairs, we recommend analysing the correlation between DPFR and the single-aspect measures (the \rel{} and \fair{} measures used for computing DPFR) across various $\alpha$'s, to get one that balances the two evaluation aspects.} 
In cases where there is no intersection, we would recommend selecting $\alpha$ where the agreement between \rel{}- and \fair{}-aligned measures to DPFR is the closest.

A caveat to this is that \rel{}- and \fair{}-aligned measures having a similar agreement to DPFR does not mean similar agreement between themselves. This is because Kendall's $\tau$ agreement is not transitive. For example, suppose we have three rankings of four models $\{A,B,C,D\}$: $rank_1=[B, A, C, D]$, $rank_2=[A, B, C, D]$, and $rank_3=[A,C,B,D]$. While two pairs of ranking have the same similarity, $\tau(rank_1, rank_2) = \tau(rank_2,rank_3) = 0.667$, the other rank pair does not have the same similarity, $\tau(rank_1,rank_3) = 0.333$. Hence, the intersection point between \rel{}- and \fair{}-aligned measures should not be interpreted as having similar model ranking for both measures.

\subsubsection*{Overall findings} In this experiment, we set out to: (i) provide insights that can guide the $\alpha$ selection for computing DPFR; and (ii) investigate cases where \joint{} evaluation approaches can serve as a good proxy for DPFR, for varying $\alpha$.

About (i), to choose an $\alpha$ that can balance relevance and fairness, we recommend analysing the DPFR correlation to the same \rel{} and \fair{} measures used in the DPFR computation, for varying $\alpha$. Similar to our experiments, we recommend selecting $\alpha$ where the \rel{} and \fair{} measures' agreement to DPFR are similar or even the same. 
About (ii), given its agreement strength and stability across various $\alpha$, avg generally makes a better proxy than both \joint{} measures and weighted-avg, especially for cases with low single-aspect score variance or short PF. However, we discourage the use of avg (or weighted-avg) to approximate DPFR for other cases, as the results can be highly unstable or unpredictable in many other settings (e.g., for different PF characteristics, model score distributions, and the changes in agreement for different $\alpha$).

\subsection{Correlation with Simple Measures (RQ5)}
\label{ss:simple}

In the previous analyses, we have compared other measures directly against DPFR. Here, we aim to have a single, fixed point of comparison for all measures, so that they can be easily compared to the same baseline. Inspired by the intuitiveness test \cite{Sakai2012IntuitivenessTest}, which assesses measures by their agreement with a simpler measure, we compare all single-aspect and \joint{} evaluation approaches (including avg and DPFR) against the same baseline: simple evaluation measures of effectiveness and of fairness; neither of which is used to compute avg or DPFR. While the main purpose of this analysis is to compare the robustness of avg and DPFR based on their correlation strength towards an evaluation aspect, the rest of the measures are included as a control to better understand the fairness-relevance relationship across different datasets.
Specifically, as a simple effectiveness measure, we use the total number of relevant items in the top-$k$ recommendation (\# relevant items). 
This is because having a greater amount of relevant items in the top-$k$ generally means better recommendation effectiveness. We expect that P@$k$ will perfectly correlate with the number of relevant items, as P@$k$ is the average ratio of the number of relevant items to the number of recommended items per user (\# unique recommended items). 
As a simple measure of fairness, we choose the total number of unique items in the top-$k$. This is related to the \fair{} measure Coverage/QF, which is the ratio of the number of unique items in the top-$k$ to the total number of items in the dataset \cite{Zhu2020FARM:APPs,Rampisela2024EvaluationStudy}.
Based on most existing recommender system individual item fairness measures, the higher the number of unique recommended items, the better the fairness \cite{Rampisela2024EvaluationStudy}. The recommender models are then ranked from best to worst according to these simple measures. We compare the rankings given by these simple measures to the ones given by existing single-aspect and \joint{} evaluation approaches. To quantify their agreement to simple measures, we compute Kendall's $\tau$. 

\subsubsection*{Comparison between avg's and DPFR's correlations to simple measures} 
\begin{figure}
    \centering
    \includegraphics[width=\linewidth]{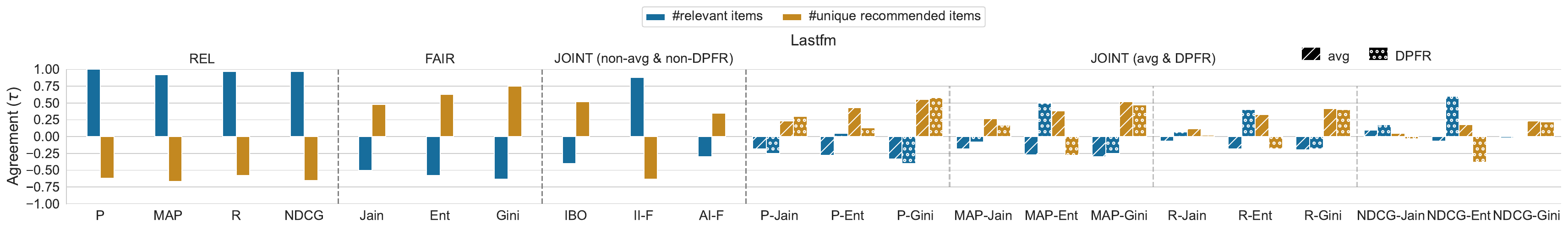}
    \includegraphics[width=\linewidth, trim=0 0 0 1cm, clip=True]{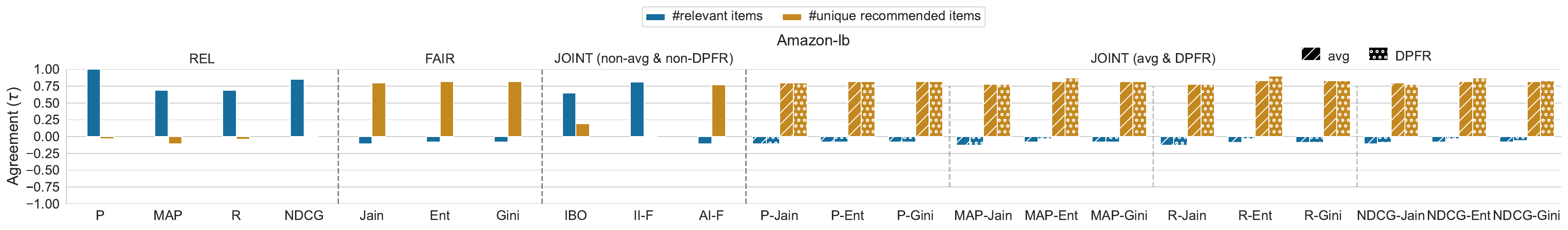}
    \includegraphics[width=\linewidth, trim=0 0 0 1cm, clip=True]{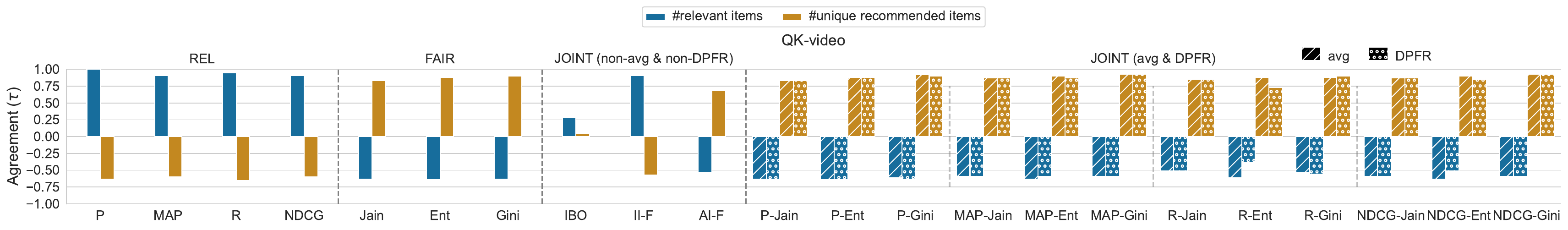}
    \includegraphics[width=\linewidth, trim=0 0 0 1cm, clip=True]{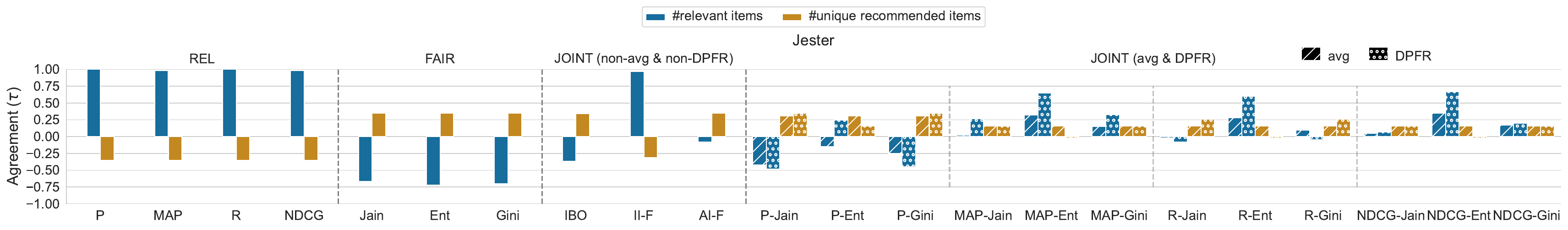}
    \includegraphics[width=\linewidth, trim=0 0 0 1cm, clip=True]{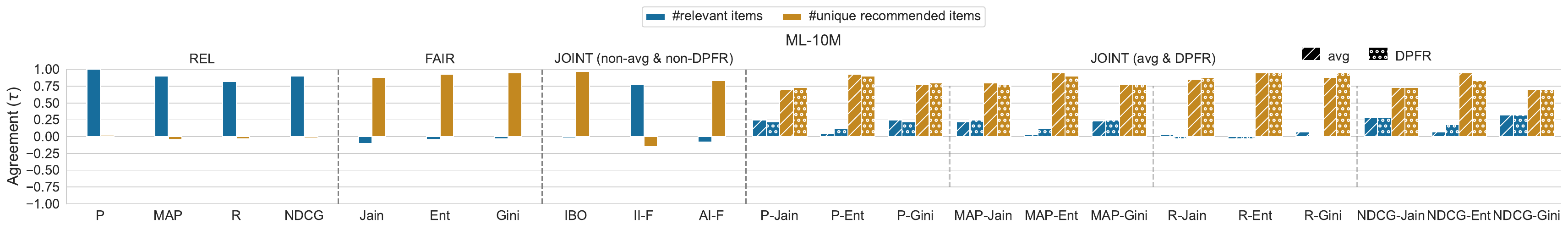}
    \includegraphics[width=\linewidth, trim=0 0 0 1cm, clip=True]{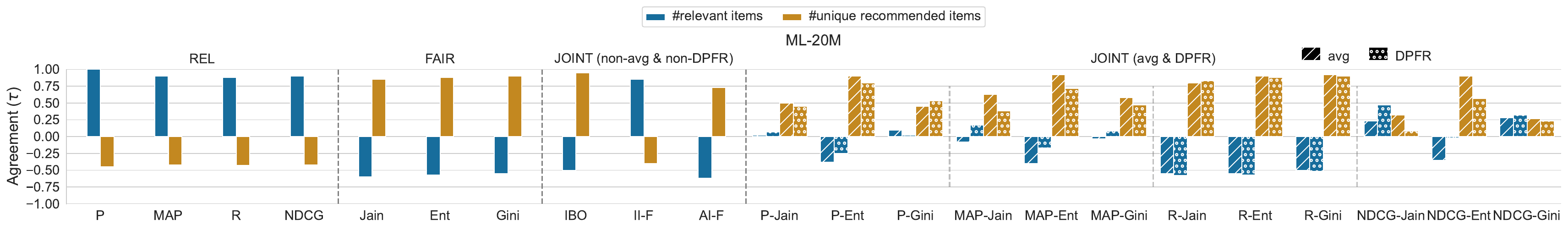}

    \caption{Agreement  (Kendall's $\tau$) of model ranking given by a simple measure of recommendation effectiveness (i.e., \#relevant items in the recommendation) and of recommendation fairness (\#unique recommended items) with the model ranking given by \rel{}, \fair{}, and \joint{} evaluation approaches (including avg and DPFR). We show all 12 \rel{}-\fair{} pairs used to compute avg and DPFR.}
    \label{fig:simple_measure}
\end{figure}

\Cref{fig:simple_measure} shows the correlation of the existing \rel, \fair, and \joint{} measures (including avg and DPFR) to simple measures. For brevity, we omit two \joint{} measures (IAA and MME) as they have similar trends with the remaining \joint{} measures (II-F and AI-F, respectively) due to their strong correlations \cite{Rampisela2024CanRelevance,Rampisela2025relevanceguidelines}. 

Firstly, we explain the overall trends in \Cref{fig:simple_measure}. Across all datasets, the overall agreement of avg and DPFR to \#unique recommended items as a simple fairness measure is higher than to \#relevant items.
Our results show that avg and DPFR in general have similar correlations to simple measures, in line with our measure correlation analysis that shows avg strongly correlates to DPFR (\Cref{ss:corr}). The similar correlation happens particularly for datasets with short PF or low \rel{}-score variance (i.e., Amazon-lb, QK-video, and ML-10M), where one evaluation aspect (i.e., fairness) dominates the joint evaluation. \textbf{When the scores for both evaluation aspects are well-distributed} (i.e., for Lastfm, Jester, ML-20M), \textbf{DPFR is dissimilar to avg in its correlation to simple measures.} 

Secondly, we look into how the single-aspect measures correlate to simple measures across different datasets, and how this affects avg's and DPFR's correlations to simple measures. In other words, we study whether: (i) the model rankings based \rel{} measures agree with the ones given by \#unique recommended items, as a simple measure of fairness; or (ii) the model rankings based on \fair{} measures agree with the ones based on \#relevant items, as a simple measure of recommendation effectiveness. 
Recommender models' ranking based on \rel{} and \fair{} scores are not correlated for some datasets: Amazon-lb and ML-10M. This can be seen from the \rel{} measures having near-zero correlation to \#unique recommended items (short or non-existent orange bars in the first column) or \fair{} measures having almost zero correlation to \#relevant items (short or non-existent blue bars in the second column).
In these cases, both avg and DPFR are strongly aligned with \#unique recommended items (i.e., they have strong positive correlations). At the same time, they have almost no relationship with \#relevant items (i.e., they have weak positive/negative correlations). The fairness dominance here is most likely due to the recommenders' \fair{} scores being more widely spread than their \rel{} scores, thus contributing more to the joint evaluation. 
On the other hand, \textbf{datasets with somewhat opposite relationship between \rel{} and \fair{}} (e.g., Lastfm, Jester, and ML-20M) \textbf{exhibit more variety in the strength and the sign of avg's and DPFR's correlation to the simple measures, indicating various trade-offs between fairness and relevance.}
An exception to this trend is QK-video. While this dataset also has strong disagreements between \rel{} and \fair{}, avg's and DPFR's correlation to simple measures are mostly uniform: strong agreement with \#unique recommended items and strong disagreement to \#relevant items. This may be due to the combination of its extremely low \rel{} score variance and very short PF.

Thirdly, we look into differences between measure pairs. Similar to previous findings (\Cref{ss:varying_alpha,sss:corr_measures}), we find that the trends for avg and DPFR computed with Jain or Gini are alike, because their PFs' characteristics are similar. In contrast, the trends differ for pairs with Ent. We also note that for Lastfm, pairs with Ent have different correlations signs for avg and DPFR, indicating different ways the two approaches balance the two evaluation aspects. Different correlation signs between avg and DPFR are also observed in Jester and ML-10M, albeit for fewer measure pairs.

Fourthly, we compare DPFR's and avg's overall agreement to simple measures in \Cref{tab:simple_measure_avg_vs_dpfr}, by counting how many times  DPFR's absolute correlation to simple measures is higher/lower than avg's. For (dis)agreement with \#relevant items, DPFR's absolute correlation is more often higher than avg's (i.e., for Lastfm, Jester, and ML-*). 
On the other hand, avg's absolute correlation with \#unique recommended items is more frequently higher than DPFR's (i.e., for Lastfm, QK-video, and ML-*). Considering both simple measures, avg's absolute correlation tends to be higher than DPFR's (i.e., for Lastfm, Amazon-lb, QK-video, and ML-20M). 
Consequently, \textbf{avg is less robust than DPFR; compared to DPFR, avg is more skewed towards one evaluation aspect while DPFR balances both better}. We note that for Amazon-lb and QK-video, the correlation strengths are mostly tied between avg and DPFR, possibly due to the overdominance of \fair{} scores in the joint evaluation, rendering similar conclusion between avg and DPFR.

\subsubsection*{Agreement for varying fairness-relevance weight.} 
We compute weighted-avg's and DPFR's agreement with simple measures for varying fairness-relevance weight ($\alpha$). We vary $\alpha \in [0,1]$ with an increment of 0.01. The idea is that given varying $\alpha$, weighted-avg's and DPFR's correlation to the simple measures should change following the different weight given to each evaluation aspect, but not too extremely. Having extreme correlation changes across $\alpha$'s could indicate instability, which is undesirable.

The results for weighted-avg and DPFR computed with all measure pairs that used Precision@$k$ (P) on Lastfm and QK-video are shown in \Cref{fig:simple_measure_alpha}; the rest of datasets and measure pairs are in \Cref{app:actual-scores}, \Cref{fig:simple_measure_alpha_app} as their trends are similar.
\textbf{Across all six datasets and varying fairness-relevance trade-off, we find that DPFR's correlation to the simple measures is more stable than weighted-avg's}. 
For Lastfm, Jester, and ML-*, weighted-avg's correlation to the simple measures vary widely for different $\alpha$'s, while DPFR's is relatively stable or changes moderately. 
For Amazon-lb and QK-video, there is a strong positive correlation between weighted-avg and \#relevant items at $\alpha=0$, which drops to zero around $\alpha\approx 0.05$ and $\alpha\approx0.1$, indicating the volatility of weighted-avg's correlation to simple measures. Similar huge changes are also found for weighted-avg's correlation to \#unique recommended items.

\subsubsection*{Overall findings}
\begin{figure}
    \centering
     \includegraphics[width=\linewidth]{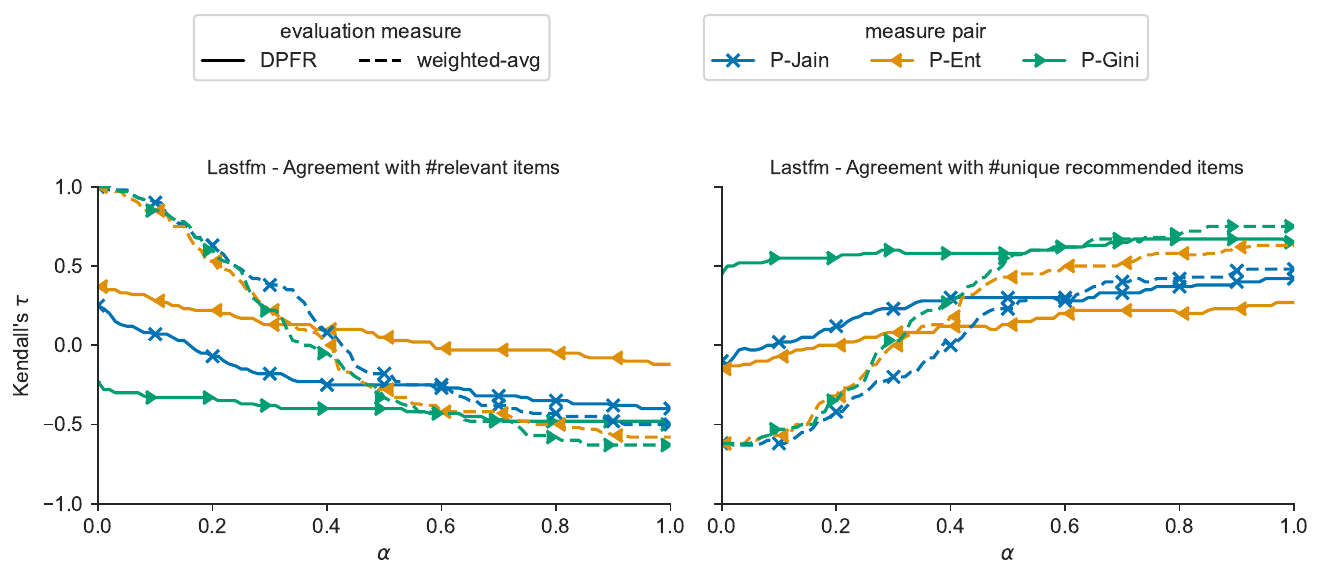}
     \includegraphics[width=\linewidth]{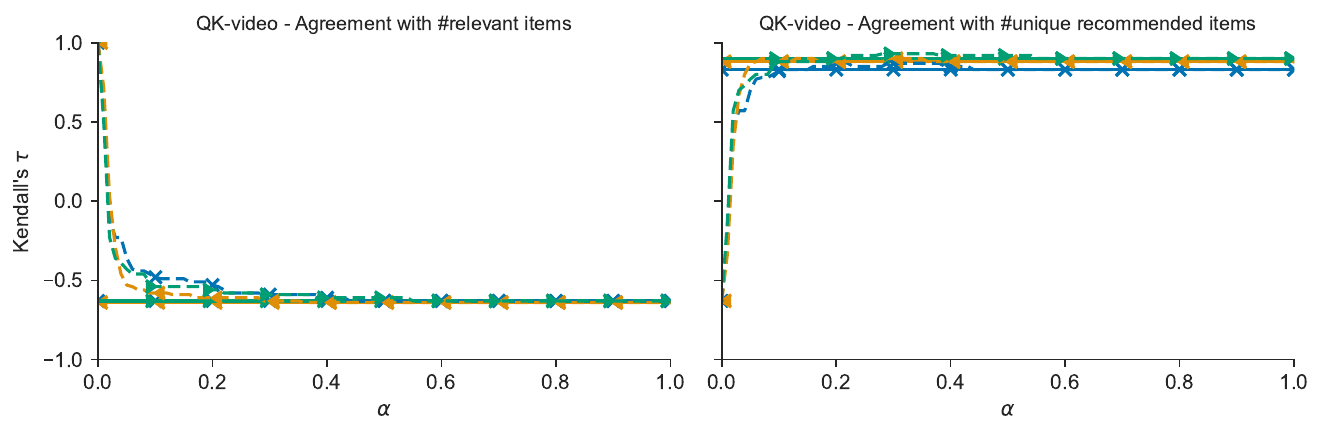}
    \caption{Agreement  (Kendall's $\tau$) of model ranking given by a simple measure of recommendation effectiveness (i.e., \#relevant items in the recommendation) and of recommendation fairness (\#unique recommended items) with the model ranking given by DPFR and weighted average (weighted-avg), for all measure pairs computed with Precision@$k$ (P) and for the datasets Lastfm and QK-video.}
    \label{fig:simple_measure_alpha}
\end{figure}

All in all, we find that \textbf{DPFR is more stable than avg and weighted-avg}, considering that: 
(i) the overall DPFR's correlation strength to simple measures is lower than avg's, excluding tied cases  (DPFR is lower than Avg for 4/6 datasets in \Cref{tab:simple_measure_avg_vs_dpfr}, `both simple measures' column); and 
(ii) DPFR's agreement to simple measures are more stable than weighted-avg's across varying fairness-relevance weight (\Cref{fig:simple_measure_alpha}). 
Both findings mean that compared to avg or weighted-avg, DPFR has a lower tendency to be skewed towards one evaluation aspect. 

Balancing fairness and relevance is not a trivial problem. In practice, averaging \rel{} and \fair{} scores does not give equal weight to each evaluation aspect. As \fair{} scores are typically more dispersed than the \rel{} ones, there are larger differences between the \fair{} scores. When the scores from the two aspects are averaged and ranked, the ranking would be affected more by the ranking of the \fair{} scores, which correlates strongly to the \#unique recommended items. 
We explain this through a simple example. Suppose that there are three models ($A$, $B$, and $C$) with the following ($\uparrow$\rel{},$\uparrow$\fair{}) scores, respectively: $A(0.94,0.1), B(0.93,0.2), C(0.92, 0.3)$. If the models are ranked based on their \rel{} scores from best to worst, we obtain the following ranking: $[A, B, C]$. When they are ranked from best to worst based on their \fair{} scores instead, we get $[C, B, A]$. Averaging the \rel{} and \fair{} scores, we have the following ranking of models, with their respective avg scores enclosed in brackets: $[C(0.61), B(0.565), A(0.52)]$. This results in a joint ranking that matches the ranking based on \fair{} scores (which would be highly similar to the ranking based on \#unique recommended items). Here, fairness overpowers relevance in what is supposedly a joint evaluation. The dominance of fairness evidently affects avg worse than DPFR (\Cref{tab:simple_measure_avg_vs_dpfr}).

These findings altogether highlight a desirable feature of DPFR: providing a joint evaluation of fairness and relevance based on dataset-dependent trade-off between the two aspects. For datasets with fairness-relevance trade-off, DPFR considers both evaluation aspects better than avg as its correlation to simple measures is less strong than avg's.
For datasets with no trade-off (i.e., where simultaneously achieving high fairness and high relevance is possible, such as in datasets with short PFs), DPFR relies more on fairness than relevance, as the recommenders' \fair{} scores vary more than their \rel{} scores. In this case, DPFR is still better than avg, which is skewed even more to the \fair{} scores.
\textbf{In short, DPFR balances both fairness and relevance evaluation aspects better than (weighted-)avg, as its correlation to the simple measures is overall lower and more stable than (weighted-)avg's.}

\begin{table}
\caption{Percentage of times the absolute correlation of simple measures to avg is higher/lower/the same than to DPFR in \Cref{fig:simple_measure}, in the avg/DPFR/Tie columns, respectively. Darker grey shades indicate a higher percentage of occurrence.}
\label{tab:simple_measure_avg_vs_dpfr}
\resizebox{\textwidth}{!}{
\begin{tabular}{lrrr|rrr|rrr}
\toprule
agreement with & \multicolumn{3}{c|}{\#relevant items} & \multicolumn{3}{c|}{\#unique rec.~items} & \multicolumn{3}{c}{both simple measures} \\
\midrule
 & Avg higher (\%)& DPFR higher (\%)& Tie (\%)& Avg higher (\%)& DPFR higher (\%)& Tie (\%)& Avg higher (\%)& DPFR higher (\%)& Tie (\%)\\
\midrule
Lastfm & \cellcolor[HTML]{E9ECEF} 41.67 & \cellcolor[HTML]{E9ECEF} 50.00 & 8.33 & \cellcolor[HTML]{CED4DA} 75.00 & 25.00 & 0.00 & \cellcolor[HTML]{CED4DA} 58.33 & \cellcolor[HTML]{E9ECEF} 37.50 & 4.17 \\
Amazon-lb & \cellcolor[HTML]{E9ECEF} 41.67 & 0.00 & \cellcolor[HTML]{CED4DA} 58.33 & 8.33 & \cellcolor[HTML]{E9ECEF} 33.33 & \cellcolor[HTML]{CED4DA} 58.33 & 25.00 & 16.67 & \cellcolor[HTML]{CED4DA} 58.33 \\
QK-video & 25.00 & 16.67 & \cellcolor[HTML]{CED4DA} 58.33 & \cellcolor[HTML]{E9ECEF} 33.33 & 8.33 & \cellcolor[HTML]{CED4DA} 58.33 & \cellcolor[HTML]{E9ECEF} 29.17 & 12.50 & \cellcolor[HTML]{CED4DA} 58.33 \\
Jester & 8.33 & \cellcolor[HTML]{ADB5BD}91.67 & 0.00 & \cellcolor[HTML]{E9ECEF} 33.33 & \cellcolor[HTML]{E9ECEF} 33.33 & \cellcolor[HTML]{E9ECEF} 33.33 & 20.83 & \cellcolor[HTML]{CED4DA}62.50 & 16.67 \\
ML-10M & 25.00 & \cellcolor[HTML]{E9ECEF} 41.67 & \cellcolor[HTML]{E9ECEF} 33.33 & \cellcolor[HTML]{E9ECEF} 41.67 & \cellcolor[HTML]{E9ECEF} 33.33 & 25.00 & \cellcolor[HTML]{E9ECEF} 33.33 & \cellcolor[HTML]{E9ECEF} 37.50 & \cellcolor[HTML]{E9ECEF} 29.17 \\
ML-20M & \cellcolor[HTML]{E9ECEF} 33.33 & \cellcolor[HTML]{CED4DA}66.67 & 0.00 & \cellcolor[HTML]{ADB5BD}83.33 & 16.67 & 0.00 & \cellcolor[HTML]{CED4DA}58.33 & \cellcolor[HTML]{E9ECEF} 41.67 & 0.00 \\
\bottomrule
\end{tabular}}
\end{table}

\subsection{Efficient DPFR (RQ6)} 
\label{ss:compare-eff}

\subsubsection{The Efficiency of the PF Generation} 
We study the efficiency of DPFR by comparing the PF, an estimated version of PF on a subset of points, and the \joint{} measures. The estimated version of PF uses 3--12 points as per \Cref{ss:compute-eff}. 
We compute the number of points in the estimated PF as \% of those in the PF, and the resulting computation times. One point in the PF translates to one round of computing all \rel{} and \fair{} measures, so fewer points mean faster. We report the estimated PF with only 3, 6, and 12 points in \Cref{tab:big-est-efficiency}. 

\begin{table}[tbp]
    \centering
    \caption{Efficiency of PF, estimated PF (Est.~PF), and \joint{} measures: \% of data points in the Est.~PF (\% pts) and computation time. 
    Both PFs compute 11 \rel{} and \fair{} measures simultaneously. 
    The times for other evaluation measures are averaged (Average per model) and summed (All models) over 16 model combinations.}
    \label{tab:big-est-efficiency}
     \begin{tabular}{lllrrrrrr}
\toprule
& & {} &  Lastfm &  Amazon-lb &  QK-video &  Jester &  ML-10M &  ML-20M \\
\midrule
\multicolumn{2}{c}{\#pts}& PF& 4882 & 847 & 499 & 16202 & 2781 & 3783 \\
\cline{1-9}
\multicolumn{2}{c}{\multirow[c]{3}{*}{\%pts}} & Est.~PF (12 pts) & 0.25 &      1.42 &     2.40 &   0.07 &   0.43 &   0.32 \\
& & Est.~PF (6 pts) &  0.12 &      0.71 &     1.20 &   0.04 &   0.22 &   0.16 \\
& & Est.~PF (3 pts) & 0.06 &      0.35 &     0.60 &   0.02 &   0.11 &   0.08\\
\midrule
\multirow[c]{14}{*}{\rotatebox[origin=c]{90}{$\downarrow$ Computation time (mins.)}}& & PF           &   19.18 &       0.56 &     10.49 &  847.42 &   28.99 &   75.77 \\
& & Est.~PF (12 pts) &    2.02 &       0.19 &      4.23 &  552.16 &    1.90 &    2.60 \\
& & Est.~PF (6 pts)  &    2.00 &       0.19 &      4.12 &  551.72 &    1.84 &    2.54 \\
& & Est.~PF (3 pts)  &    2.01 &      0.19 &     4.07 &  552.26 &   1.82 &   2.52\\
\cline{2-9}
& \multirow[c]{5}{*}{\rotatebox[origin=c]{90}{\small Average/model}} & IBO    &    <0.3s &     <0.3s &      0.01 &    0.01 &   <0.3s &     0.01  \\
& & MME    &    2.04 &       0.03 &     19.51 &    0.09 &   15.25 &    89.13 \\
& & IAA    &    <0.3s &     <0.3s &      0.01 &    0.02 &    0.01 &     0.02 \\
& & II-F   &    <0.3s &     <0.3s &      0.01 &    0.01 &    0.01 &     0.02 \\
& & AI-F   &    <0.3s &     <0.3s &      0.01 &    0.01 &   <0.3s &     0.01 \\
\cline{2-9}
& \multirow[c]{5}{*}{\rotatebox[origin=c]{90}{\small All models}} & IBO  &    0.02 &     <0.3s &      0.10 &    0.12 &    0.06 &     0.15 \\
& & MME  &   32.63 &       0.49 &    312.14 &    1.38 &  244.03 &  1426.10 \\
& & IAA  &    0.03 &     <0.3s &      0.10 &    0.36 &    0.13 &     0.30  \\
& & II-F &    0.04 &     <0.3s &      0.16 &    0.14 &     0.1 &     0.25 \\
& & AI-F &   0.03 &     <0.3s &      0.11 &    0.13 &    0.07 &     0.17 \\
\bottomrule
\end{tabular}
\end{table}

\begin{table}[tbp]
    \centering
    \caption{Effectiveness of estimated PF (Est.~PF). Dist is the average distance between midpoints in the PF and Est.~PF over 12 measure pairs; low distance is desirable as the Est.~PF's reference point is close to the PF's, indicating higher effectiveness of the estimation. Minimum agreement (Min $\tau$) is the Kendall's $\tau$ correlation between DPFR with PF and Est.~PF; higher $\tau$ indicates more similar ranking of models given by PF and Est.~PF, thus higher effectiveness of the estimation.}
    \label{tab:big-est-effectiveness}
    \begin{tabular}{lllrrrrrr}
    \toprule
    & & {} &  Lastfm &  Amazon-lb &  QK-video &  Jester &  ML-10M &  ML-20M \\
    \midrule
    \multicolumn{2}{c}{\multirow[c]{3}{*}{$\uparrow$ 
    Min $\tau$}} & Est.~PF (12 pts)& 0.95 &       1.00 &      1.00 &    0.98 &    0.98 &    0.97\\
    \multicolumn{2}{c}{} &  Est.~PF (6 pts)&  0.90 &       0.97 &      1.00 &    0.98 &    0.95 &    0.92\\
    \multicolumn{2}{c}{} &  Est.~PF (3 pts)&  0.78 &       0.98 &      1.00 &    1.00 &    0.97 &    0.75\\
    \midrule
    \multicolumn{2}{c}{\multirow[c]{3}{*}{$\downarrow$ Dist.}} & Est.~PF (12 pts)& 0.01 &      0.02 &     0.00 &   0.00 &   0.01 &   0.01 \\
    \multicolumn{2}{c}{} &  Est.~PF (6 pts)& 0.03 &      0.05 &     0.00 &   0.00 &   0.02 &   0.02 \\
    \multicolumn{2}{c}{} &  Est.~PF (3 pts)& 0.03 &      0.05 &     0.00 &   0.00 &   0.03 &   0.05 \\
    \bottomrule
    \end{tabular}
\end{table}

The PF (\Cref{fig:pareto-pair-plot}) has hundreds to tens of thousands of points each, while the estimated PF only contains 0.02--2.40\% of the points, which means reduced computational complexity for the PF generation. In terms of actual computation time (\Cref{tab:big-est-efficiency}), computing the PF with \textsc{Oracle2Fair} take 0.56--75.77 mins to compute, but only 0.19--4.07 mins for the PFs estimated with 3 points for all datasets except Jester. For Jester, it takes $\sim$14 hours and the estimation only takes $\sim$9 hours. 
However, this is expected for Jester as it has 62K users in the test split, as opposed to the 3.5K or fewer in the other datasets (i.e., see \Cref{tab:data_split} in \Cref{app:stat}). 
While computing the estimated PFs is on average slower than computing the joint measures IBO, IAA, II-F, and AI-F, it is expected as the (estimated) PFs compute 11 measures simultaneously. Yet, in most cases (except Amazon-lb and Jester), the estimated PFs are still faster to compute than the time to compute MME for one recommender model per dataset, let alone to compute MME for all models. For ML-20M, computing the estimated PF is even up to 35 times faster than computing MME of one model.  

\subsubsection{The Effectiveness of Efficient DPFR} 
To what extent is the DPFR from the efficiently generated PF (estimated PF) a reasonable proxy for fairness-relevance joint evaluation using the PF, in terms of giving a similar ordering of models? 
We compare the DPFR from the PF and the estimated PF using Kendall's $\tau$ correlations. Further, as DPFR is computed based on an $\alpha$-based reference point lying on the PF, to quantify possible accuracy loss of the estimated PF, we also compute the error of the midpoint estimation. This error is the Euclidean distance between the reference point in the PF and the estimated PF (i.e., the midpoint in our case), as per \cite{Wang2016AEngineering}. For brevity, we report the effectiveness of the estimated PF with only 3, 6, and 12 points in \Cref{tab:big-est-effectiveness}; see \Cref{app:actual-scores}, \Cref{tab:corr_est} for extended results.

We first analyse the error of the midpoint estimation. For the 12 measure pairs and 6 datasets, the midpoint coordinates on average do not move much: the distance is 0.00--0.05, even when the PF is only estimated with 3 points. 
Ergo, the correlations between the rank ordering of models given by the DPFR of PF and its estimation are still equivalent ($\tau \geq 0.9$) when estimated with 6 or 12 points \cite{Maistro2021PrincipledRankings,Voorhees2001EvaluationDocuments}. Even the 3-point estimation maintains high agreement ($\tau \in [0.75,1]$), with only 5 cases having $\tau<0.9$ across 6 datasets and 12 measure pairs. 
Therefore, it is possible to only compute a small number of points in the PF, e.g., 6 or 12 points, and still make a reliable PF estimation, evidenced by the small shift of the PF midpoint and the rank ordering of the models remaining equivalent ($\tau \geq 0.9$), if not identical ($\tau=1$) for all measure pairs and datasets.

\section{Guidelines for Evaluation with DPFR}
\label{s:guidelines}

Based on our experimental findings (\Cref{s:experiments}), we summarise factors to consider when selecting single-aspect and multi-aspect evaluation measures/approaches in \Cref{tab:guidelines}.
We also provide detailed guidelines on using DPFR. 
For specific guidelines on \fair{} measures and relevance-aware fairness measures for individual items, we refer the reader to \cite{Rampisela2024EvaluationStudy} and \cite{Rampisela2025relevanceguidelines}, respectively.

\begin{table*}
    \caption{Factors to consider when selecting single-aspect measures (\rel/\fair{} measures) and multi-aspect evaluation approaches: relevance-aware fairness measures (\joint{}), avg, weighted-avg, Efficient DPFR, and DPFR. Here, \joint{} measures refer exclusively to IBO/IAA/II-F/MME/AI-F. Efficient DPFR refers to DPFR computed with an estimated version of PF (\Cref{ss:compare-eff}). All statements below represent desirable features. YES means the statement applies, NO means the statement does not apply, and N/A means the statement is not applicable to the measure/approach.}
    \label{tab:guidelines}
    \centering
    \resizebox{\textwidth}{!}{
    \begin{tabular}{l|cccccc}
    \toprule
    \toprule
         & \rel{}/\fair{} & \joint{} & avg & weighted-avg & Efficient DPFR & DPFR \\
    \midrule
    \multicolumn{7}{l}{\textit{Factors related to multi-aspect evaluation}} \\
    \midrule
     This measure/approach can be used with existing single-aspect measures  &N/A&NO&YES&YES&YES&YES \\
    This measure/approach has a tunable fairness-relevance trade-off weight &N/A&NO&NO&YES&YES&YES \\
     This measure/approach is based on the dataset's possible fairness-relevance trade-offs  &N/A&NO&NO&NO&YES&YES \\
    This measure/approach generally gives equivalent model ranking to DPFR &NO&NO&NO&NO&YES& N/A \\
    \midrule
    \multicolumn{7}{l}{\textit{Other factors to consider}} \\
    \midrule
    This measure/approach can be used for single-aspect evaluation &YES&NO*&NO &YES\textsuperscript{\textdagger} & NO & NO\\
    This measure/approach is efficient to compute &YES&YES$\textsuperscript{\textdaggerdbl}$&YES&YES&YES&NO \\
    \bottomrule
    \bottomrule
    \multicolumn{7}{l}{\small Remarks:}\\
    \multicolumn{7}{l}{\small*even if they can give similar ranking to single-aspect measures, they should not be used for single-aspect evaluation}\\
    \multicolumn{7}{l}{\small\textsuperscript{\textdagger}for $\alpha\in\{0,1\}$}\\
    \multicolumn{7}{l}{\small\textsuperscript{\textdaggerdbl}except for MME}\\
    \bottomrule
    \end{tabular} 
    }
\end{table*}

\subsubsection*{Selecting relevance-fairness measure pairs} DPFR can technically be used with existing \rel{} and \fair{} measures, but not all combinations of measure pairs would result in a meaningful Pareto Frontier (PF), as explained in \Cref{sss:no-fit}. We recommend pairing Precision@$k$, MAP@$k$, Recall@$k$, or NDCG@$k$ as \rel{} measures with Jain's Index (Jain), Entropy (Ent), or Gini Index (Gini) as \fair{} measures. These pairs are more likely to produce PFs with trade-offs between fairness and relevance than other \rel{}-\fair{} pairs. Pairing MAP-Ent or NDCG-Ent in particular leads to distinguishable conclusions from simply averaging or taking the best model based on single-aspect measures. 
Moreover, using the individual measure in the pair for single-aspect evaluation has other benefits.
First, the use of rank-based \rel{} measures enables more fine-grained recommendation effectiveness assessment. Second, Ent as a \fair{} measure is less strict than Jain or Gini \cite{Rampisela2024EvaluationStudy}, which means that compared to Jain/Gini, it is less likely to compress its scores to one end of its range. As a result, it is easier to identify differences in recommender model \fair{} scores.

\subsubsection*{Interpreting and selecting PF reference point with and without \texorpdfstring{$\alpha$}{$\alpha$}}
Firstly, DPFR at $\alpha=0$ or $\alpha=1$ should not be interpreted as a replacement for \rel{} or \fair{} measures. This is based on the findings that single-aspect measures do not give equivalent ranking of models to DPFR at $\alpha\in\{0,1\}$ (\Cref{ss:varying_alpha}). DPFR is intended as a joint evaluation measure that accounts for both relevance and fairness. 

Secondly, to select $\alpha$, we encourage analysing how $\alpha$ affects DPFR's correlation to the single-aspect measures used in its computation (\Cref{ss:varying_alpha}). If the intention is to select an $\alpha$ that balances fairness and relevance, one can select $\alpha$ where the \rel{} and \fair{} measures' correlation to DPFR are the most similar. In the case where the generated PF is short, or the recommenders' scores have low variance in one evaluation aspect, $\alpha$ may not have much effect on the final ranking given by DPFR.

As an alternative of selecting $\alpha$, one can always choose to directly use the points in the PF. Suppose that the goal is to have a recommender model with an NDCG score of at least $0.8$. One can then search the PF for the best possible \fair{} score that can be obtained for NDCG $\geq0.8$ and use the \rel{}-\fair{} scores as a reference point for DPFR computation.

\subsubsection*{Proxies for DPFR} We acknowledge that DPFR is not the most efficient joint fairness-relevance evaluation approach. As such, we have provided Efficient DPFR (\Cref{ss:compute-eff}), which requires less computational time and serves as a good proxy for DPFR (\Cref{ss:compare-eff}). 

In restricted cases where either the recommenders' \rel{} or \fair{} scores do not vary much, simple average's alignment to our Pareto-optimal based DPFR may surprisingly be better than weighted average's (even for various fairness-relevance trade-off weights) or other \joint{} measures', though still worse than Efficient DPFR's. 
However, it can be hard to determine how low the score variance should be, for the simple average to successfully estimate the ranking of models given by DPFR.
Another case where simple average may align with DPFR is when the dataset distribution does not have much trade-off between the achievable fairness and relevance (i.e., the PF is short). Nevertheless, obtaining this information would require generating the (estimated) PF first. Even then, more research is needed to understand how the PF length affects the alignment of simple average to DPFR.
For all the above-mentioned reasons, we recommend using Efficient DPFR over simple average to estimate DPFR.

\section{Conclusions}

We have proposed DPFR, a novel joint evaluation approach for Recommender System (RS) fairness and relevance. DPFR has several advantages. Firstly, it can be used with various pairs of existing single-aspect measures. Additionally, it also follows the Pareto optimality concept, facilitating a grounded joint evaluation based on how close the recommendation is to the best possible fairness-relevance trade-offs, according to the ground truth. As such, DPFR is modular, tractable, and intuitive. Through a series of experiments, we find that no existing joint evaluation approaches can be a reliable proxy for DPFR. Further, our results show that DPFR is the most robust among existing joint evaluation approaches, as it is the least likely to be skewed towards one evaluation aspect. We conclude with a series of guidelines to guide the use of DPFR, by providing tips on how to best select the measure pairs and the reference point in the DPFR computation, as well as advising how to efficiently and effectively approximate DPFR. 

In this work, we designed DPFR for individual item fairness. Future work could extend DPFR to other RS fairness concepts (e.g., group fairness or user fairness \cite{Rampisela2025StairwayFairness}) and possibly to other RS evaluation aspects that are adjacent to fairness, such as diversity and popularity bias.

\begin{acks}
The work is supported by the Algorithms, Data, and Democracy project (ADD-project), funded by Villum Foundation and Velux Foundation. 
Qiuchi Li contributed to the idea of computing the reference point. We thank DIKU IR Lab members and the anonymous reviewers, who have provided helpful feedback to improve earlier versions of the manuscript.
\end{acks}

\bibliographystyle{ACM-Reference-Format}
\bibliography{references}

\appendix
\section{Extended Dataset Statistics}\label{app:stat}
\Cref{tab:data_split} presents the statistics of each dataset split. For several datasets (e.g., Amazon-lb and ML-*), the number of users in the test split is significantly less than the number of users in the train split. \Cref{tab:rel_item_stat} presents the statistics of items in the test split, per user.

\begin{table}[htbp]
    \centering
        \caption{Number of [users, items, and interactions] in the train, validation, and test split after preprocessing. 
    } 
    \label{tab:data_split}
\begin{tabular}{llll}
\toprule
{} &                   train &                     val &                    test \\
\midrule
Lastfm    &     [1842, 2821, 42758] &     [1831, 2448, 14248] &     [1836, 2476, 14237] \\
Amazon-lb &       [1054, 552, 8860] &        [470, 204, 1811] &        [437, 209, 1726] \\
QK-video  &     [4656, 6245, 34345] &      [3470, 4095, 8726] &      [3514, 4101, 8706] \\
Jester    &   [63724, 100, 1294511] &    [62137, 100, 427623] &    [62167, 100, 427926] \\
ML-10M    &  [49378, 6838, 4944064] &    [2695, 7828, 296914] &    [1523, 7880, 121707] \\
ML-20M    &  [89917, 8719, 9882504] &    [4987, 10742, 472243] &    [2178, 13935, 233394] \\
\bottomrule
\end{tabular}
\end{table}

\begin{table}[htbp]
    \caption{Statistics of items in the test split, per user, i.e., the number of relevant items per user.}
    \label{tab:rel_item_stat}
    \centering
\begin{tabular}{lrrrr}
\toprule
{} &    mean &  min &  median &   max \\
\midrule
Lastfm    &    7.75 &    1 &       8 &    19 \\
Amazon-lb &    3.95 &    1 &       3 &    16 \\
QK-video  &    2.48 &    1 &       2 &    16 \\
Jester    &    6.88 &    1 &       6 &    29 \\
ML-10M    &   79.91 &    1 &      46 &  1632 \\
ML-20M    &  107.16 &    1 &      53 &  2266 \\
\bottomrule
\end{tabular}
\end{table}

\section{Algorithms for Generating Pareto Frontier}\label{app:algo}
We present the pseudocodes of the algorithms for generating the Pareto Frontier: the Oracle (\Cref{alg:oracle}) and  \textsc{Oracle2Fair} (\Cref{alg:oracle2fair}). We provide the worst-case time complexity analysis for both algorithms and discuss a possible edge case. 
We denote as $k$ the recommendation cut-off, as $m$ the number of users, as $n$ the number of items, as $H$ the maximum number of items in a user's interaction history ($H_u$) across all users, and as $R$ the maximum number of items in a user's test split ($R_u^*$) across all users. We use the binary logarithm ($\log_2{}$). For brevity, we omit lines with $O(1)$, and we state the reasoning for each line (L).

\SetKwComment{Comment}{/* }{ */}
\RestyleAlgo{ruled}
\LinesNumbered

\begin{algorithm*}[htbp]
{\small
\caption{Oracle \\ Create recommendations with the highest relevance}\label{alg:oracle}

\KwData{ \\
$I$: all items in the dataset; \\
$H_u$: items in train-val split for each user $u \in U$;  \\
$R_u^*$: items in test split (relevant items) for each user $u \in U$; \\
$k$: number of recommended items
}
\KwResult{\\
$rec$: most relevant recommendation\\\\
$result$: a list of relevance and fairness scores\\
$itemNotInRec$: items that are not in the recommendation}

\Comment{Handle users with exactly $|R_u^*|=k$}
\lForEach{$u \in U$ where $|R_u^*| = k$}{{$rec[u] \gets R_u^*$}}

\Comment{Handle users with $|R_u^*| > k$}

\For{$K=k+1$ \KwTo $max(|R_u^*|)$}{\label{ln:startgtk}
    $userWithK \gets$ get users where $|R_u^*|=K$ \\
    \ForEach{$u \in userWithK$}{
        $takenItem[u] \gets {R_u^* \cap rec}$ \\
        $weight[u] \gets sum(countInRec(takenItem[u]))$ \\
        }
    $sortUserWithK \gets$ sort $userWithK$ by the least weight \\
    $tempRec[u] \gets R_u^* \setminus takenItem[u]$ \\
    keep only max $k$ items in $tempRec[u]$ \\

    \ForEach{$u \in sortUserWithK$}{
        $rec[u] \gets tempRec[u]$ \\
        $numItemToAdd \gets k - |tempRec[u]|$ \\
        sort $takenItem[u]$ by the least item count \\
        $rec[u].append(takenItem[u][:numItemToAdd])$  
    }
}  \label{ln:endgtk}

\Comment{Handle users with $|R_u^*| < k$}
$remainUser \gets$ get users where $|R_u^*|<k$ \\ \label{ln:startltk}
\lForEach{$u \in remainUser$}{$rec[u] \gets |R_u^*|$} 
$itemNotInRec \gets I \setminus rec$ \\
\ForEach{$u \in remainUser$}{
    \While{$|rec[u]|<k$ and $itemNotInRec \neq \emptyset$}{
        \For{$item \in itemNotInRec$}{
         \If{$item \notin H_u$}{
         $rec[u].append(item)$ \\
         $itemNotInRec \gets itemNotInRec\setminus \{item\}$
         }
        }    
    }

    \If{$|rec[u]|<k$}{
        \While{$|rec[u]|<k$}{

        $candItem \gets$ least popular item in $rec$ that is not in $H_u \cup R_u^* \cup rec[u]$ \\
        $rec[u].append(candItem)$
    }
    }
} \label{ln:endltk}
$result \gets calculateScores(rec)$
}
\end{algorithm*}

\begin{algorithm*}[htbp]
\caption{\textsc{Oracle2Fair} \\After recommending maximally relevant items, iteratively change the recommendation list to increase fairness until maximum fairness is reached}\label{alg:oracle2fair}
\KwData{$H_u, R_u^*,I, k$}
\KwResult{\\
$rec$: most fair possible recommendation;\\
$result$: a list of relevance and fairness scores}

$rec, result, itemNotInRec \gets Oracle(I, H_u, R_u^*)$

\Comment{Get the most popular item in the recommendations and its frequency count}
 $newMostPop \gets mostPop \gets getMostPopItem(rec)$  \\ \label{ln:updatecountbegin}
 $newCntPop \gets cntPop \gets cnt(mostPop, rec)$ \\

$uWithMostPop \gets$ all users with $mostPop \in rec[u]$ \\
sort $uWithMostPop$ by largest index of $mostPop$ in $rec[u]$ \\ \label{ln:updatecountend}
\For{$i \in itemNotInRec$}{
\lIf{$cntPop=1$}{break}    
\If{$newMostPop \neq mostPop$}{
    $mostPop \gets newMostPop$ \\ \label{ln:mostpopstart}
    update $uWithMostPop$ following $mostPop$ \\ \label{ln:mostpopend}
}
\If{$newMostPop = mostPop$}{
    $candU \gets$ all $u$ in $uWithMostPop$ where $i \notin H_u$  \\
    \If{$\exists u \in candU$ with $i \in R_u^*$ \label{ln:recommendstart}}{
        recommend $i$ to the top $u$ from $candU$ with $i \in R_u^*$
    }
    \lElse{recommend $i$ to the top $u$ from $candU$}
    reorder $rec[u]$ so all relevant items are at the top \\
    $result.append(calculateScores(rec))$ \\ \label{ln:recommendend}
    $itemNotInRec \gets itemNotInRec \setminus \{i\}$ \\
    $newMostPop \gets getMostPopItem(rec)$ \\
    $newCntPop \gets cnt(mostPop, rec)$ 
    }
  }
\Else{
    do lines \ref{ln:updatecountbegin}--\ref{ln:updatecountend} \\ \label{ln:dolines}
    $i \gets leastPop \gets getLeastPopItem(rec)$ \\ \label{ln:getleastpop}
    $m, n \gets |U|, |I|$ \\
    \While{$cntPop > \lceil km/n \rceil$}{
    \lIf{$newMostPop \neq mostPop$}{do lines \ref{ln:mostpopstart}--\ref{ln:mostpopend}}
    \If{$newMostPop = mostPop$}{
    $candU \gets$ all $u$ in $uWithMostPop$ where $i \notin H_u \cup rec[u]$  \\
    do lines \ref{ln:recommendstart}--\ref{ln:recommendend} \\
    do lines \ref{ln:dolines}--\ref{ln:getleastpop}
    }
    }
}

\end{algorithm*}

\subsection{Time Complexity of the Oracle (\Cref{alg:oracle})}
\label{app:analyse-oracle}

The line-by-line time complexity analysis of Oracle (\Cref{alg:oracle}) is as follows:

\begin{itemize}
    \item L1: $O(km)$, creating list of size $k$ for $m$ users
    \item L2--17: $O(R^2m + Rm \log{m} + Rmk\log{k})$, resulting from at most $R$ iterations of:
    \begin{itemize}
        \item L3: $O(m)$, linear search on list of size $m$
        \item L4--7: $O(mR)$, at most $m$ times looking up at most $R$ values
        \item L8: $O(m \log{m})$, sorting a list of size $m$
        \item L11--16: $O(mk \log{k})$, at most $m$ times sorting list of size $k$
    \end{itemize}
    
    \item  L18: $O(m)$, linear search on list of size $m$
    \item  L19: $O(m)$, $m$ assignment operations
    \item  L21--36: $O(k^2m^2)$, resulting from at most $m$ iterations of: 
    \begin{itemize}
        \item L22--29: $O(kH)$, at most $k$ times of linear search on list of size $H$
        \item L30--L35: $O(k^2m)$, $k$ times of counting in a list of size $km$.
    
        In most cases, the number of users $m >> H$, so the time complexity of this block is dominated by $O(k^2m)$.
    \end{itemize}
    \item L37: $O(km)$ computing relevance/fairness measures for $m$ users based on recommendation list of size $k$

\end{itemize}

Overall, the time complexity of Oracle is dominated by that of L2 and L21--36. Hence, the time complexity of Oracle is $O(R^2m + Rm \log{m} + Rmk \log{k} + k^2m^2)$.

\subsection{Time Complexity of \textsc{Oracle2Fair} (\Cref{alg:oracle2fair})}

The line-by-line time complexity analysis of \textsc{Oracle2Fair} (\Cref{alg:oracle2fair}) is as follows:

\begin{itemize}
    \item L1: $O(R^2m + Rm \log{m} + Rmk \log{k} + k^2m^2)$ (derived in \Cref{app:analyse-oracle})
    \item L2--4: $O(km)$, $m$ times of counting and linear search on list of size $k$
    \item L5: $O(m \log{m})$, sorting a list of size $m$ (e.g., with Tim Sort)
    \item L6--24: $O(max(Hmn, kmn))$, resulting from at most $n$ iterations of: 
    \begin{itemize}
        \item L10: $O(km)$, $m$ times of linear search on list of size $k$
        \item L13: $O(Hm)$, $m$ times linear search on list of size at most $H$
        \item L14: $O(km)$
        \item L18: $O(k \log{k})$, sorting a list of size $k$
        \item L21--L22: $O(km)$; the term $O (k \log{k})$ is dominated by $O(km)$, as typically the cut-off $k$ is much smaller than the number of users $m$ (and hence $\log{k} < m$).
    \end{itemize}

     \item L26: $O(km + m \log{m})$, combining L2--4 and L5
     \item L27: $O(km)$, $m$ times counting on list of size $k$
     \item L29--34: $O(max(Hkm^2,k^2m^2)+km^2\log{m})$, resulting from at most $km$ iterations (derived below) of:
     \begin{itemize}
         \item L30: $O(km)$, from L10
         \item L32: $O(Hm)$, $m$ times linear search on list of size at most $H$
         \item L33: $O(km)$, from L14 and L18
         \item L34: $O(km+m \log{m})$, from L26--27
     \end{itemize}
\end{itemize}

We estimate the worst-case complexity of the number of iterations of the code block in L29--34, by assuming that the initial recommendation is the unfairest. The unfairest recommendation happens when the same $k$ unique items are recommended to all $m$ users. 
This code block aims to reduce the max frequency count of all items to $\left\lceil km/n \right\rceil$. Hence, the number of iterations is $\sum^{k} (m-\left\lceil km/n \right\rceil) \leq k(m-\frac{km}{n}-1) = km-\frac{k^2m}{n}-k \leq km$.

All in all, the time complexity of \textsc{Oracle2Fair} is dominated by that of L1, L6--24, and L29--34. Hence, the overall time complexity of \textsc{Oracle2Fair} is $O(R^2m + Rm \log{m} + Rmk \log{k} +  k^2m^2 + km^2\log{m} + kmn )$ if $k\geq H$, or $O(R^2m + Rm \log{m} + Rmk \log{k} +  Hkm^2 + km^2\log{m} + Hmn )$ otherwise. To simplify the time complexity, further assumptions need to be made for one or more variables.

\subsection{Possible Edge Cases}
There might be edge cases, for example, datasets where the train/val set of each user contains almost all items in the dataset (each user has rated/clicked most items in the dataset). In this case, if we do not want to re-recommend items in the train/val set to users, some items may have item counts more than
$\left\lceil km/n \right\rceil$ at the end of the process, and \Cref{alg:oracle2fair} might not halt. However, such datasets are rare in recommender systems.

\section{Modifications to the GS Algorithm}\label{app:gs}
The original GS algorithm \cite{Wang2022ProvidingSystems} increases individual item fairness within clusters of similar items. The item similarity is determined based on the item embedding. As our experiments and the \fair{} measures do not deal with the additional constraint of item similarity, we consider all items as similar. Therefore, we only have a single cluster of items.

On top of that, we also modify GS to increase computational efficiency. In the original GS algorithm, for each pair of candidate items for replacement $i$ and candidate items to be replaced $i'$, the algorithm finds all users that have $i$ in the original recommendation list. The algorithm then computes the loss in relevance (computed using predicted relevance value) if item $i$ is replaced by $i'$. Until this point, our modified algorithm does the same. The difference is that we save each $i, i', u$, and the loss associated, while the original algorithm only saves the information for the one user $u^*$, whose recommendation list will suffer the least loss when we replace $i$ with $i'$. The original GS then proceeds to make the replacement, update the pool of candidate items for replacement and to be replaced, and go through the entire process again. Initially, we found that with the GS algorithm, around 20\% of the initial recommendations are replaced during the process, meaning that for Amazon-lb, there are at least $437 \times 10 \times 0.2 \geq 800$ iterations of the process (\Cref{tab:data_split}). The number of iterations is much bigger for ML-10M, which has more than three times the  recommendation slots as Amazon-lb, and hence it is very costly to use the GS algorithm as is. 

Our modified GS utilises the saved information earlier. After going through all pairs of $(i,i')$, we sort the saved list from the smallest to the largest loss, and (attempt to) perform the replacement using the first $P$ pairs, where $P$ is 25\% of the number of recommendation slots. During the replacement process, if the item that is supposed to be replaced no longer exists in the user's recommendation list, we simply skip the replacement.

\section{Extended Results}
\label{app:actual-scores}

\begin{figure*}[tbp]
    \centering
    \includegraphics[width=.48\columnwidth]{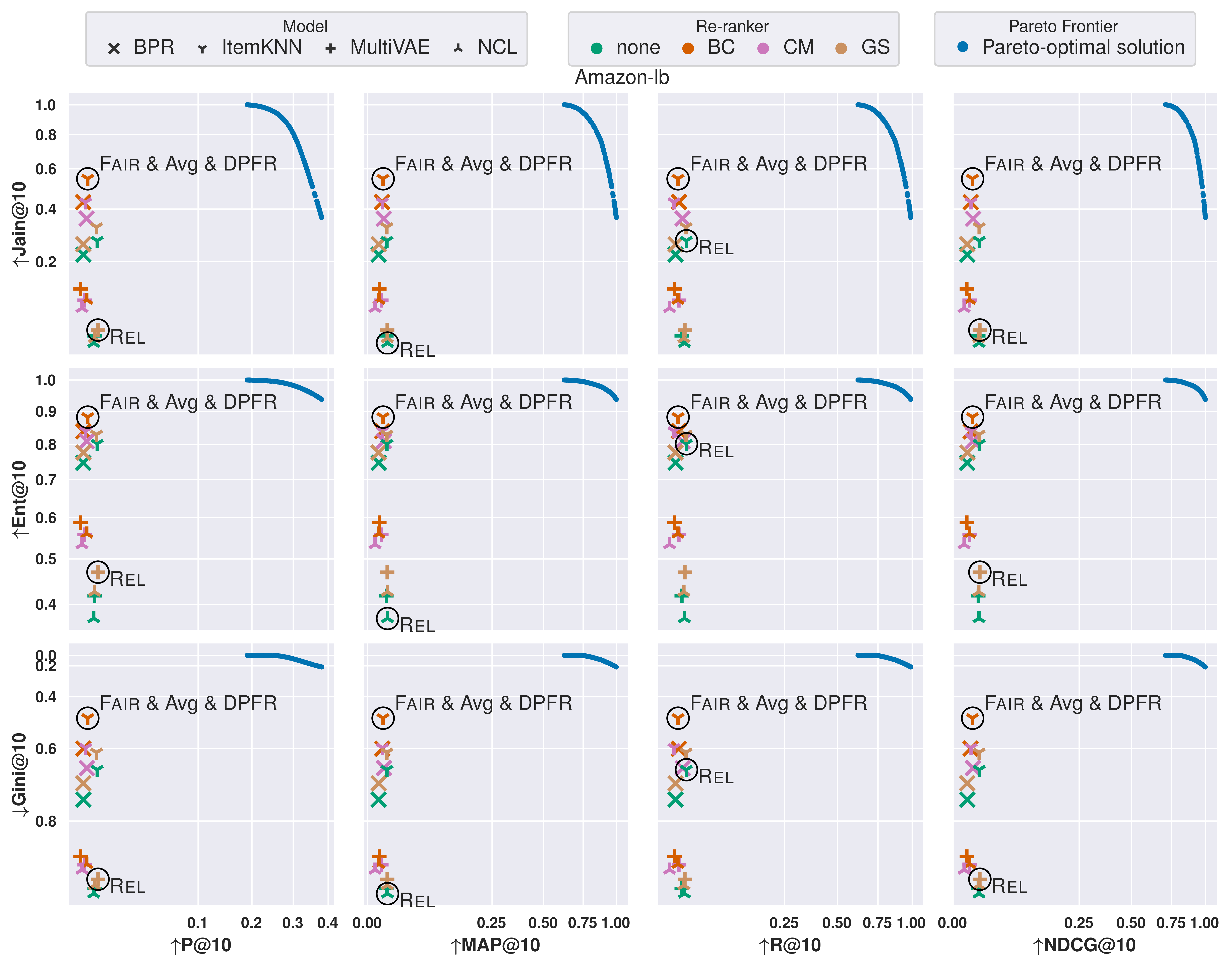}
    \includegraphics[width=.48\columnwidth]{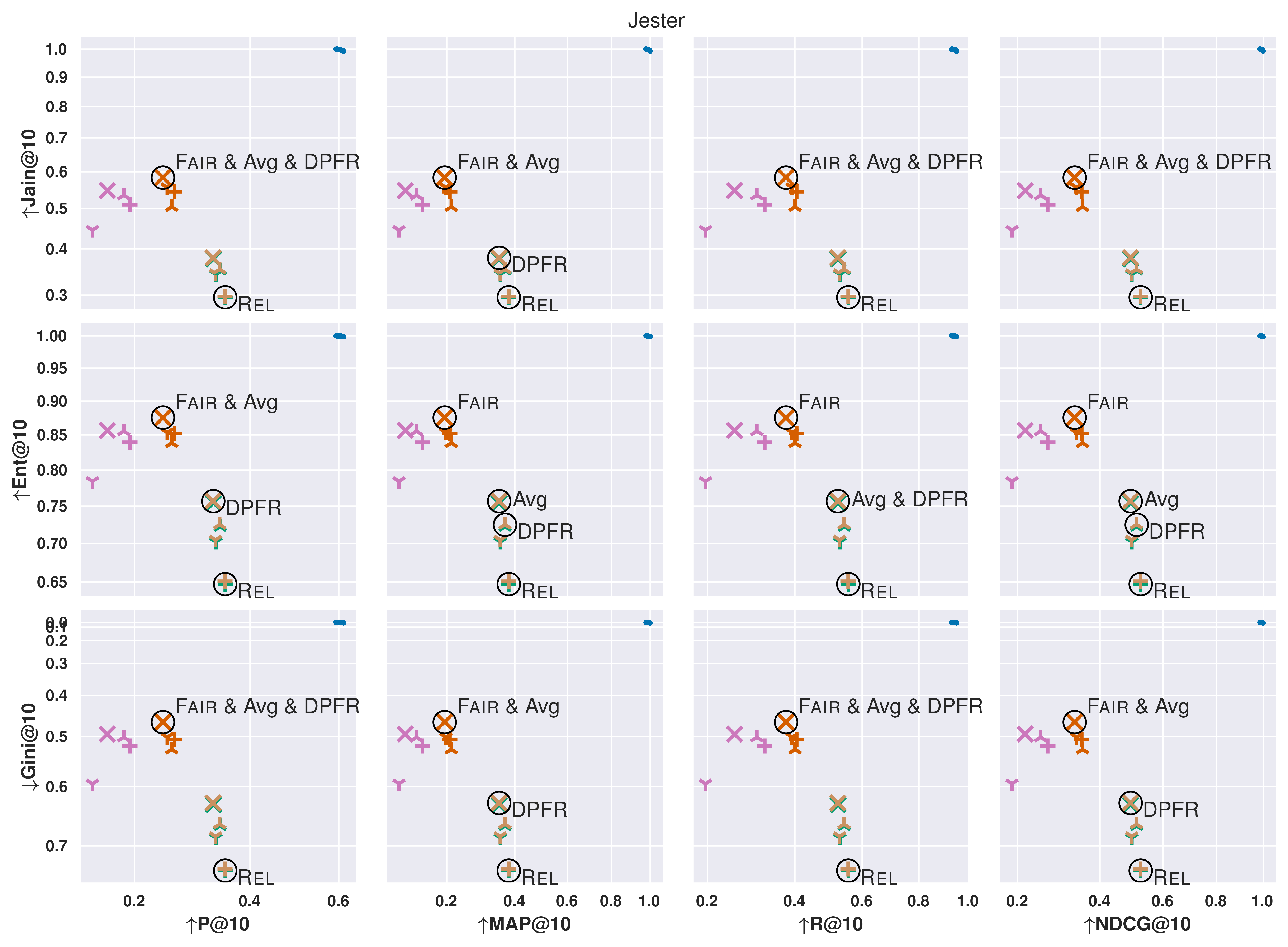}
    \includegraphics[width=.48\columnwidth]{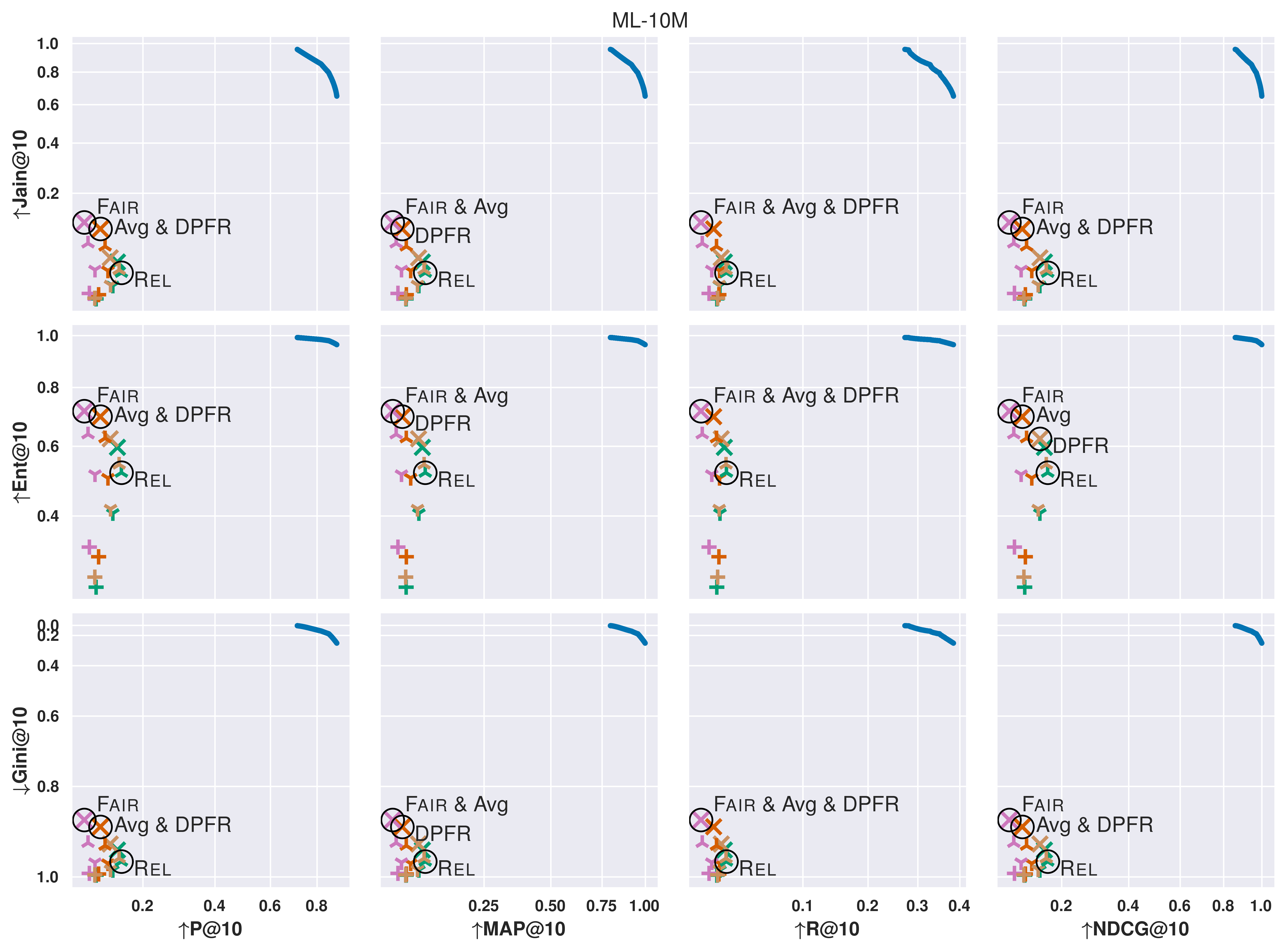}
    \includegraphics[width=.48\columnwidth]{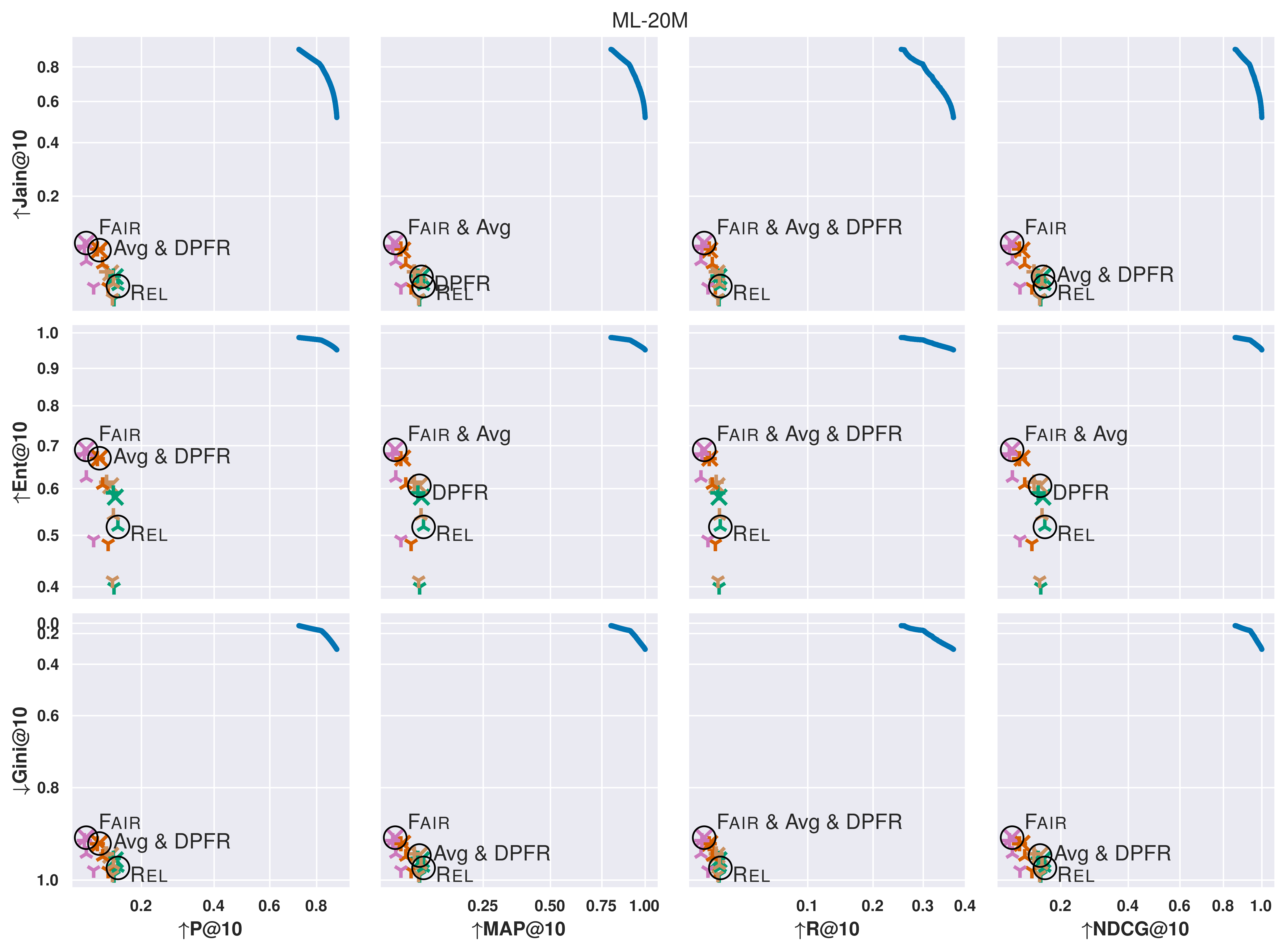}
    
    \caption{Pareto Frontier of fairness and relevance (in blue), together with recommender model scores for Amazon-lb, Jester, and ML-*. \fair{} measures are on the $y$-axis and \rel{} measures are on the $x$-axis. We implement exponential-like scales to enhance the visibility of the model plots. The texts \rel{}, \fair{}, avg, and DPFR denote the best model based on each evaluation approach.}
    \label{fig:app-pairplot}

\end{figure*}

\begin{figure*}
    \centering
    \includegraphics[width=0.48\linewidth]{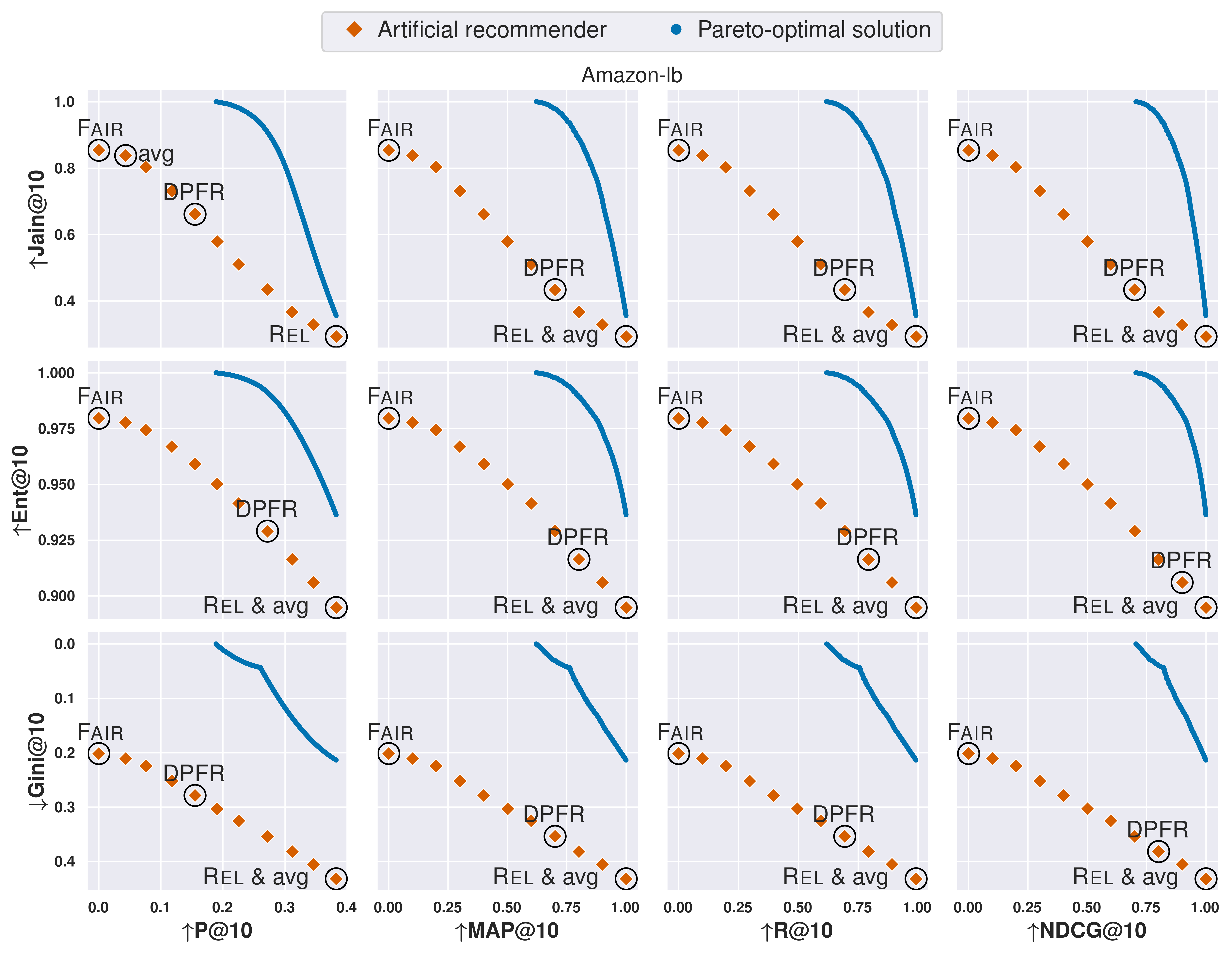}
    \includegraphics[width=0.48\linewidth]{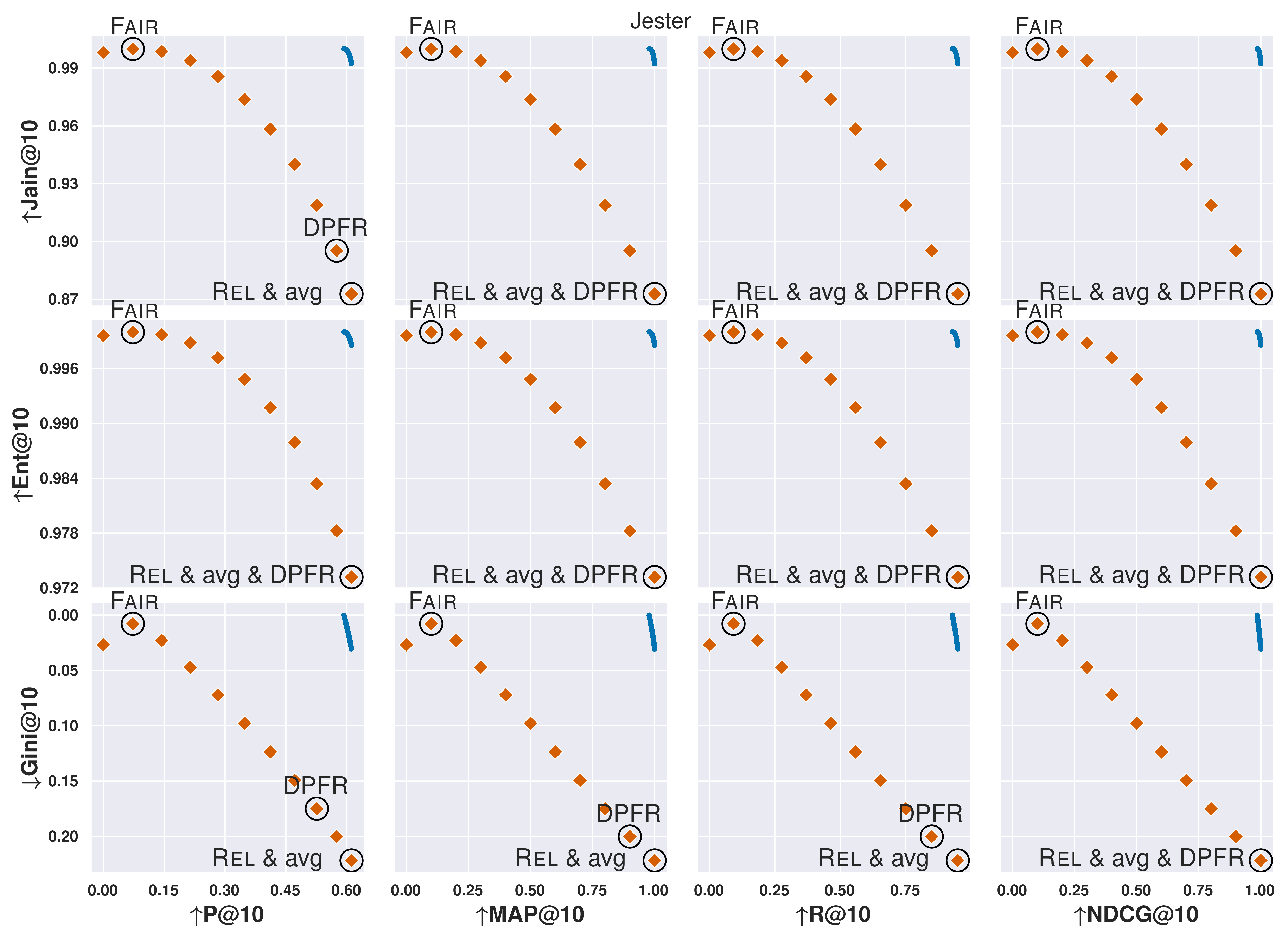}
    \includegraphics[width=0.48\linewidth]{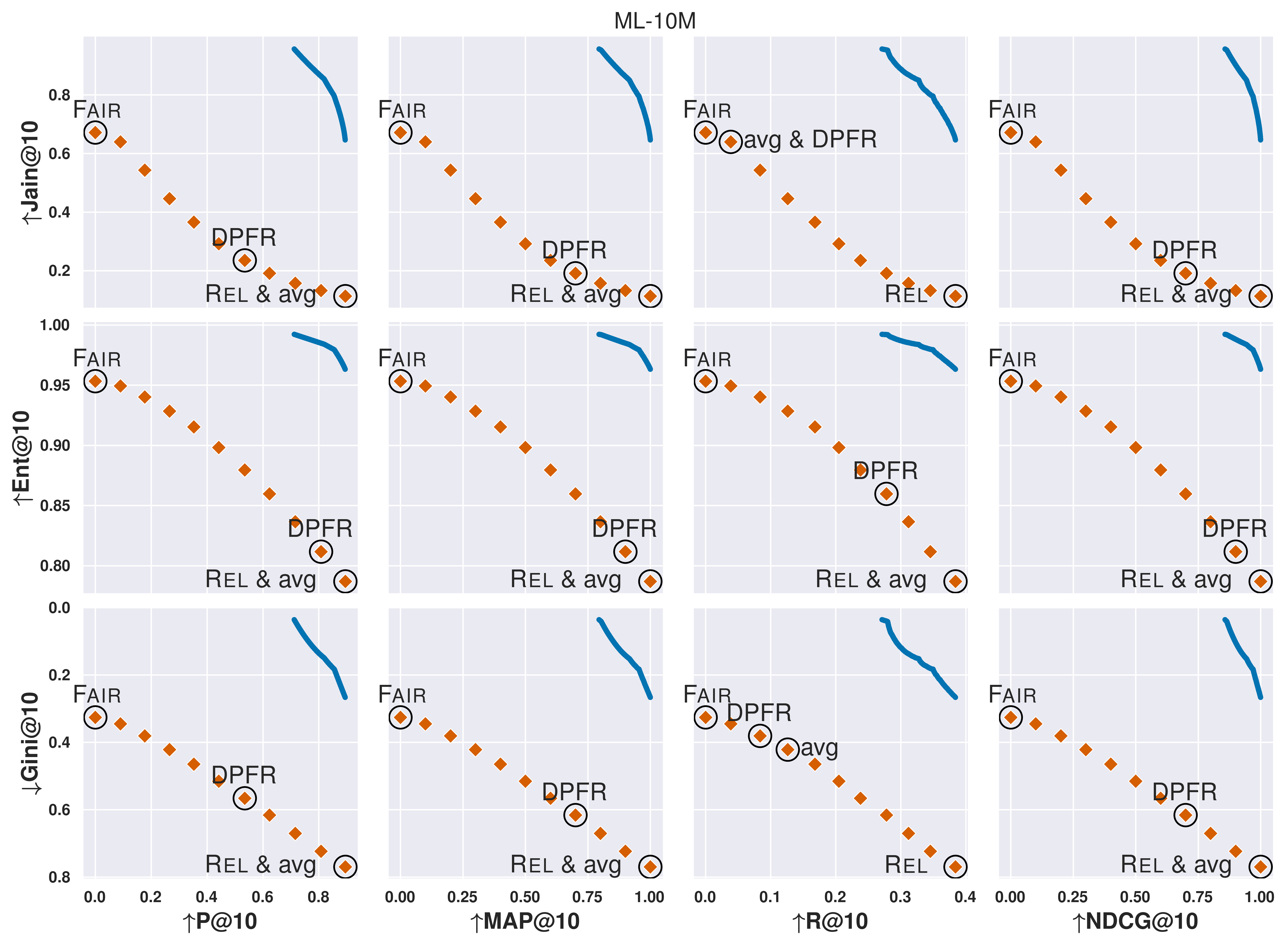}
    \includegraphics[width=0.48\linewidth]{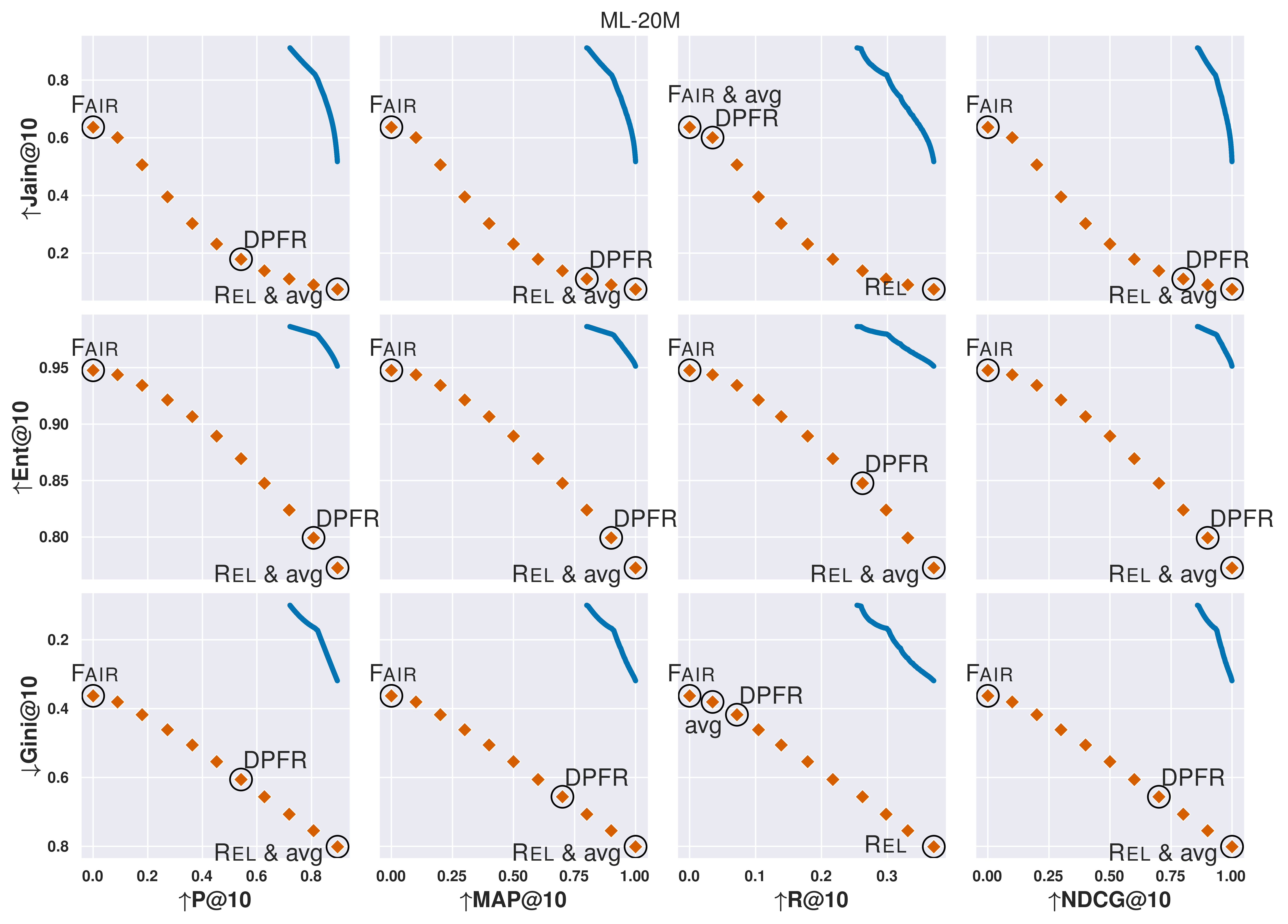}
    \caption{Pareto Frontier of fairness and relevance (in blue) and artificial recommender scores (in orange) for Amazon-lb, Jester, ML-10M, and ML-20M. The best runs based on \rel{}, \fair{}, avg, or DPFR are circled.}
    \label{fig:artificial_app}
\end{figure*}

In \Cref{tab:app-gradient}, we present the gradient values of the PF, used in determining which pair of measures are suitable for DPFR. 
In \Cref{fig:app-pairplot}, we show the Pareto Frontier (PF) of fairness and relevance together with scores from recommenders trained on real-world data in \Cref{tab:base-rerank-all-1,tab:base-rerank-all-2} for Amazon-lb, Jester, and ML-*.  
We show the PF and scores from artificial recommendation in \Cref{fig:artificial_app}. 

\begin{figure}
    \centering   
    \includegraphics[width=\columnwidth]{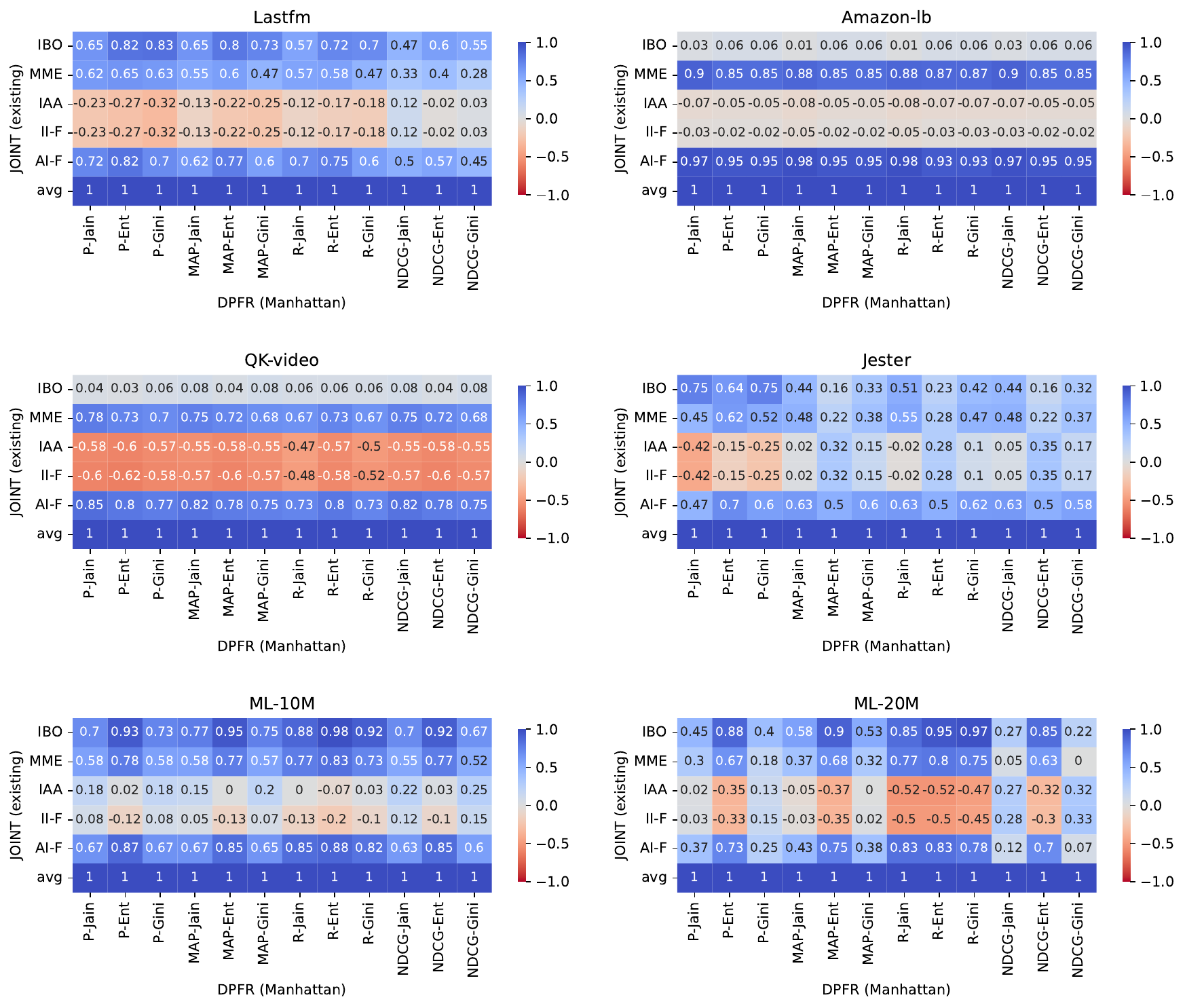}

    \caption{Kendall's $\tau$ correlation heatmap between the rank ordering of RSs based on existing \joint{} evaluation approaches (including the average of \rel{} and \fair{} scores, avg), and DPFR (computed with Manhattan distance).}
    \label{fig:corr_manhattan}
\end{figure}

In \Cref{tab:dpfr-scores,tab:dpfr-scores2}, we present the DPFR scores (computed with Euclidean distance) for all datasets. Similarly, \Cref{tab:dpfr-scores-manhattan,tab:dpfr-scores2-manhattan} present the DPFR scores computed with Manhattan distance for all datasets. \Cref{fig:corr_manhattan} presents the Kendall's $\tau$ correlation between the rank ordering of RSs between DPFR (Manhattan) and joint evaluation approaches.

\begin{figure}
    \centering

    \includegraphics[width=0.85\linewidth]{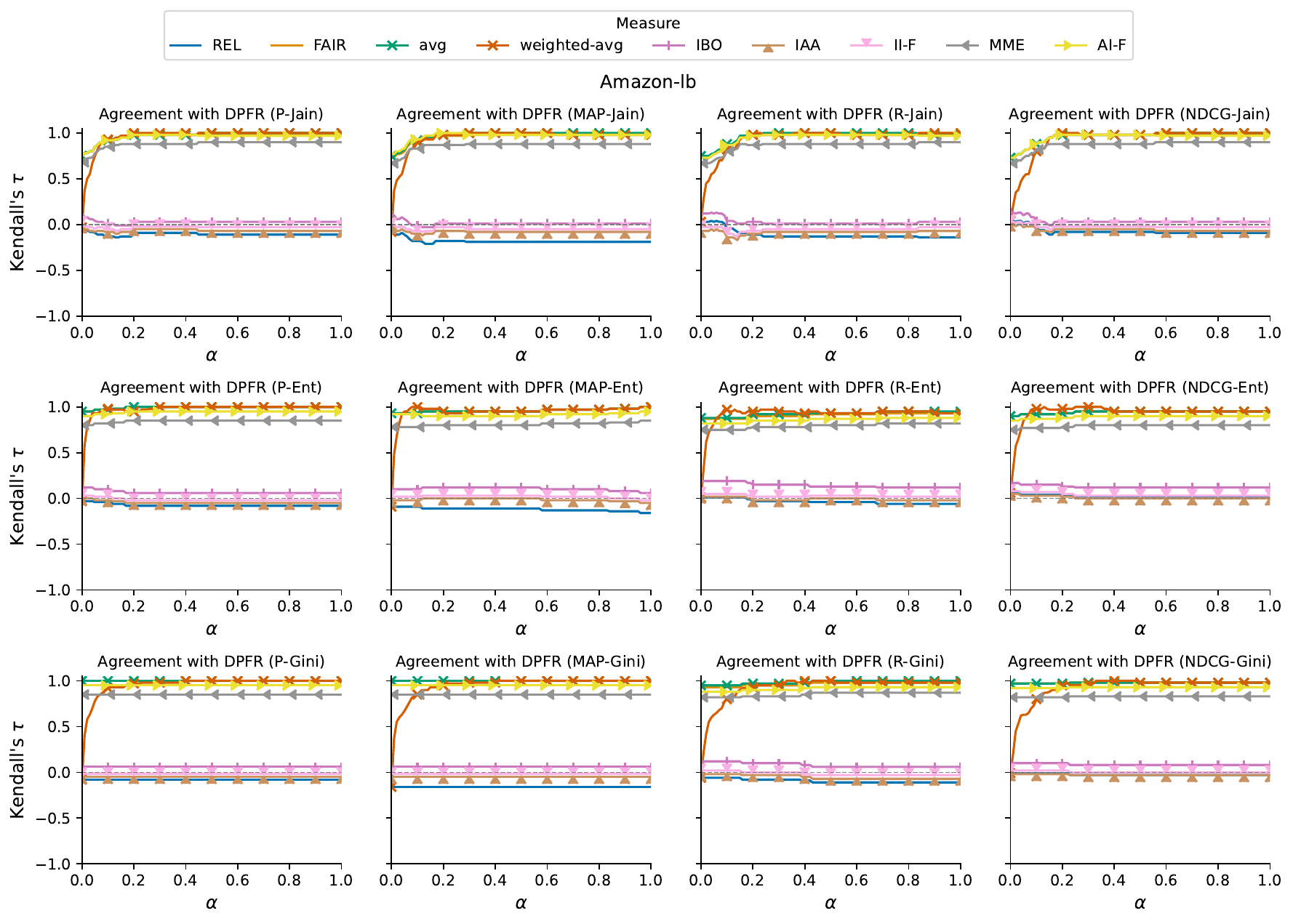}
    \includegraphics[width=0.85\linewidth]{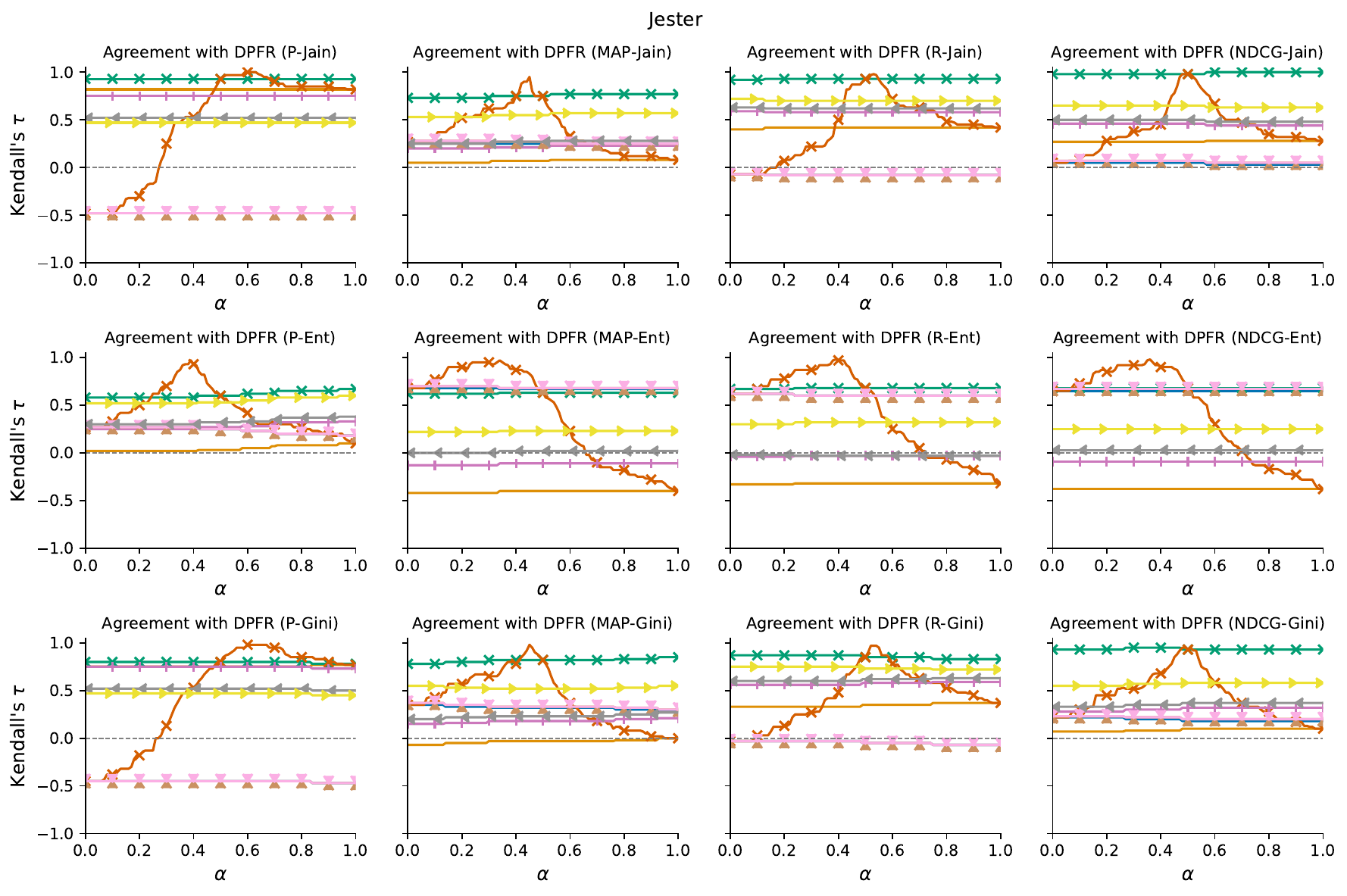}

    \caption{Agreement (Kendall's $\tau$) between DPFR and \rel{}, \fair{}, and \joint{} evaluation approaches (including avg and weighted-avg) when varying $\alpha$ for Amazon-lb and Jester. Note that for \rel{}, \fair{}, and avg, we only show the agreement of the measures that are part of the DPFR \rel{}-\fair{} measure pair. For example, in DPFR (P-Jain), \rel{} is P and \fair{} is Jain. Dashed grey lines denote no agreement ($\tau=0$).
    }
    \label{fig:corr_alpha_app1}
\end{figure}

\begin{figure}
    \centering

    \includegraphics[width=0.85\linewidth]{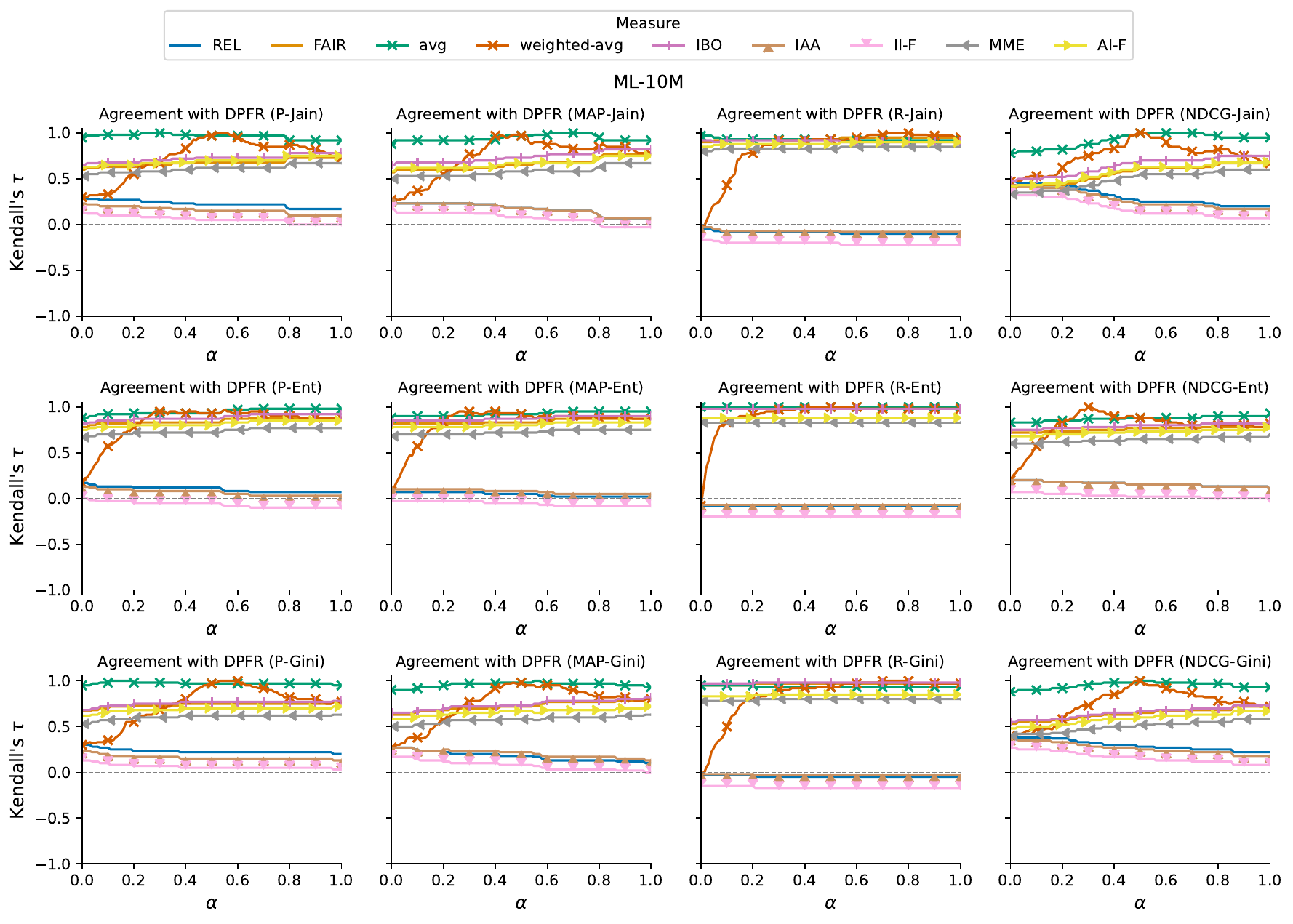}
    \includegraphics[width=0.85\linewidth]{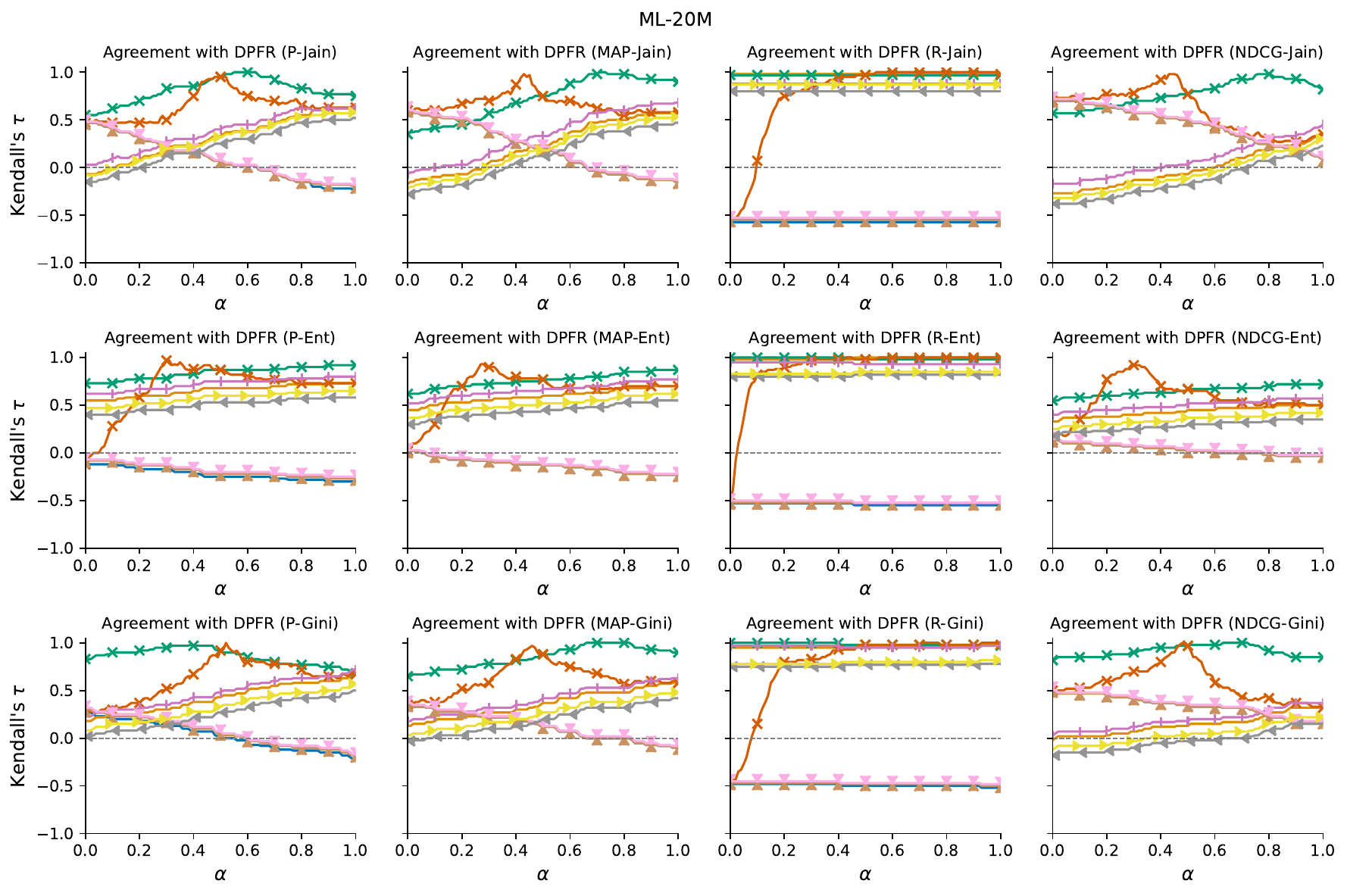}

    \caption{Agreement (Kendall's $\tau$) between DPFR and \rel{}, \fair{}, and \joint{} evaluation approaches (including avg and weighted-avg) when varying $\alpha$ for ML-10M and ML-20M. Note that for \rel{}, \fair{}, and avg, we only show the agreement of the measures that are part of the DPFR \rel{}-\fair{} measure pair. For example, in DPFR (P-Jain), \rel{} is P and \fair{} is Jain. Dashed grey lines denote no agreement ($\tau=0$).
    }
    \label{fig:corr_alpha_app2}
\end{figure}

\begin{figure}
    \centering
    \includegraphics[width=.55\linewidth]{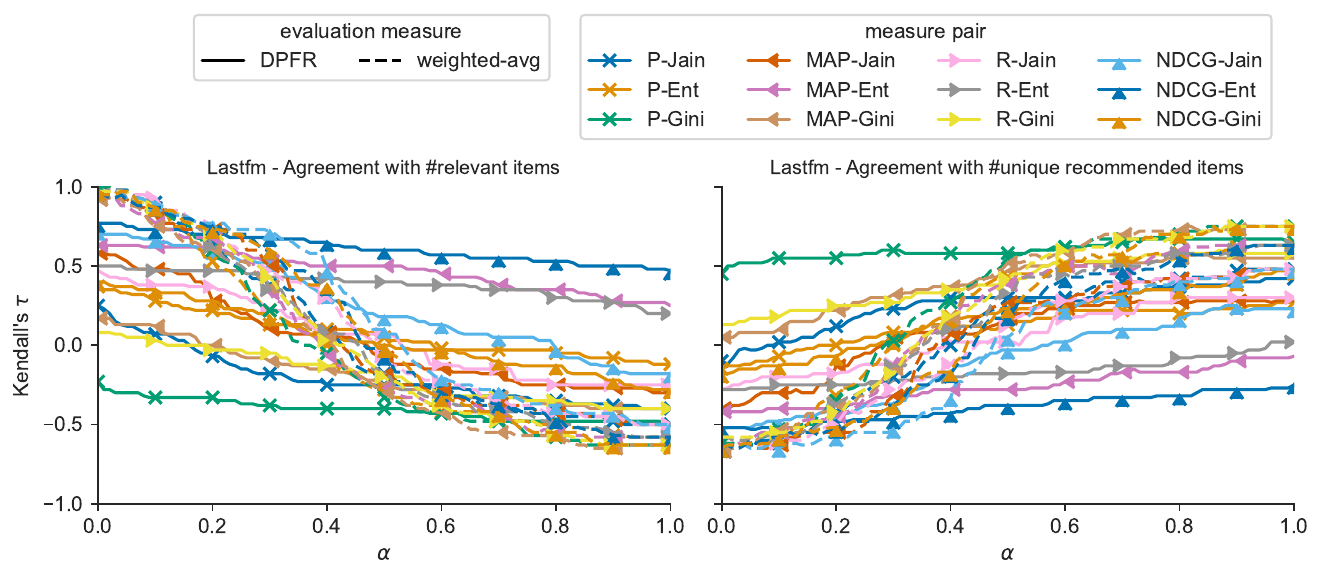}
     \includegraphics[width=.55\linewidth]{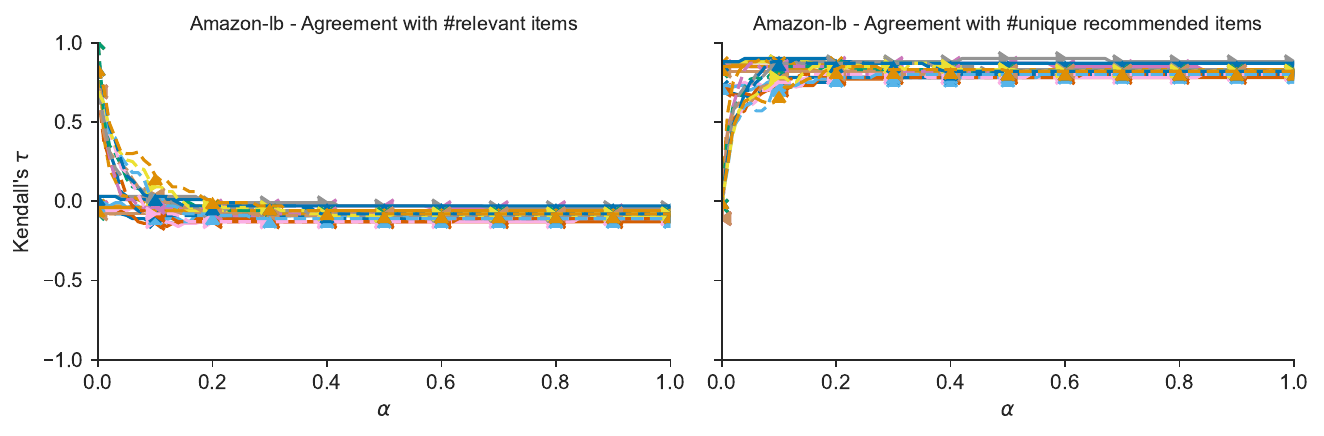}
     \includegraphics[width=.55\linewidth]{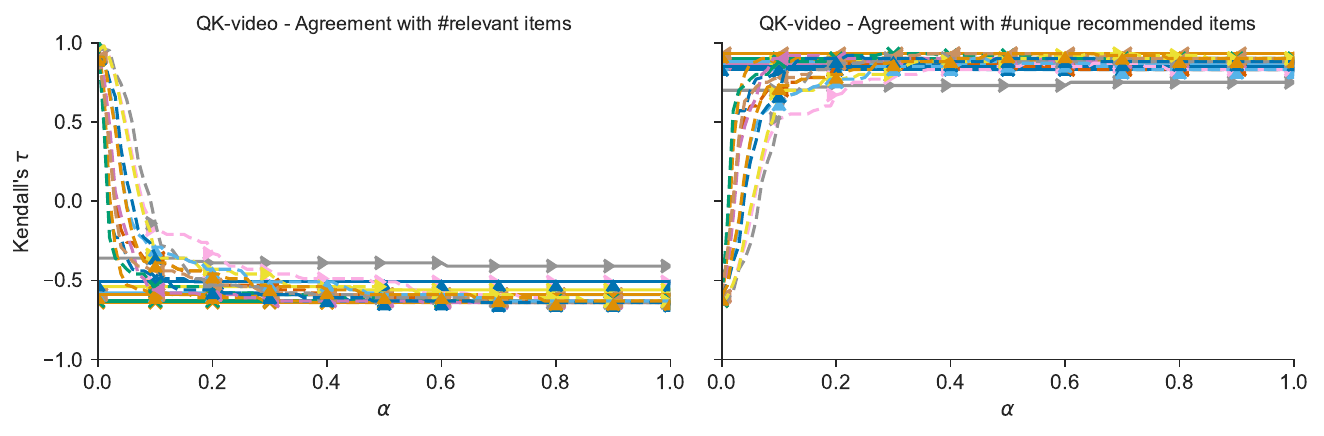}
     \includegraphics[width=.55\linewidth]{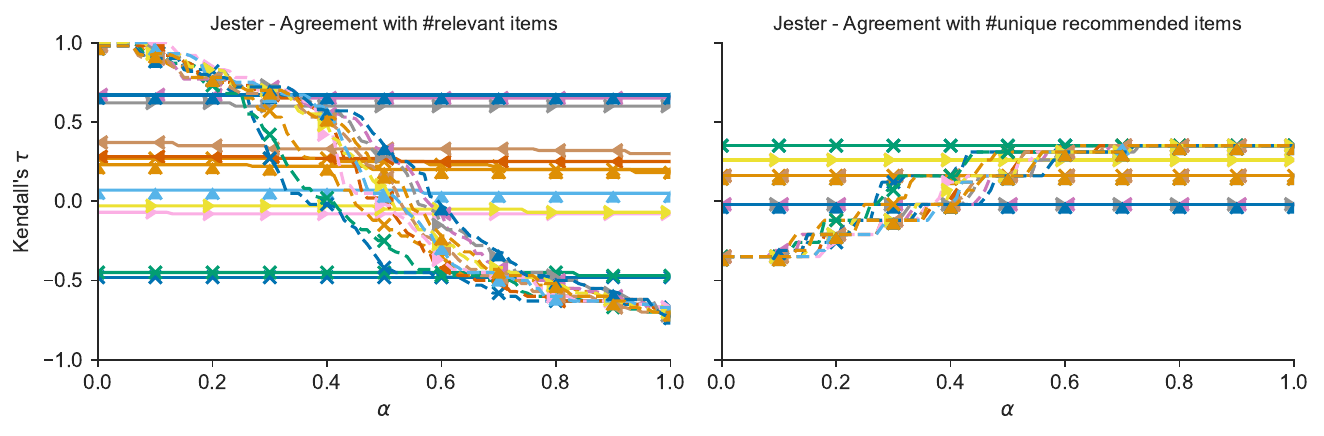}
     \includegraphics[width=.55\linewidth]{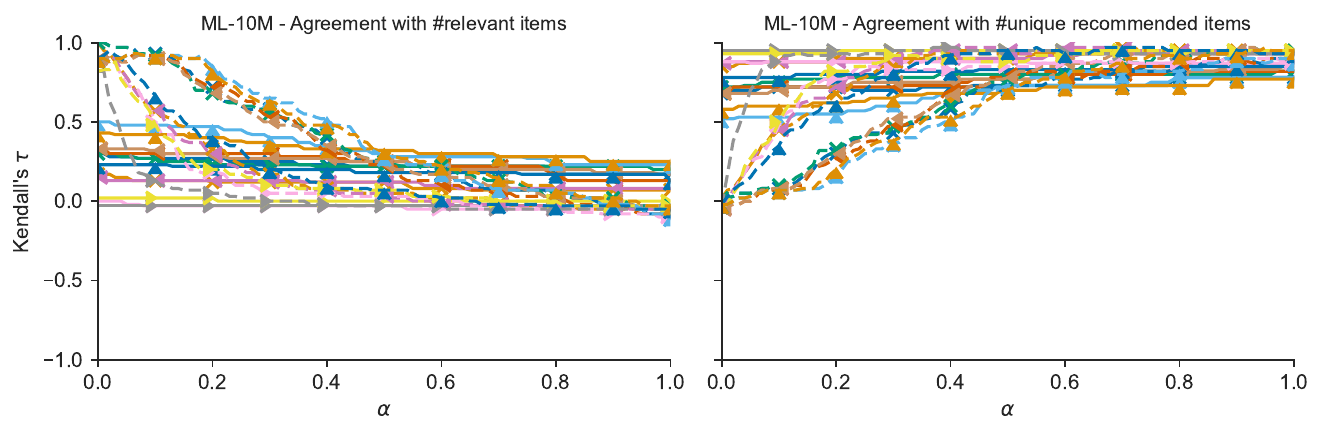}
     \includegraphics[width=.55\linewidth]{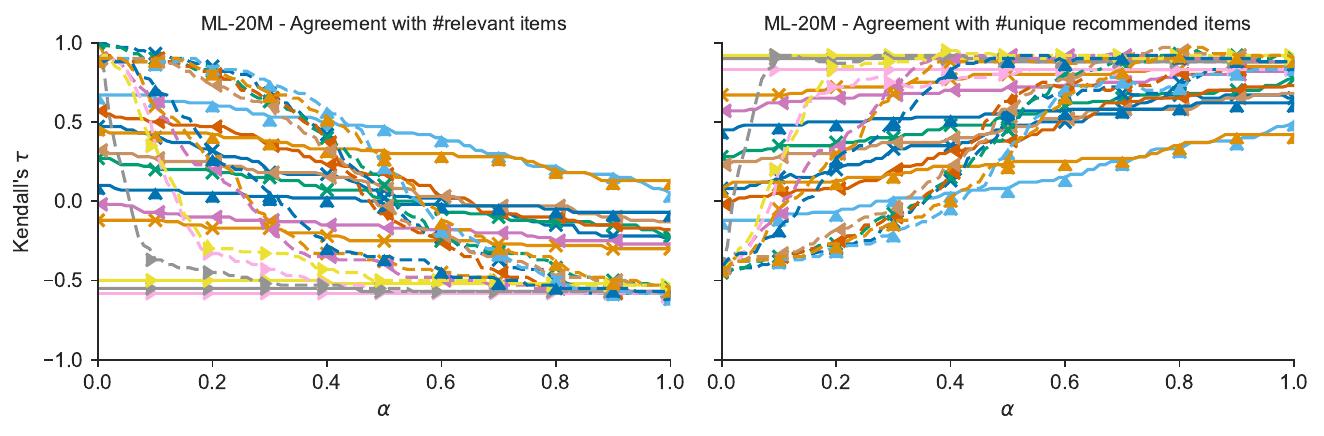}

    \caption{Agreement  (Kendall's $\tau$) of model ranking given by a simple measure of recommendation effectiveness (i.e., \#relevant items in the recommendation) and of recommendation fairness (\#unique recommended items) with the model ranking given by DPFR and weighted average (weighted-avg), for all datasets. We show all 12 \rel{}-\fair{} pairs that are used to compute DPFR and weighted-avg.
    }
    \label{fig:simple_measure_alpha_app}
\end{figure}

We show the agreement between DPFR and existing single- and multi-aspect evaluation approaches across various $\alpha$'s in \Cref{fig:corr_alpha_app1} (for Amazon-lb and Jester) and in \Cref{fig:corr_alpha_app2} (for ML-*).
The agreement of model ranking given by DPFR and weighted average to the ones given by simple measures across various $\alpha$'s is presented in \Cref{fig:simple_measure_alpha_app}.
In \Cref{tab:corr_est} we present the Kendall's $\tau$ correlation scores of the DPFR from estimated PF and the PF.

\begin{table*}[tbp]
\centering
\caption{The gradient values of the PF, based on the extreme points (starting and ending points). We consider a gradient to be `good' if it is not zero or undefined (-).}\label{tab:app-gradient} 
\begin{tabular}{lrrrrrrrl}
\toprule
{} & Lastfm & Amazon-lb & QK-video & Jester &  ML-10M &  ML-20M &  \# good &    conclusion \\
\midrule
HR-Ent    &   -97.57 &      -1.86 &     -0.31 &      - &  -14.74 &   -6.95 &        5 &  inconsistent \\
HR-FSat   & -1439.17 &     -19.92 &      0.00 &      - &  -30.48 &  -18.97 &        4 &  inconsistent \\
HR-Gini   &   561.63 &       6.23 &      3.71 &      - &  117.19 &   43.44 &        5 &  inconsistent \\
HR-Jain   &  -979.86 &     -18.77 &     -5.80 &      - & -157.73 &  -78.22 &        5 &  inconsistent \\
HR-QF     &     0.00 &       0.00 &      0.00 &      - &  -30.48 &  -18.97 &        2 &  inconsistent \\
\midrule
MAP-Ent   &    -0.17 &      -0.17 &     -0.03 &  -0.07 &   -0.14 &   -0.18 &        6 &   always good \\
MAP-FSat  &    -2.46 &      -1.81 &      0.00 & -44.47 &   -0.29 &   -0.48 &        5 &  inconsistent \\
MAP-Gini  &     0.96 &       0.56 &      0.34 &   1.42 &    1.12 &    1.10 &        6 &   always good \\
MAP-Jain  &    -1.68 &      -1.70 &     -0.54 &  -0.37 &   -1.51 &   -1.98 &        6 &   always good \\
MAP-QF    &     0.00 &       0.00 &      0.00 &    0.0 &   -0.29 &   -0.48 &        2 &  inconsistent \\
\midrule
MRR-Ent   &   -97.57 &      -1.86 &     -0.31 &      - &  -14.74 &   -6.95 &        5 &  inconsistent \\
MRR-FSat  & -1439.17 &     -19.92 &      0.00 &      - &  -30.48 &  -18.97 &        4 &  inconsistent \\
MRR-Gini  &   561.63 &       6.23 &      3.71 &      - &  117.19 &   43.44 &        5 &  inconsistent \\
MRR-Jain  &  -979.86 &     -18.77 &     -5.80 &      - & -157.73 &  -78.22 &        5 &  inconsistent \\
MRR-QF    &     0.00 &       0.00 &      0.00 &      - &  -30.48 &  -18.97 &        2 &  inconsistent \\
\midrule
NDCG-Ent  &    -0.24 &      -0.22 &     -0.04 &   -0.1 &   -0.20 &   -0.25 &        6 &   always good \\
NDCG-FSat &    -3.50 &      -2.32 &      0.00 & -68.56 &   -0.42 &   -0.68 &        5 &  inconsistent \\
NDCG-Gini &     1.37 &       0.73 &      0.47 &    2.2 &    1.62 &    1.55 &        6 &   always good \\
NDCG-Jain &    -2.38 &      -2.19 &     -0.73 &  -0.57 &   -2.18 &   -2.79 &        6 &   always good \\
NDCG-QF   &     0.00 &       0.00 &      0.00 &    0.0 &   -0.42 &   -0.68 &        2 &  inconsistent \\
\midrule
P-Ent     &    -0.20 &      -0.33 &     -0.07 &  -0.08 &   -0.16 &   -0.20 &        6 &   always good \\
P-FSat    &    -2.95 &      -3.53 &      0.00 & -51.41 &   -0.33 &   -0.55 &        5 &  inconsistent \\
P-Gini    &     1.15 &       1.10 &      0.89 &   1.65 &    1.26 &    1.26 &        6 &   always good \\
P-Jain    &    -2.01 &      -3.33 &     -1.40 &  -0.43 &   -1.70 &   -2.27 &        6 &   always good \\
P-QF      &     0.00 &       0.00 &      0.00 &    0.0 &   -0.33 &   -0.55 &        2 &  inconsistent \\
\midrule
R-Ent     &    -0.17 &      -0.17 &     -0.03 &  -0.07 &   -0.26 &   -0.30 &        6 &   always good \\
R-FSat    &    -2.57 &      -1.83 &      0.00 & -48.04 &   -0.53 &   -0.82 &        5 &  inconsistent \\
R-Gini    &     1.00 &       0.57 &      0.34 &   1.54 &    2.04 &    1.88 &        6 &   always good \\
R-Jain    &    -1.75 &      -1.73 &     -0.54 &   -0.4 &   -2.75 &   -3.39 &        6 &   always good \\
R-QF      &     0.00 &       0.00 &      0.00 &    0.0 &   -0.53 &   -0.82 &        2 &  inconsistent \\
\bottomrule
\end{tabular}
\end{table*}

\begin{table*}[tbp]
\caption{$\downarrow$DPFR scores (computed with Euclidean distance) at $k=10$ of the recommender models for Lastfm, Amazon-lb, and QK-video, without and with re-ranking the top $k'=25$ items using Borda Count (BC), COMBMNZ (CM), and Greedy Substitution (GS). The best score per measure pair is in bold.}
\label{tab:dpfr-scores}
\resizebox{0.95\textwidth}{!}{
\begin{tabular}{ll*{4}{r}|*{4}{r}|*{4}{r}|*{4}{r}}
\toprule
 &  model & \multicolumn{4}{c|}{ItemKNN} & \multicolumn{4}{c|}{BPR} & \multicolumn{4}{c|}{MultiVAE} & \multicolumn{4}{c}{NCL} \\ 
\midrule
 &  reranking & - & BC & CM & GS & - & BC & CM & GS & - & BC & CM & GS & - & BC & CM & GS \\
\midrule
\multirow[c]{12}{*}{\rotatebox[origin=c]{90}{Lastfm}} & P-Jain & 0.853 & 0.819 & 0.861 & 0.853 & 0.837 & 0.784 & 0.824 & 0.834 & 0.805 & \bfseries 0.728 & 0.776 & 0.799 & 0.813 & 0.735 & 0.773 & 0.809 \\
 & P-Ent & 0.585 & 0.524 & 0.571 & 0.567 & 0.567 & 0.525 & 0.566 & 0.551 & 0.510 & 0.501 & 0.550 & 0.505 & 0.524 & \bfseries 0.497 & 0.544 & 0.513 \\
 & P-Gini & 0.869 & 0.802 & 0.821 & 0.850 & 0.871 & 0.820 & 0.841 & 0.856 & 0.811 & \bfseries 0.737 & 0.759 & 0.789 & 0.835 & 0.756 & 0.775 & 0.813 \\
 & MAP-Jain & 1.044 & 1.042 & 1.072 & 1.042 & 1.029 & 1.015 & 1.040 & 1.026 & 1.005 & \bfseries 0.974 & 1.004 & 0.998 & 1.008 & 0.979 & 1.003 & 1.003 \\
 & MAP-Ent & 0.811 & 0.802 & 0.830 & 0.797 & 0.797 & 0.803 & 0.824 & 0.784 & 0.759 & 0.791 & 0.815 & \bfseries 0.753 & 0.764 & 0.787 & 0.813 & 0.755 \\
 & MAP-Gini & 1.038 & 1.009 & 1.022 & 1.021 & 1.039 & 1.025 & 1.036 & 1.025 & 0.991 & \bfseries 0.961 & 0.972 & 0.970 & 1.007 & 0.975 & 0.986 & 0.987 \\
 & R-Jain & 0.968 & 0.947 & 1.004 & 0.969 & 0.952 & 0.917 & 0.970 & 0.954 & 0.923 & \bfseries 0.875 & 0.936 & 0.923 & 0.927 & \bfseries 0.875 & 0.930 & 0.928 \\
 & R-Ent & 0.720 & 0.683 & 0.747 & 0.708 & 0.703 & 0.686 & 0.741 & 0.696 & \bfseries 0.656 & 0.675 & 0.738 & 0.660 & 0.664 & 0.665 & 0.729 & 0.661 \\
 & R-Gini & 0.968 & 0.918 & 0.955 & 0.953 & 0.969 & 0.935 & 0.971 & 0.959 & 0.914 & \bfseries 0.868 & 0.907 & 0.900 & 0.932 & 0.878 & 0.918 & 0.917 \\
 & NDCG-Jain & 0.989 & 0.995 & 1.046 & 0.989 & 0.973 & 0.968 & 1.013 & 0.972 & 0.948 & 0.933 & 0.985 & 0.945 & 0.949 & \bfseries 0.932 & 0.978 & 0.947 \\
 & NDCG-Ent & 0.760 & 0.758 & 0.813 & 0.747 & 0.743 & 0.762 & 0.805 & 0.733 & 0.703 & 0.756 & 0.807 & 0.702 & 0.706 & 0.746 & 0.797 & \bfseries 0.700 \\
 & NDCG-Gini & 1.000 & 0.976 & 1.009 & 0.984 & 1.000 & 0.994 & 1.023 & 0.987 & 0.949 & 0.934 & 0.966 & \bfseries 0.932 & 0.964 & 0.943 & 0.974 & 0.947 \\
\cline{1-18}
\multirow[c]{12}{*}{\rotatebox[origin=c]{90}{Amazon-lb}} & P-Jain & 0.529 & \bfseries 0.352 & 0.415 & 0.489 & 0.571 & 0.415 & 0.464 & 0.542 & 0.733 & 0.656 & 0.679 & 0.726 & 0.742 & 0.678 & 0.694 & 0.736 \\
 & P-Ent & 0.340 & \bfseries 0.307 & 0.324 & 0.327 & 0.376 & 0.325 & 0.339 & 0.358 & 0.636 & 0.493 & 0.517 & 0.590 & 0.678 & 0.515 & 0.537 & 0.629 \\
 & P-Gini & 0.633 & \bfseries 0.486 & 0.575 & 0.587 & 0.708 & 0.578 & 0.630 & 0.669 & 0.896 & 0.834 & 0.850 & 0.878 & 0.906 & 0.849 & 0.860 & 0.890 \\
 & MAP-Jain & 0.998 & \bfseries 0.904 & 0.936 & 0.974 & 1.025 & 0.936 & 0.962 & 1.007 & 1.129 & 1.078 & 1.093 & 1.123 & 1.134 & 1.093 & 1.105 & 1.130 \\
 & MAP-Ent & 0.830 & \bfseries 0.818 & 0.825 & 0.825 & 0.848 & 0.825 & 0.830 & 0.840 & 0.990 & 0.906 & 0.919 & 0.961 & 1.017 & 0.919 & 0.933 & 0.985 \\
 & MAP-Gini & 0.993 & \bfseries 0.906 & 0.958 & 0.963 & 1.045 & 0.959 & 0.991 & 1.018 & 1.180 & 1.135 & 1.147 & 1.166 & 1.187 & 1.146 & 1.156 & 1.175 \\
 & R-Jain & 0.984 & \bfseries 0.893 & 0.927 & 0.960 & 1.015 & 0.924 & 0.948 & 0.997 & 1.119 & 1.069 & 1.082 & 1.112 & 1.123 & 1.082 & 1.096 & 1.119 \\
 & R-Ent & 0.817 & \bfseries 0.809 & 0.818 & 0.811 & 0.839 & 0.815 & 0.818 & 0.831 & 0.981 & 0.898 & 0.909 & 0.950 & 1.007 & 0.909 & 0.925 & 0.975 \\
 & R-Gini & 0.981 & \bfseries 0.897 & 0.951 & 0.951 & 1.037 & 0.949 & 0.980 & 1.010 & 1.172 & 1.127 & 1.138 & 1.157 & 1.178 & 1.137 & 1.149 & 1.166 \\
 & NDCG-Jain & 1.022 & \bfseries 0.937 & 0.966 & 1.000 & 1.051 & 0.967 & 0.990 & 1.034 & 1.148 & 1.102 & 1.116 & 1.142 & 1.153 & 1.116 & 1.127 & 1.149 \\
 & NDCG-Ent & 0.868 & \bfseries 0.859 & 0.866 & 0.863 & 0.889 & 0.867 & 0.871 & 0.881 & 1.023 & 0.945 & 0.957 & 0.994 & 1.049 & 0.956 & 0.970 & 1.018 \\
 & NDCG-Gini & 1.028 & \bfseries 0.947 & 0.997 & 1.000 & 1.083 & 0.999 & 1.029 & 1.056 & 1.211 & 1.170 & 1.181 & 1.197 & 1.218 & 1.180 & 1.190 & 1.206 \\
\cline{1-18}
\multirow[c]{12}{*}{\rotatebox[origin=c]{90}{QK-video}} & P-Jain & 0.563 & \bfseries 0.297 & 0.468 & 0.487 & 0.941 & 0.700 & 0.658 & 0.922 & 1.007 & 0.983 & 0.989 & 1.005 & 0.999 & 0.945 & 0.950 & 0.996 \\
 & P-Ent & 0.245 & \bfseries 0.236 & 0.243 & 0.241 & 0.335 & 0.258 & 0.254 & 0.310 & 0.623 & 0.499 & 0.510 & 0.588 & 0.542 & 0.403 & 0.400 & 0.504 \\
 & P-Gini & 0.521 & \bfseries 0.327 & 0.495 & 0.456 & 0.833 & 0.649 & 0.611 & 0.789 & 1.002 & 0.978 & 0.981 & 0.996 & 0.985 & 0.931 & 0.925 & 0.972 \\
 & MAP-Jain & 1.106 & \bfseries 0.996 & 1.061 & 1.070 & 1.332 & 1.180 & 1.157 & 1.318 & 1.379 & 1.365 & 1.371 & 1.378 & 1.371 & 1.338 & 1.343 & 1.369 \\
 & MAP-Ent & 0.979 & \bfseries 0.977 & 0.979 & 0.978 & 0.995 & 0.980 & 0.980 & 0.987 & 1.125 & 1.066 & 1.073 & 1.106 & 1.079 & 1.024 & 1.025 & 1.061 \\
 & MAP-Gini & 1.082 & \bfseries 1.003 & 1.070 & 1.052 & 1.254 & 1.147 & 1.127 & 1.225 & 1.372 & 1.358 & 1.362 & 1.367 & 1.357 & 1.324 & 1.322 & 1.347 \\
 & R-Jain & 1.097 & \bfseries 0.982 & 1.047 & 1.060 & 1.312 & 1.162 & 1.145 & 1.301 & 1.355 & 1.345 & 1.357 & 1.356 & 1.343 & 1.314 & 1.326 & 1.342 \\
 & R-Ent & 0.969 & 0.962 & 0.964 & 0.968 & 0.969 & \bfseries 0.959 & 0.966 & 0.964 & 1.096 & 1.040 & 1.056 & 1.079 & 1.043 & 0.993 & 1.002 & 1.027 \\
 & R-Gini & 1.073 & \bfseries 0.989 & 1.056 & 1.043 & 1.233 & 1.129 & 1.114 & 1.207 & 1.348 & 1.338 & 1.348 & 1.346 & 1.329 & 1.300 & 1.304 & 1.321 \\
 & NDCG-Jain & 1.106 & \bfseries 0.996 & 1.060 & 1.070 & 1.326 & 1.178 & 1.156 & 1.314 & 1.373 & 1.361 & 1.369 & 1.372 & 1.362 & 1.332 & 1.340 & 1.361 \\
 & NDCG-Ent & 0.980 & \bfseries 0.976 & 0.978 & 0.979 & 0.988 & 0.977 & 0.980 & 0.982 & 1.118 & 1.061 & 1.071 & 1.099 & 1.069 & 1.017 & 1.021 & 1.051 \\
 & NDCG-Gini & 1.083 & \bfseries 1.002 & 1.069 & 1.053 & 1.248 & 1.145 & 1.127 & 1.221 & 1.366 & 1.354 & 1.360 & 1.362 & 1.349 & 1.319 & 1.318 & 1.340 \\
\bottomrule

\end{tabular}}
\end{table*}

\begin{table*}[tbp]
\caption{$\downarrow$DPFR scores (computed with Euclidean distance) at $k=10$ of the recommender models for Jester and ML-*, without and with re-ranking the top $k'=25$ items using Borda Count (BC), COMBMNZ (CM), and Greedy Substitution (GS). The best score per measure pair is in bold.}
\label{tab:dpfr-scores2}
\resizebox{0.95\textwidth}{!}{
\begin{tabular}{ll*{4}{r}|*{4}{r}|*{4}{r}|*{4}{r}}
\toprule
 &  model & \multicolumn{4}{c|}{ItemKNN} & \multicolumn{4}{c|}{BPR} & \multicolumn{4}{c|}{MultiVAE} & \multicolumn{4}{c}{NCL} \\ 
\midrule
 &  reranking & - & BC & CM & GS & - & BC & CM & GS & - & BC & CM & GS & - & BC & CM & GS \\
\midrule
\multirow[c]{12}{*}{\rotatebox[origin=c]{90}{Jester}} & P-Jain & 0.709 & 0.566 & 0.719 & 0.707 & 0.679 & \bfseries 0.550 & 0.631 & 0.677 & 0.747 & 0.569 & 0.638 & 0.745 & 0.698 & 0.604 & 0.625 & 0.696 \\
 & P-Ent & 0.401 & 0.382 & 0.507 & 0.400 & 0.367 & 0.381 & 0.462 & \bfseries 0.366 & 0.433 & 0.372 & 0.440 & 0.430 & 0.381 & 0.381 & 0.442 & 0.380 \\
 & P-Gini & 0.724 & 0.602 & 0.739 & 0.722 & 0.675 & \bfseries 0.578 & 0.651 & 0.673 & 0.766 & 0.598 & 0.651 & 0.763 & 0.704 & 0.619 & 0.642 & 0.701 \\
 & MAP-Jain & 0.915 & 0.908 & 1.048 & 0.914 & 0.894 & 0.897 & 0.987 & \bfseries 0.892 & 0.932 & 0.905 & 0.976 & 0.930 & 0.898 & 0.923 & 0.974 & 0.897 \\
 & MAP-Ent & 0.704 & 0.805 & 0.914 & 0.703 & 0.687 & 0.804 & 0.889 & 0.686 & 0.704 & 0.795 & 0.859 & 0.703 & 0.681 & 0.794 & 0.868 & \bfseries 0.680 \\
 & MAP-Gini & 0.927 & 0.930 & 1.062 & 0.926 & 0.891 & 0.915 & 1.001 & \bfseries 0.889 & 0.947 & 0.924 & 0.985 & 0.945 & 0.903 & 0.933 & 0.986 & 0.901 \\
 & R-Jain & 0.774 & 0.704 & 0.927 & 0.773 & 0.748 & \bfseries 0.700 & 0.821 & 0.746 & 0.802 & 0.702 & 0.787 & 0.800 & 0.759 & 0.732 & 0.787 & 0.758 \\
 & R-Ent & 0.507 & 0.566 & 0.773 & 0.506 & 0.483 & 0.576 & 0.699 & \bfseries 0.482 & 0.521 & 0.555 & 0.636 & 0.519 & 0.484 & 0.563 & 0.651 & 0.483 \\
 & R-Gini & 0.788 & 0.733 & 0.943 & 0.787 & 0.745 & \bfseries 0.723 & 0.837 & 0.743 & 0.819 & 0.727 & 0.798 & 0.817 & 0.765 & 0.745 & 0.801 & 0.763 \\
 & NDCG-Jain & 0.823 & 0.793 & 0.977 & 0.822 & 0.798 & \bfseries 0.782 & 0.900 & 0.797 & 0.845 & 0.788 & 0.878 & 0.844 & 0.807 & 0.810 & 0.878 & 0.805 \\
 & NDCG-Ent & 0.579 & 0.673 & 0.833 & 0.578 & 0.558 & 0.673 & 0.791 & 0.557 & 0.586 & 0.660 & 0.745 & 0.584 & 0.556 & 0.661 & 0.758 & \bfseries 0.555 \\
 & NDCG-Gini & 0.837 & 0.818 & 0.992 & 0.835 & 0.795 & 0.802 & 0.914 & \bfseries 0.793 & 0.862 & 0.810 & 0.887 & 0.860 & 0.812 & 0.821 & 0.890 & 0.810 \\
\cline{1-18}
\multirow[c]{12}{*}{\rotatebox[origin=c]{90}{ML-10M}} & P-Jain & 1.072 & 1.067 & 1.082 & 1.075 & 1.047 & \bfseries 1.023 & 1.031 & 1.051 & 1.098 & 1.094 & 1.104 & 1.099 & 1.052 & 1.044 & 1.059 & 1.053 \\
 & P-Ent & 0.884 & 0.831 & 0.844 & 0.880 & 0.765 & \bfseries 0.749 & 0.766 & 0.764 & 1.023 & 0.974 & 0.970 & 1.010 & 0.801 & 0.771 & 0.791 & 0.791 \\
 & P-Gini & 1.080 & 1.075 & 1.088 & 1.081 & 1.042 & \bfseries 1.025 & 1.032 & 1.041 & 1.106 & 1.102 & 1.111 & 1.107 & 1.055 & 1.050 & 1.065 & 1.053 \\
 & MAP-Jain & 1.172 & 1.172 & 1.182 & 1.173 & 1.149 & \bfseries 1.134 & 1.135 & 1.150 & 1.193 & 1.192 & 1.201 & 1.194 & 1.155 & 1.153 & 1.161 & 1.154 \\
 & MAP-Ent & 0.995 & 0.955 & 0.962 & 0.991 & 0.893 & \bfseries 0.886 & 0.893 & 0.888 & 1.119 & 1.076 & 1.073 & 1.106 & 0.924 & 0.906 & 0.915 & 0.914 \\
 & MAP-Gini & 1.173 & 1.173 & 1.182 & 1.174 & 1.139 & \bfseries 1.130 & \bfseries 1.130 & 1.135 & 1.195 & 1.193 & 1.202 & 1.195 & 1.152 & 1.153 & 1.160 & 1.149 \\
 & R-Jain & 0.860 & 0.846 & 0.847 & 0.859 & 0.835 & 0.780 & \bfseries 0.768 & 0.829 & 0.868 & 0.866 & 0.867 & 0.867 & 0.847 & 0.813 & 0.810 & 0.844 \\
 & R-Ent & 0.653 & 0.570 & 0.563 & 0.643 & 0.492 & 0.423 & \bfseries 0.415 & 0.472 & 0.807 & 0.748 & 0.730 & 0.788 & 0.554 & 0.473 & 0.468 & 0.534 \\
 & R-Gini & 0.889 & 0.877 & 0.876 & 0.888 & 0.850 & 0.805 & \bfseries 0.793 & 0.838 & 0.900 & 0.897 & 0.898 & 0.899 & 0.872 & 0.843 & 0.840 & 0.866 \\
 & NDCG-Jain & 1.140 & 1.142 & 1.156 & 1.142 & 1.115 & \bfseries 1.107 & 1.114 & 1.118 & 1.168 & 1.166 & 1.180 & 1.169 & 1.119 & 1.125 & 1.138 & 1.119 \\
 & NDCG-Ent & 0.972 & 0.932 & 0.944 & 0.968 & 0.863 & 0.866 & 0.881 & \bfseries 0.860 & 1.103 & 1.060 & 1.062 & 1.091 & 0.894 & 0.884 & 0.901 & 0.885 \\
 & NDCG-Gini & 1.151 & 1.153 & 1.166 & 1.153 & 1.114 & \bfseries 1.112 & 1.119 & \bfseries 1.112 & 1.180 & 1.177 & 1.190 & 1.180 & 1.127 & 1.134 & 1.147 & 1.124 \\
\cline{1-18}
\multirow[c]{12}{*}{\rotatebox[origin=c]{90}{ML-20M}} & P-Jain & 1.014 & 1.015 & 1.033 & 1.016 & 0.998 & \bfseries 0.992 & 1.000 & 1.001 & 1.000 & 0.993 & 1.006 & 1.006 & 1.002 & 1.004 & 1.020 & 1.006 \\
 & P-Ent & 0.885 & 0.842 & 0.858 & 0.880 & 0.776 & \bfseries 0.761 & 0.773 & 0.771 & 0.775 & 0.765 & 0.780 & 0.774 & 0.807 & 0.783 & 0.800 & 0.801 \\
 & P-Gini & 1.056 & 1.057 & 1.073 & 1.057 & 1.031 & \bfseries 1.026 & 1.033 & 1.030 & 1.031 & 1.029 & 1.039 & 1.033 & 1.040 & 1.042 & 1.057 & 1.042 \\
 & MAP-Jain & 1.118 & 1.124 & 1.137 & 1.119 & \bfseries 1.103 & \bfseries 1.103 & 1.104 & \bfseries 1.103 & 1.107 & 1.105 & 1.111 & 1.109 & 1.106 & 1.114 & 1.123 & 1.108 \\
 & MAP-Ent & 0.996 & 0.963 & 0.973 & 0.991 & 0.900 & 0.894 & 0.898 & \bfseries 0.892 & 0.902 & 0.897 & 0.905 & 0.897 & 0.926 & 0.913 & 0.921 & 0.919 \\
 & MAP-Gini & 1.151 & 1.156 & 1.167 & 1.151 & 1.127 & 1.129 & 1.129 & \bfseries 1.124 & 1.130 & 1.131 & 1.136 & 1.129 & 1.135 & 1.143 & 1.151 & 1.135 \\
 & R-Jain & 0.775 & 0.768 & 0.769 & 0.775 & 0.757 & 0.723 & \bfseries 0.712 & 0.753 & 0.756 & 0.721 & 0.718 & 0.752 & 0.766 & 0.744 & 0.741 & 0.764 \\
 & R-Ent & 0.645 & 0.573 & 0.568 & 0.634 & 0.490 & 0.426 & \bfseries 0.413 & 0.471 & 0.483 & 0.426 & 0.421 & 0.465 & 0.542 & 0.471 & 0.462 & 0.523 \\
 & R-Gini & 0.836 & 0.829 & 0.829 & 0.836 & 0.808 & 0.778 & \bfseries 0.768 & 0.800 & 0.805 & 0.779 & 0.774 & 0.797 & 0.822 & 0.803 & 0.799 & 0.818 \\
 & NDCG-Jain & 1.081 & 1.089 & 1.107 & 1.083 & \bfseries 1.065 & 1.072 & 1.078 & 1.066 & 1.072 & 1.075 & 1.086 & 1.075 & 1.068 & 1.083 & 1.097 & 1.071 \\
 & NDCG-Ent & 0.971 & 0.939 & 0.955 & 0.966 & 0.871 & 0.873 & 0.883 & \bfseries 0.865 & 0.876 & 0.878 & 0.891 & 0.873 & 0.898 & 0.892 & 0.906 & 0.892 \\
 & NDCG-Gini & 1.121 & 1.129 & 1.144 & 1.122 & 1.097 & 1.105 & 1.110 & \bfseries 1.095 & 1.102 & 1.109 & 1.118 & 1.102 & 1.105 & 1.119 & 1.132 & 1.105 \\
\bottomrule

\end{tabular}}
\end{table*}
\begin{table*}[tbp]
\caption{$\downarrow$DPFR scores (computed with Manhattan distance) at $k=10$ of the recommender models for Lastfm, Amazon-lb, and QK-video, without and with re-ranking the top $k'=25$ items using Borda Count (BC), COMBMNZ (CM), and Greedy Substitution (GS). The best score per measure pair is in bold.}
\label{tab:dpfr-scores-manhattan}
\resizebox{0.95\textwidth}{!}{
\begin{tabular}{ll*{4}{r}|*{4}{r}|*{4}{r}|*{4}{r}}
\toprule
 &  model & \multicolumn{4}{c|}{ItemKNN} & \multicolumn{4}{c|}{BPR} & \multicolumn{4}{c|}{MultiVAE} & \multicolumn{4}{c}{NCL} \\ 
\midrule
 &  reranking & - & BC & CM & GS & - & BC & CM & GS & - & BC & CM & GS & - & BC & CM & GS \\
\midrule
\multirow[c]{12}{*}{\rotatebox[origin=c]{90}{Lastfm}} & P-Jain & 1.181 & 1.146 & 1.212 & 1.181 & 1.158 & 1.103 & 1.163 & 1.159 & 1.122 & \bfseries 1.030 & 1.097 & 1.118 & 1.128 & 1.038 & 1.093 & 1.127 \\
 & P-Ent & 0.827 & 0.714 & 0.764 & 0.799 & 0.800 & 0.711 & 0.756 & 0.774 & 0.706 & \bfseries 0.639 & 0.687 & 0.687 & 0.733 & 0.647 & 0.691 & 0.710 \\
 & P-Gini & 1.185 & 1.116 & 1.154 & 1.165 & 1.185 & 1.140 & 1.179 & 1.171 & 1.116 & \bfseries 1.039 & 1.073 & 1.095 & 1.141 & 1.060 & 1.094 & 1.119 \\
 & MAP-Jain & 1.476 & 1.468 & 1.508 & 1.474 & 1.455 & 1.423 & 1.457 & 1.450 & 1.420 & \bfseries 1.348 & 1.388 & 1.407 & 1.425 & 1.360 & 1.389 & 1.417 \\
 & MAP-Ent & 1.105 & 1.019 & 1.043 & 1.074 & 1.079 & 1.014 & 1.033 & 1.047 & 0.986 & \bfseries 0.941 & 0.960 & 0.958 & 1.012 & 0.952 & 0.970 & 0.982 \\
 & MAP-Gini & 1.467 & 1.424 & 1.437 & 1.444 & 1.469 & 1.447 & 1.461 & 1.449 & 1.401 & \bfseries 1.344 & 1.351 & 1.371 & 1.423 & 1.369 & 1.378 & 1.396 \\
 & R-Jain & 1.366 & 1.339 & 1.418 & 1.368 & 1.344 & 1.295 & 1.367 & 1.348 & 1.304 & \bfseries 1.227 & 1.306 & 1.306 & 1.309 & 1.230 & 1.301 & 1.312 \\
 & R-Ent & 0.998 & 0.892 & 0.956 & 0.972 & 0.971 & 0.889 & 0.945 & 0.948 & 0.874 & \bfseries 0.822 & 0.881 & 0.859 & 0.900 & 0.825 & 0.884 & 0.880 \\
 & R-Gini & 1.360 & 1.298 & 1.350 & 1.342 & 1.360 & 1.321 & 1.373 & 1.349 & 1.288 & \bfseries 1.225 & 1.271 & 1.272 & 1.311 & 1.242 & 1.292 & 1.294 \\
 & NDCG-Jain & 1.398 & 1.404 & 1.472 & 1.398 & 1.375 & 1.361 & 1.420 & 1.374 & 1.341 & \bfseries 1.295 & 1.362 & 1.335 & 1.342 & 1.299 & 1.356 & 1.339 \\
 & NDCG-Ent & 1.046 & 0.973 & 1.025 & 1.017 & 1.018 & 0.971 & 1.014 & 0.990 & 0.926 & 0.905 & 0.952 & \bfseries 0.904 & 0.948 & 0.909 & 0.955 & 0.923 \\
 & NDCG-Gini & 1.409 & 1.380 & 1.421 & 1.389 & 1.408 & 1.405 & 1.443 & 1.393 & 1.342 & \bfseries 1.310 & 1.344 & 1.318 & 1.361 & 1.328 & 1.364 & 1.338 \\
\cline{1-18}
\multirow[c]{12}{*}{\rotatebox[origin=c]{90}{Amazon-lb}} & P-Jain & 0.741 & \bfseries 0.468 & 0.584 & 0.689 & 0.793 & 0.584 & 0.656 & 0.757 & 0.978 & 0.893 & 0.918 & 0.969 & 0.987 & 0.917 & 0.935 & 0.981 \\
 & P-Ent & 0.470 & \bfseries 0.392 & 0.436 & 0.443 & 0.529 & 0.437 & 0.466 & 0.499 & 0.855 & 0.689 & 0.718 & 0.802 & 0.902 & 0.715 & 0.742 & 0.847 \\
 & P-Gini & 0.850 & \bfseries 0.680 & 0.786 & 0.798 & 0.935 & 0.789 & 0.848 & 0.892 & 1.135 & 1.071 & 1.088 & 1.115 & 1.146 & 1.086 & 1.099 & 1.128 \\
 & MAP-Jain & 1.349 & \bfseries 1.076 & 1.192 & 1.297 & 1.402 & 1.192 & 1.264 & 1.366 & 1.586 & 1.501 & 1.527 & 1.578 & 1.594 & 1.527 & 1.545 & 1.589 \\
 & MAP-Ent & 0.995 & \bfseries 0.917 & 0.960 & 0.968 & 1.054 & 0.961 & 0.989 & 1.025 & 1.380 & 1.214 & 1.243 & 1.328 & 1.425 & 1.241 & 1.269 & 1.371 \\
 & MAP-Gini & 1.384 & \bfseries 1.213 & 1.319 & 1.332 & 1.470 & 1.322 & 1.381 & 1.426 & 1.668 & 1.604 & 1.621 & 1.649 & 1.678 & 1.621 & 1.634 & 1.661 \\
 & R-Jain & 1.333 & \bfseries 1.064 & 1.182 & 1.281 & 1.389 & 1.178 & 1.249 & 1.353 & 1.573 & 1.489 & 1.513 & 1.563 & 1.580 & 1.513 & 1.535 & 1.575 \\
 & R-Ent & 0.981 & \bfseries 0.907 & 0.952 & 0.954 & 1.044 & 0.950 & 0.977 & 1.015 & 1.369 & 1.204 & 1.231 & 1.315 & 1.413 & 1.230 & 1.260 & 1.359 \\
 & R-Gini & 1.369 & \bfseries 1.203 & 1.311 & 1.317 & 1.459 & 1.310 & 1.367 & 1.415 & 1.657 & 1.594 & 1.609 & 1.635 & 1.665 & 1.609 & 1.625 & 1.649 \\
 & NDCG-Jain & 1.370 & \bfseries 1.100 & 1.216 & 1.318 & 1.427 & 1.216 & 1.288 & 1.391 & 1.608 & 1.526 & 1.551 & 1.599 & 1.616 & 1.551 & 1.570 & 1.611 \\
 & NDCG-Ent & 1.034 & \bfseries 0.959 & 1.002 & 1.007 & 1.097 & 1.003 & 1.031 & 1.068 & 1.419 & 1.257 & 1.285 & 1.366 & 1.465 & 1.283 & 1.311 & 1.410 \\
 & NDCG-Gini & 1.428 & \bfseries 1.260 & 1.367 & 1.376 & 1.518 & 1.370 & 1.428 & 1.474 & 1.713 & 1.653 & 1.669 & 1.693 & 1.723 & 1.668 & 1.682 & 1.706 \\
\cline{1-18}
\multirow[c]{12}{*}{\rotatebox[origin=c]{90}{QK-video}} & P-Jain & 0.748 & \bfseries 0.415 & 0.641 & 0.664 & 1.144 & 0.895 & 0.851 & 1.125 & 1.211 & 1.188 & 1.196 & 1.210 & 1.201 & 1.148 & 1.155 & 1.198 \\
 & P-Ent & 0.302 & \bfseries 0.255 & 0.297 & 0.286 & 0.473 & 0.344 & 0.331 & 0.438 & 0.808 & 0.673 & 0.686 & 0.771 & 0.718 & 0.561 & 0.557 & 0.677 \\
 & P-Gini & 0.700 & \bfseries 0.463 & 0.670 & 0.626 & 1.029 & 0.839 & 0.798 & 0.984 & 1.204 & 1.181 & 1.186 & 1.198 & 1.184 & 1.131 & 1.127 & 1.171 \\
 & MAP-Jain & 1.493 & \bfseries 1.160 & 1.386 & 1.408 & 1.883 & 1.639 & 1.595 & 1.863 & 1.950 & 1.930 & 1.939 & 1.949 & 1.938 & 1.891 & 1.899 & 1.935 \\
 & MAP-Ent & 1.044 & \bfseries 0.997 & 1.039 & 1.027 & 1.209 & 1.086 & 1.072 & 1.173 & 1.544 & 1.413 & 1.426 & 1.506 & 1.452 & 1.301 & 1.297 & 1.411 \\
 & MAP-Gini & 1.442 & \bfseries 1.205 & 1.413 & 1.368 & 1.765 & 1.580 & 1.539 & 1.720 & 1.940 & 1.921 & 1.926 & 1.933 & 1.919 & 1.871 & 1.867 & 1.905 \\
 & R-Jain & 1.482 & \bfseries 1.146 & 1.371 & 1.398 & 1.855 & 1.618 & 1.581 & 1.839 & 1.916 & 1.902 & 1.920 & 1.918 & 1.898 & 1.858 & 1.875 & 1.898 \\
 & R-Ent & 1.033 & \bfseries 0.983 & 1.024 & 1.017 & 1.182 & 1.064 & 1.058 & 1.150 & 1.510 & 1.384 & 1.407 & 1.475 & 1.412 & 1.268 & 1.273 & 1.373 \\
 & R-Gini & 1.432 & \bfseries 1.191 & 1.398 & 1.358 & 1.738 & 1.559 & 1.525 & 1.696 & 1.906 & 1.892 & 1.907 & 1.903 & 1.879 & 1.838 & 1.843 & 1.868 \\
 & NDCG-Jain & 1.493 & \bfseries 1.159 & 1.385 & 1.408 & 1.875 & 1.636 & 1.594 & 1.857 & 1.941 & 1.924 & 1.936 & 1.940 & 1.926 & 1.883 & 1.894 & 1.924 \\
 & NDCG-Ent & 1.044 & \bfseries 0.997 & 1.039 & 1.028 & 1.202 & 1.083 & 1.072 & 1.167 & 1.535 & 1.407 & 1.423 & 1.499 & 1.441 & 1.294 & 1.293 & 1.400 \\
 & NDCG-Gini & 1.443 & \bfseries 1.205 & 1.412 & 1.369 & 1.759 & 1.578 & 1.539 & 1.714 & 1.931 & 1.915 & 1.923 & 1.926 & 1.908 & 1.864 & 1.863 & 1.895 \\
\bottomrule
\end{tabular}}
\end{table*}

\begin{table*}[tbp]
\caption{$\downarrow$DPFR scores (computed with Manhttan distance) at $k=10$ of the recommender models for Jester and ML-*, without and with re-ranking the top $k'=25$ items using Borda Count (BC), COMBMNZ (CM), and Greedy Substitution (GS). The best score per measure pair is in bold.}
\label{tab:dpfr-scores2-manhattan}
\resizebox{0.95\textwidth}{!}{
\begin{tabular}{ll*{4}{r}|*{4}{r}|*{4}{r}|*{4}{r}}
\toprule
 &  model & \multicolumn{4}{c|}{ItemKNN} & \multicolumn{4}{c|}{BPR} & \multicolumn{4}{c|}{MultiVAE} & \multicolumn{4}{c}{NCL} \\ 
\midrule
 &  reranking & - & BC & CM & GS & - & BC & CM & GS & - & BC & CM & GS & - & BC & CM & GS \\
\midrule
\multirow[c]{12}{*}{\rotatebox[origin=c]{90}{Jester}} & P-Jain & 0.925 & 0.796 & 1.013 & 0.924 & 0.895 & \bfseries 0.775 & 0.892 & 0.894 & 0.956 & 0.796 & 0.899 & 0.954 & 0.909 & 0.841 & 0.883 & 0.907 \\
 & P-Ent & 0.567 & 0.499 & 0.674 & 0.565 & 0.519 & \bfseries 0.484 & 0.583 & 0.517 & 0.604 & 0.489 & 0.570 & 0.601 & 0.539 & 0.507 & 0.562 & 0.537 \\
 & P-Gini & 0.942 & 0.840 & 1.039 & 0.940 & 0.891 & \bfseries 0.813 & 0.920 & 0.889 & 0.975 & 0.833 & 0.915 & 0.973 & 0.915 & 0.859 & 0.905 & 0.913 \\
 & MAP-Jain & 1.294 & 1.235 & 1.443 & 1.292 & 1.264 & \bfseries 1.210 & 1.329 & 1.262 & 1.314 & 1.237 & 1.334 & 1.313 & 1.270 & 1.274 & 1.321 & 1.268 \\
 & MAP-Ent & 0.935 & 0.938 & 1.104 & 0.933 & 0.887 & 0.919 & 1.020 & \bfseries 0.885 & 0.962 & 0.929 & 1.004 & 0.959 & 0.900 & 0.939 & 1.000 & 0.897 \\
 & MAP-Gini & 1.311 & 1.279 & 1.469 & 1.309 & 1.260 & \bfseries 1.248 & 1.358 & 1.257 & 1.334 & 1.274 & 1.351 & 1.332 & 1.276 & 1.292 & 1.344 & 1.274 \\
 & R-Jain & 1.067 & 0.990 & 1.297 & 1.066 & 1.038 & \bfseries 0.979 & 1.137 & 1.036 & 1.088 & 0.990 & 1.106 & 1.087 & 1.045 & 1.035 & 1.100 & 1.043 \\
 & R-Ent & 0.708 & 0.693 & 0.958 & 0.706 & 0.661 & 0.687 & 0.827 & \bfseries 0.659 & 0.736 & 0.682 & 0.776 & 0.733 & 0.674 & 0.700 & 0.779 & 0.672 \\
 & R-Gini & 1.084 & 1.035 & 1.324 & 1.082 & 1.034 & \bfseries 1.016 & 1.166 & 1.032 & 1.108 & 1.027 & 1.123 & 1.106 & 1.051 & 1.053 & 1.123 & 1.049 \\
 & NDCG-Jain & 1.153 & 1.100 & 1.359 & 1.152 & 1.123 & \bfseries 1.078 & 1.230 & 1.121 & 1.173 & 1.098 & 1.218 & 1.171 & 1.129 & 1.137 & 1.209 & 1.127 \\
 & NDCG-Ent & 0.795 & 0.803 & 1.020 & 0.792 & 0.746 & 0.786 & 0.921 & \bfseries 0.744 & 0.821 & 0.791 & 0.888 & 0.818 & 0.759 & 0.802 & 0.888 & 0.757 \\
 & NDCG-Gini & 1.170 & 1.144 & 1.385 & 1.168 & 1.119 & \bfseries 1.115 & 1.258 & 1.116 & 1.192 & 1.135 & 1.234 & 1.190 & 1.135 & 1.154 & 1.232 & 1.133 \\
\cline{1-18}
\multirow[c]{12}{*}{\rotatebox[origin=c]{90}{ML-10M}} & P-Jain & 1.513 & 1.507 & 1.529 & 1.516 & 1.478 & \bfseries 1.447 & 1.457 & 1.485 & 1.551 & 1.544 & 1.559 & 1.553 & 1.483 & 1.475 & 1.498 & 1.485 \\
 & P-Ent & 1.246 & 1.159 & 1.171 & 1.240 & 1.048 & \bfseries 0.979 & 0.986 & 1.035 & 1.446 & 1.377 & 1.371 & 1.428 & 1.117 & 1.042 & 1.058 & 1.097 \\
 & P-Gini & 1.518 & 1.513 & 1.534 & 1.521 & 1.467 & \bfseries 1.449 & 1.459 & 1.468 & 1.558 & 1.551 & 1.565 & 1.559 & 1.484 & 1.481 & 1.504 & 1.481 \\
 & MAP-Jain & 1.657 & 1.657 & 1.670 & 1.659 & 1.624 & 1.597 & \bfseries 1.595 & 1.625 & 1.687 & 1.685 & 1.697 & 1.688 & 1.633 & 1.627 & 1.637 & 1.632 \\
 & MAP-Ent & 1.388 & 1.306 & 1.310 & 1.380 & 1.192 & 1.126 & \bfseries 1.121 & 1.172 & 1.580 & 1.515 & 1.506 & 1.560 & 1.264 & 1.191 & 1.194 & 1.242 \\
 & MAP-Gini & 1.658 & 1.659 & 1.672 & 1.660 & 1.610 & 1.595 & \bfseries 1.593 & 1.605 & 1.691 & 1.688 & 1.699 & 1.690 & 1.629 & 1.629 & 1.639 & 1.624 \\
 & R-Jain & 1.117 & 1.103 & 1.106 & 1.117 & 1.088 & 1.032 & \bfseries 1.023 & 1.084 & 1.127 & 1.124 & 1.128 & 1.126 & 1.101 & 1.067 & 1.069 & 1.098 \\
 & R-Ent & 0.883 & 0.787 & 0.781 & 0.873 & 0.691 & 0.597 & \bfseries 0.584 & 0.666 & 1.054 & 0.989 & 0.972 & 1.034 & 0.767 & 0.667 & 0.661 & 0.743 \\
 & R-Gini & 1.139 & 1.126 & 1.128 & 1.139 & 1.095 & 1.051 & \bfseries 1.042 & 1.084 & 1.151 & 1.148 & 1.152 & 1.150 & 1.118 & 1.091 & 1.092 & 1.112 \\
 & NDCG-Jain & 1.613 & 1.615 & 1.634 & 1.616 & 1.577 & \bfseries 1.560 & 1.567 & 1.580 & 1.652 & 1.648 & 1.668 & 1.653 & 1.583 & 1.588 & 1.606 & 1.583 \\
 & NDCG-Ent & 1.359 & 1.279 & 1.289 & 1.351 & 1.160 & \bfseries 1.104 & 1.108 & 1.142 & 1.559 & 1.494 & 1.492 & 1.540 & 1.229 & 1.168 & 1.178 & 1.208 \\
 & NDCG-Gini & 1.627 & 1.630 & 1.649 & 1.630 & 1.576 & \bfseries 1.571 & 1.578 & 1.573 & 1.668 & 1.664 & 1.683 & 1.669 & 1.593 & 1.604 & 1.621 & 1.589 \\
\cline{1-18}
\multirow[c]{12}{*}{\rotatebox[origin=c]{90}{ML-20M}} & P-Jain & 1.434 & 1.436 & 1.461 & 1.436 & 1.411 & \bfseries 1.401 & 1.410 & 1.416 & 1.415 & 1.403 & 1.419 & 1.422 & 1.416 & 1.419 & 1.441 & 1.423 \\
 & P-Ent & 1.249 & 1.177 & 1.194 & 1.240 & 1.065 & \bfseries 1.005 & 1.007 & 1.048 & 1.059 & 1.010 & 1.020 & 1.048 & 1.123 & 1.061 & 1.072 & 1.109 \\
 & P-Gini & 1.489 & 1.491 & 1.516 & 1.491 & 1.456 & \bfseries 1.451 & 1.461 & 1.455 & 1.456 & 1.455 & 1.470 & 1.460 & 1.467 & 1.473 & 1.494 & 1.470 \\
 & MAP-Jain & 1.577 & 1.583 & 1.599 & 1.578 & 1.553 & 1.546 & \bfseries 1.544 & 1.553 & 1.559 & 1.547 & 1.554 & 1.561 & 1.559 & 1.565 & 1.575 & 1.562 \\
 & MAP-Ent & 1.390 & 1.322 & 1.331 & 1.380 & 1.205 & 1.149 & \bfseries 1.140 & 1.184 & 1.202 & 1.153 & 1.153 & 1.185 & 1.265 & 1.206 & 1.205 & 1.246 \\
 & MAP-Gini & 1.627 & 1.634 & 1.649 & 1.628 & 1.593 & 1.592 & 1.591 & \bfseries 1.588 & 1.597 & 1.595 & 1.600 & 1.594 & 1.605 & 1.614 & 1.624 & 1.605 \\
 & R-Jain & 1.017 & 1.010 & 1.014 & 1.017 & 0.997 & 0.963 & \bfseries 0.952 & 0.994 & 0.997 & 0.961 & 0.960 & 0.993 & 1.007 & 0.985 & 0.985 & 1.005 \\
 & R-Ent & 0.868 & 0.786 & 0.782 & 0.856 & 0.686 & 0.602 & \bfseries 0.584 & 0.662 & 0.677 & 0.603 & 0.595 & 0.654 & 0.749 & 0.662 & 0.651 & 0.726 \\
 & R-Gini & 1.076 & 1.069 & 1.072 & 1.075 & 1.045 & 1.016 & \bfseries 1.006 & 1.037 & 1.043 & 1.017 & 1.014 & 1.035 & 1.060 & 1.042 & 1.042 & 1.056 \\
 & NDCG-Jain & 1.527 & 1.536 & 1.559 & 1.528 & \bfseries 1.502 & 1.505 & 1.510 & 1.504 & 1.511 & 1.508 & 1.521 & 1.515 & 1.508 & 1.523 & 1.541 & 1.511 \\
 & NDCG-Ent & 1.359 & 1.294 & 1.309 & 1.349 & 1.173 & 1.126 & \bfseries 1.124 & 1.154 & 1.173 & 1.132 & 1.139 & 1.158 & 1.232 & 1.183 & 1.189 & 1.215 \\
 & NDCG-Gini & 1.586 & 1.596 & 1.618 & 1.587 & 1.551 & 1.559 & 1.565 & \bfseries 1.548 & 1.558 & 1.564 & 1.576 & 1.558 & 1.562 & 1.581 & 1.598 & 1.563 \\
\bottomrule
\end{tabular}}
\end{table*}

\begin{table*}[tbp]
    \centering
    \caption{
    Range of agreement ($\tau$) between estimated PF and PF across 12 measure pairs, using the estimated PF with 3--12 points.}
    \label{tab:corr_est}
    \resizebox{0.72\textwidth}{!}{
\begin{tabular}{lrrrrrr}
\toprule
{\#pts} &      Lastfm &   Amazon-lb &    QK-video &      Jester &      ML-10M &      ML-20M \\
\midrule
3  &  0.78--1.00 &  0.98--1.00 &  1.00--1.00 &  1.00--1.00 &  0.97--1.00 &  0.75--1.00 \\
4  &  0.88--1.00 &  0.98--1.00 &  0.98--1.00 &  0.98--1.00 &  0.98--1.00 &  0.93--1.00 \\
5  &  0.78--1.00 &  0.98--1.00 &  1.00--1.00 &  1.00--1.00 &  0.97--1.00 &  0.92--1.00 \\
6  &  0.90--1.00 &  0.97--1.00 &  1.00--1.00 &  0.98--1.00 &  0.95--1.00 &  0.92--1.00 \\
7  &  0.88--1.00 &  1.00--1.00 &  1.00--1.00 &  1.00--1.00 &  0.98--1.00 &  0.93--1.00 \\
8  &  0.90--1.00 &  0.98--1.00 &  1.00--1.00 &  0.98--1.00 &  1.00--1.00 &  0.95--1.00 \\
9  &  0.98--1.00 &  1.00--1.00 &  1.00--1.00 &  1.00--1.00 &  0.97--1.00 &  0.98--1.00 \\
10 &  0.88--1.00 &  1.00--1.00 &  1.00--1.00 &  0.98--1.00 &  1.00--1.00 &  0.95--1.00 \\
11 &  0.92--1.00 &  1.00--1.00 &  1.00--1.00 &  1.00--1.00 &  0.98--1.00 &  0.97--1.00 \\
12 &  0.95--1.00 &  1.00--1.00 &  1.00--1.00 &  0.98--1.00 &  0.98--1.00 &  0.97--1.00 \\
\bottomrule
\end{tabular}}
\end{table*}

\section{Further Discussions}
\label{app:discussion}

\subsection{The Impact of Replacing Frequently Recommended Items}
In this work, we replace frequently recommended items in two cases: during the Pareto Frontier generation (PF) with the {\sc Oracle2Fair} algorithm (\Cref{ss:generation}) and as part of the fair re-rankers (\Cref{s:experiments}). In both cases, none of the replacements significantly affect the overall recommendation performance:

\subsubsection*{1. Replacement in the {\sc Oracle2Fair} algorithm.} The {\sc Oracle2Fair} algorithm is used to generate the recommendation lists whose scores make up the PF (not the scores from the model-generated recommendation lists, that we compare to a point in the PF). The replacements are done on lists separate from the model's recommendation lists. Therefore, the replacements in the {\sc Oracle2Fair} algorithm do not affect the recommendation performance based on relevance and fairness measures. 

\subsubsection*{2. Replacement in the fair re-rankers.} We look at recommendation performance based on NDCG (relevance) and Gini (fairness). In all six datasets, when comparing the best non-reranked model to its reranked versions (e.g., for Lastfm, it is NCL vs NCL-BC, NCL-CM, and NCL-GS), the decrease in NDCG is not more than 0.26, and the decrease is even below 0.15 in all datasets excluding Jester (\Cref{tab:base-rerank-all-1,tab:base-rerank-all-2}). For Jester, the 0.26 decrease in NDCG is exchanged for an improvement in Gini by 0.218. Hence, we believe that the impact of item replacement is reasonable.

\subsection{Using Rank-based Fairness Measures}

Suppose we have two \fair{} measures, e.g., set-based Gini (Gini) and rank-based Gini (Gini-w). In \cite{Rampisela2024EvaluationStudy}, the absolute scores of Gini and Gini-w do not differ considerably and the two measures correlate strongly with Kendall's $\tau \geq 0.9$. Hence, generating the PF with rank-based fairness measures such as Gini-w is not expected to result in significantly different conclusions from the set-based version.

\end{document}